\documentclass[aps,prd,twocolumn,epsf,groupaddress,nofootinbib]{revtex4}
\usepackage{bm}
\usepackage{epsfig}
\usepackage{amsfonts,amsmath,amsthm,amssymb}
\usepackage{latexsym}
\usepackage{mathrsfs}
\usepackage{mathcomp}
\usepackage{dsfont}
\usepackage{amsbsy}

\newcommand{\postscript}[2]{\setlength{\epsfxsize}{#2\hsize}
   \centerline{\epsfbox{#1}}}

\usepackage[usenames,dvipsnames]{xcolor}
\definecolor{orange}{cmyk}{0,0.5,1,0}
\definecolor{rossoCP3}{cmyk}{0,.88,.77,.40}
\definecolor{graa}{rgb}{0.8,0.8,0.8}
\definecolor{blaa}{rgb}{0.2,0.2,0.6}

\begin{document}

\title{\color{rossoCP3} Symmetries and the fundamental forces of Nature}

\author{Luis A. Anchordoqui}
\affiliation{Department of Physics and Astronomy,  Lehman College, City University of New York, NY 10468, USA \\
Department of Physics, Graduate Center, City University  of New York, 365 Fifth Avenue, NY 10016, USA\\
Department of Astrophysics, American Museum of Natural History, Central Park West  79 St., NY 10024, USA
}

\date{Fall 2015}
\begin{abstract}
 \noindent  
 Lecture script of a one-semester course that aims to develop an understanding and appreciation of fundemental concepts in modern physics for students who are comfortable with calculus. This document contains the first six lessons that will walk you through the special theory of relativity.
\end{abstract}
\maketitle

\section{Galilean Ivariance}

Almost anyone who has awaited to depart from a pier and has had the boat slowly start to leave has had the sensation that it is the pier that has moved away. This simple physiological phenomenon has its basis in a generic law of physics, which was first expressed by Galileo and hence is known as Galilean invariance~\cite{Galileo}. It is, perhaps, the most stunning and far reaching of all of the laws of physics. It is impossible to over emphasize its importance; it is the basis of our understanding of motion in spacetime. The simplest statement of the law is that there is no experiment which can be implemented to measure a uniform velocity. Since we can only know what can be measured, we can never know how fast we are moving. There is no speedometer on the starship Enterprise.

Stated this boldly, the idea is very counter to our experience. This is because what we generally measure  is not a velocity in space, but is our velocity relative to Earth. Relative velocities are detectable. We note the amount of street that passes below our car, or feel the flow of the air that moves over our face and infer a speed but we do not know how fast the Earth is moving and thus do not know what our absolute velocity is. We do know that the Earth moves around the Sun and thus we can find our velocity relative to the Sun. We know that the Sun is moving in our Galaxy and even that our Galaxy is moving relative to other nearby galaxies and thus can determine our velocity relative to the local cluster of galaxies. We are also able to infer our velocity relative to the place that we occupied in the early universe, our motion relative the cosmic microwave background (CMB)~\cite{Penzias:1965wn}, but again we cannot know whether that place had a velocity.\footnote{The CMB is an all-pervasive, blackbody, background light from the Big Bang that has cooled to a present temperature of  2.726~K.}  

The inability to detect velocity is one of the most mysterious and counter intuitive concepts that has ever been layed down. Consider a remote and empty part of the universe. Without stars or galaxies around there are no discernible forces and therefore a released particle moves in a straight line with constant velocity. This is one of Newton's laws~\cite{Newton:1687} and was his way of enunciating Galilean invariance. From now on we will assume that this empty region is ``space.'' We envision this space as the stable background structure introduced by Newton against which motion takes place. Nowadays, it is generally easy to convince someone that this space obeys the Copernican principle; namely, it is not centered on some special place like the Earth~\cite{Copernicus}. It is also not so difficult to convince someone that this concept could be extended to the general Copernican principle: in an empty universe there is no special place that could be called the center. This viewpoint, that there is a structure called space which is stable and has no special places in it, is better stated as the fact the  universe is homogeneous. Expressed in the same way that the statement of Galilean invariance above, we can say that there is no experiment that can be performed in space that can distinguish one place from another. This is the definition of a homogeneous space. It should be obvious that if you cannot distinguish between places that you cannot have a center or a boundary. These are special places and this is contrary to the viewpoint that all places are the same.

\begin{figure}[tbp] \postscript{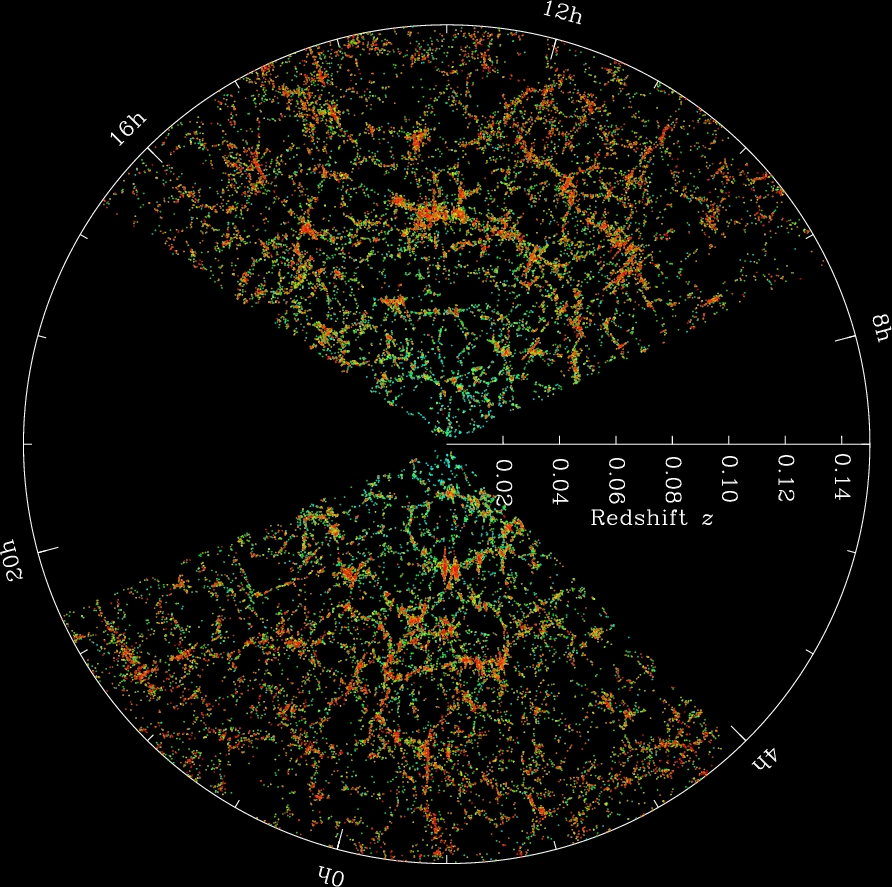}{0.85} 
\caption{Slices through the Sloan Digital Sky Survey (SDSS) 3-dimensional map of the distribution of galaxies~\cite{Ahn:2012fh}. Earth is at the center, and each point represents a galaxy, typically containing about 100 billion stars. Galaxies are colored according to the ages of their stars, with the redder, more strongly clustered points showing galaxies that are made of older stars. The outer circle is at a distance of two billion light years. The region between the wedges was not mapped by the SDSS because dust in our own Galaxy obscures the view of the distant universe in these directions.} 
\label{sdss}
\end{figure}

\begin{figure}[tbp] 
\postscript{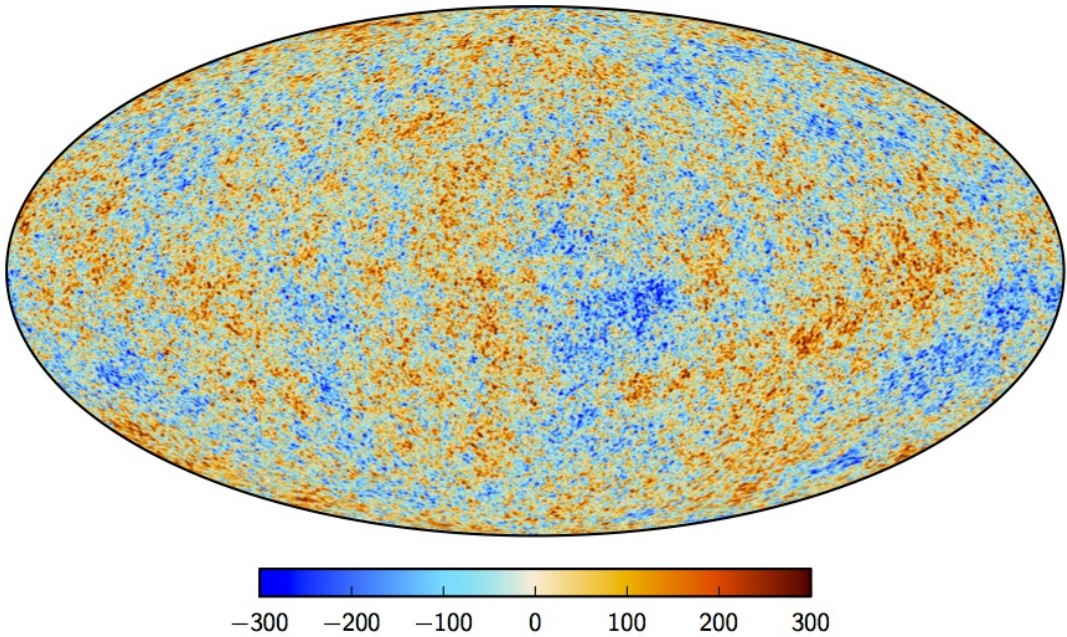}{0.85} 
\caption{
The CMB over the entire sky, color-coded to represent differences in temperature from the average 2.726~K: the color scale ranges from $+300~\mu{\rm K}$ (red) to $300~\mu{\rm K}$ (dark blue), representing slightly hotter and colder spots (and also variations in density.) Results are from the WMAP satellite~\cite{Bennett:2012zja} and the Planck mission~\cite{Adam:2015rua}.}
\label{wmap-planck}
\end{figure}

It may look that the assumptions made herein about the nature of space are so obvious that the universe must obey them. However, we know this is never the case and one has to  test any hypothesis. Of course one may argue that this in not a hypothesis that is testable since we cannot be anywhere other than where we are. The best test of this hypothesis is that we find that the laws of physics as we propound them here on Earth are found to be applicable everywhere, including  distant space. Stars in remote galaxies evolve in the same way as nearby stars.   We can also look at distributions of matter such as galaxies. Again, there is no indication that the universe is not homogeneous; see Fig.~\ref{sdss}. A related concept is isotropy. This is the idea that space is the same in all directions. This hypothesis has been tested very precisely by the observed distribution of the CMB; see Fig.~\ref{wmap-planck}.

\begin{figure}[tbp]
\postscript{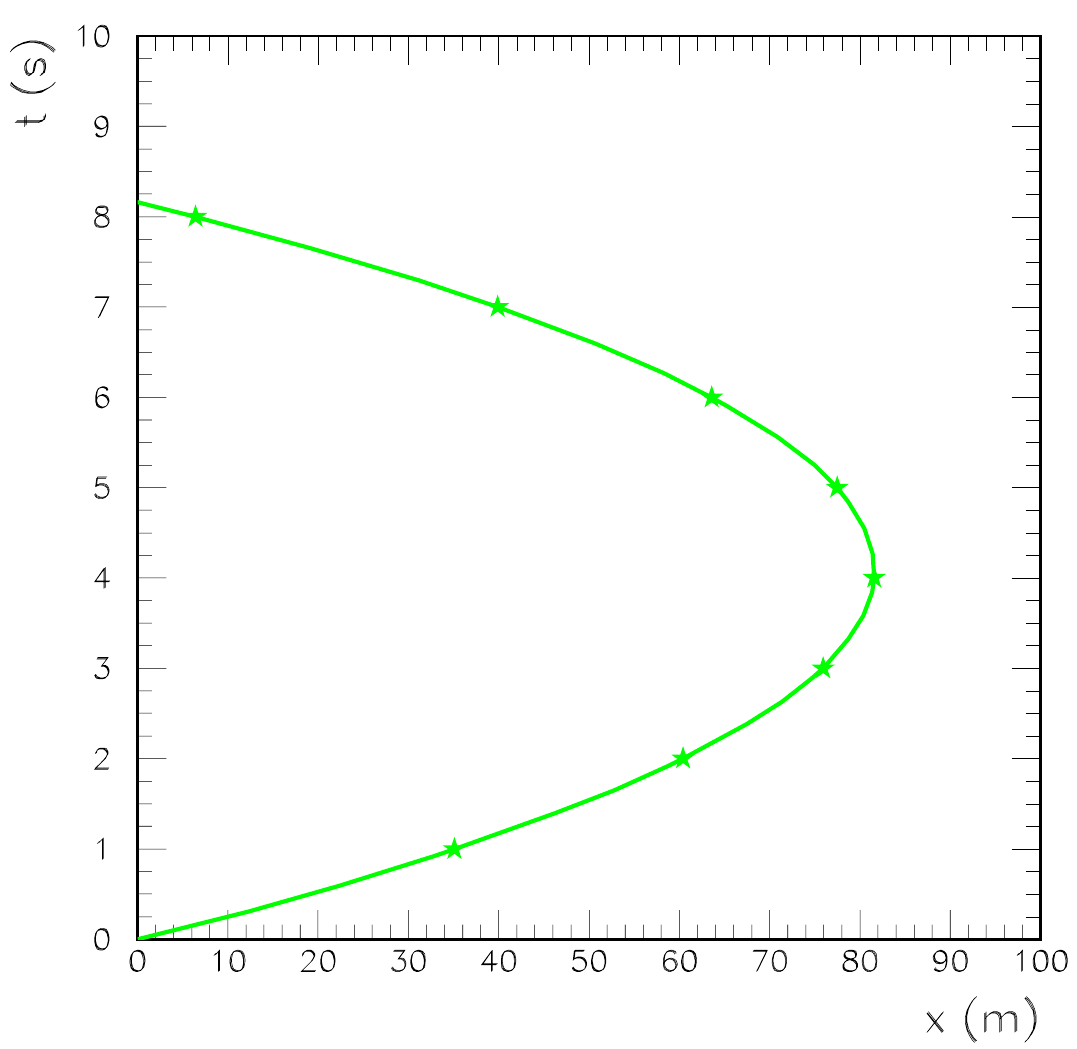}{0.85}
\caption{Spacetime diagram of a freely falling object, lanunched upward at 40~m/s.}
\label{fig:f1}
\end{figure}

In the classical world we describe the state of a system by the motion of its particles. As a point particle moves around, we can use Cartesian coordinates $(x, y, z)$ to describe where the particle is at any given time $t$. Equipping ourselves with a notebook, we determine the motion of a particle by writing down the $(x,y,z)$ components of the particle's position and the time on our watch $t$ at which the particle was in that position. For example, say we are only interested in the $x$-component of its position as it moved around. We would then make a two column table listing the pairs of times and positions, and this table would give us information on the motion of the particle. Note that the position $x$ and the time $t$ are on an equal footing, and hence it takes a set of pairs of values $(x,t)$ to describe the $x$-component of the particle's trajectory. In Fig.~\ref{fig:f1} we show a graph of $t$ versus $x$ (it is conventional to display $x$ on the horizontal axis). This is our first spacetime diagram. Each point on the line represents an ``event'' in the history of the particle's motion. Taken all together, the sequence of events trace out a continuous curve, called a ``worldline'', with a single value of $x$ for each $t$. The velocity of the particle is $dx/dt$, which is one over the gradient of the line in our diagram, so steeper lines represent slower motion, and stationary particles have vertical worldlines.

In our construction of the spacetime diagram we made a tacit assumption about the observer making the measurements. Simplistically, we might claim that the observer is at rest. However, we have seen that the concept of absolute rest is pretty meaningless. It is more useful to specify whether the observer is accelerating or not. Indeed, we can make measurements to ascertain if some object is accelerating, since the acceleration of massive bodies originates in forces that can be measured. Therefore, any observer can make an experiment to see whether or not he is accelerating. A non–accelerating observer is called an inertial observer. Throughout this course all our observers will be inertial.

Consider the case where one observer is moving at a constant velocity with respect to another. Neither observer is accelerating, so both are inertial. If we now impose Galilean invariance, each must have the same rules of physics and, thus, observe a universe that is homogeneous and isotropic. Yet, they are moving toward each other. It is not intuitive that space and time can be constructed consistently in this way but they are. Each observer, one  making measurements with $(x, y, z)$ and $t$ and the other with $(x', y', z')$ and $t'$, will conclude that the universe is homogeneous and isotropic. Not only that, but there is no experiment that they can perform which could yield a different result. 

It should be clear that although the different  observers must have the same results for any experiment, they will each describe the other’s experiment differently. For example,  the coordinates of events seen by the two observers are  different. How do the coordinates of the same event written down by the two different observers differ? It is instructive to consider an example. Imagine that our two observers are a tourist on top of {\it Ponte di Rialto} and a gondolieri passing under it at velocity $v$ in a direction aligned with the $x$-axis of the tourist on the bridge. We will assume for now that the two observers have identical synchronized watches, and that at $t=0$, Vinnie the gondolieri is directly under the bridge, so that the origins of the coordinate systems of the two observers coincide at time zero. Let us denote the coordinates of the events seen by Brittany on the bridge by $t$ and $x$, and the components of the same event seen by Vinnie as $t'$ and $x'$. It is common practice to refer to these two observers as the unprimed observer (Brittany) and the primed observer (Vinnie). In this case the primed observer is moving to the right, in the direction of increasing $x$, with respect to the the unprimed observer.

\begin{figure}[tbp] \postscript{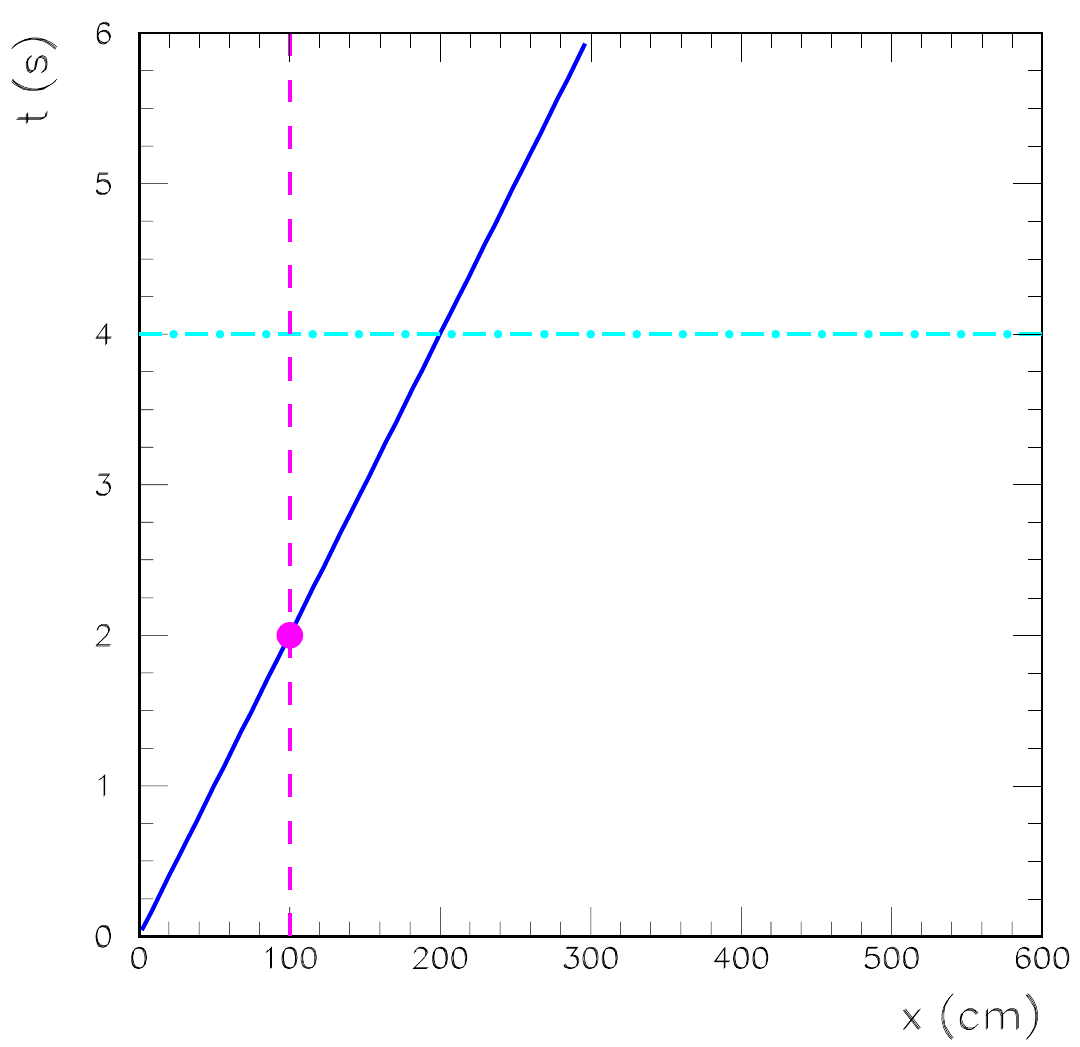}{0.85} 
\caption{The worldline of the bottle of champagne in the coordinate system of the unprimed observer on the bridge.}
\label{fig:f2}
\end{figure}

Suppose there is a bottle of champagne positioned  next to Vinnie's sit. What is the worldline of this bottle for the two observers? For Vinnie, the bottle is not moving. Thus, we can conclude that in his coordinate system the bottle is at the origin at all times, so that $x'= 0$ for all $t'$. What about the worldline of the same bottle in the coordinate system of the unprimed observer? For Brittany, the bottle is moving in the direction of increasing $x$ at velocity $v$, so that $x = vt$. How then are $x$ and $t$ related to $x'$ and $t'$? For a start, we decided that both observers have identical synchronised watches, so $t = t'$. Now, to make $x' = 0$ consistent with $x = vt$ we must have $x' = x-vt$. Finally, we take the location of Brittany at $z=0$, that is we neglect the height of the bridge so that the origins of the two coordinate systems coincide at $t = 0$. Because there is no motion of the two observers parallel to the $y$ or $z$ axes subsequently, the different components of the bottle worldline in the two coordinate systems are related through the following relations:
\begin{eqnarray}
t' &=& t \,,  \nonumber\\
x' &=&  x -vt \,, \nonumber  \\
y' &=& y \,, \nonumber \\
z' &=& z\, . 
\label{eq:Galileo}
\end{eqnarray}
These are the Galilean transformations, the rules that indicate how to translate one of the observer's observations to the other observer's observations.  In Fig.~\ref{fig:f2} we show a spacetime diagram for the motion of the bottle in the coordinate system of the unprimed observer. The worldline has a constant slope equal to $1/v$, where $v$ is the gondola's velocity, and it passes through the origin since we decided that Vinnie and the bottle of champagne passed under the bridge at time $t = 0$. For definiteness, let us suppose that the gondola is moving at $v = 50~{\rm cm/ s}$, and consider its motion for 6 seconds after it passes under the bridge.

Though we have drawn these kinds of plots a lot of times, it is instructive to think more about them now. Consider first the set of points for which $x = 100~{\rm cm}$. All these points lie on dashed vertical line passing through the $x$-axis. Because the gondola is only 100~m after the bridge at a single instant, the worldline intersects this vertical line once, so there is a single event where the gondola is 100~m after the bridge, which is indicated by a pink circle on the plot. Let us now  make this same argument algebraically. The vertical dotted line has the equation $x = 100$. To figure out the equation for the worldline of the gondola we begin with the standard equation for a straight line in an $x$-$y$ plane, $y = mx+c$, where $m$ is the gradient of the line, and $c$ is the intercept with the $y$ axis. Next, we relate to the current set of axes, where $y$ is replaced by $t$, and the gradient is $1/v$. Finally, the $y$-axis intercept becomes the $t$-axis intercept, which is zero because we have decided to define zero seconds as the time when Vinnie passes under the bridge. Therefore the equation of the worldline of the gondola is $t= x/50$. To find out where this line intersects the vertical dashed line, substitute in the equation of this line, which is $x = 100~{\rm cm}$, and you get $t = 100/50~{\rm s}$, or $t = 2~{\rm s}$.  

\begin{figure}[tbp]
\postscript{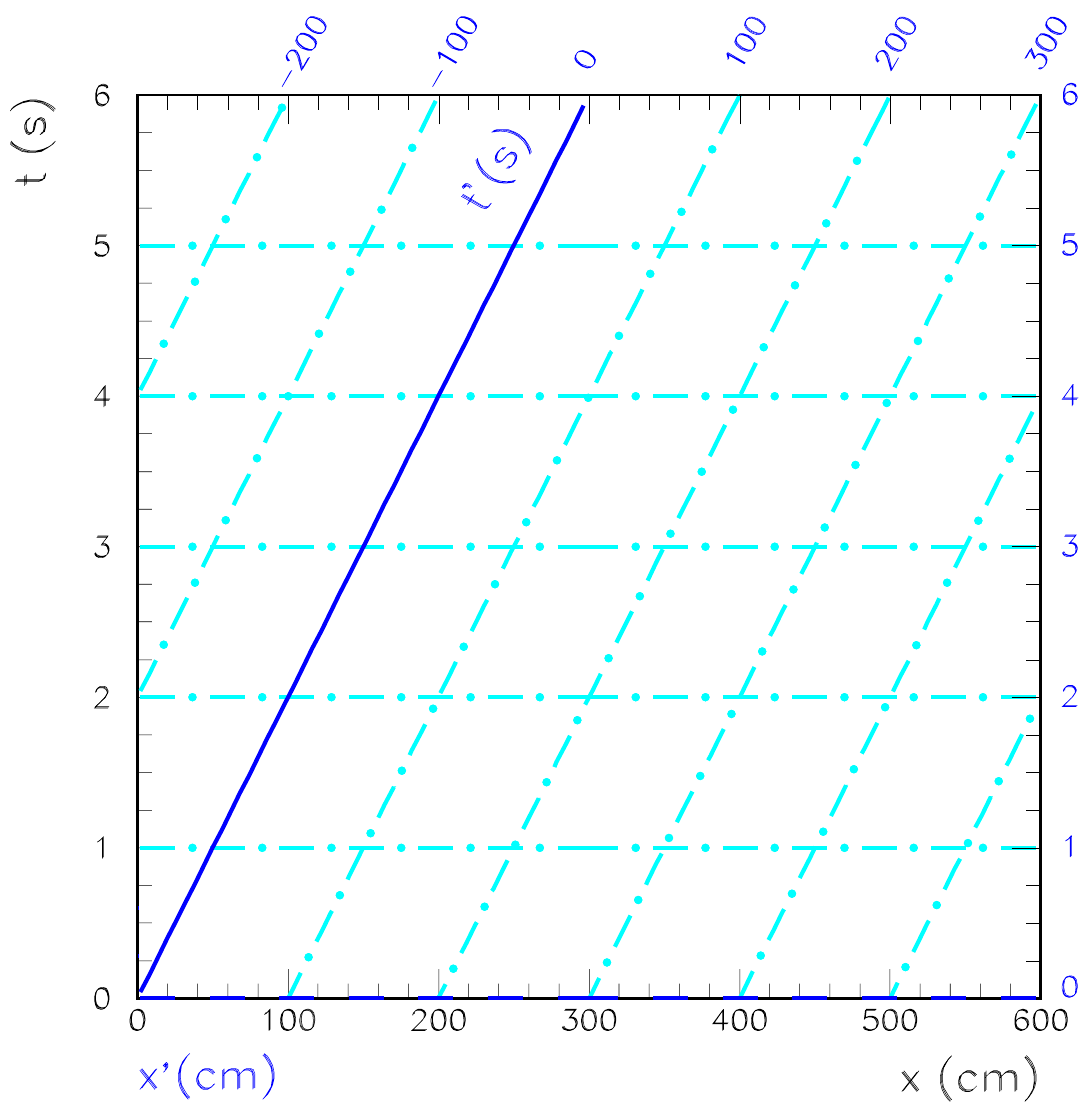}{0.85}
\caption{Spacetime  diagram for the moving gondola in the coordinate system of the observer on the bridge, with lines of constant $t'$ and lines of constant $x'$ overlaid.}
\label{fig:f3}
\end{figure}

To relate this spacetime diagram to what Vinnie sees,  we have to figure out how his coordinate system overlays onto the spacetime diagram of Brittany. This is easy, but probably unfamiliar. What we would like is a grid to overlay on the spacetime diagram we have drawn, from which we can read off the coordinates of events along the worldline, as they would be observed by Vinnie. Let us figure out how to draw grid lines of constant time first. In the unprimed coordinates, these lines are horizontal. For example, we have drawn in Fig.~\ref{fig:f2} a dot-dashed horizontal line that represents the set of events for which $t = 4~{\rm s}$. But from the Galilean transformations, we know that $t' = t$, so that this horizontal line represents also the set of events for which $t' = 4~{\rm s}$. Therefore,  the lines of constant $t'$ are coincident with the lines of constant $t$. Next, let us figure out where the $x'$ axis is. The $x$ axis is coincident with the line of constant $t = 0$. Similiarly, the $x'$ axis will be coincident with the line of constant $t' = 0$, which is coincident with the line of constant $t = 0$. Therefore the $x$ axis and the $x'$ axis lie on top of each other. We can therefore draw half of the grid we require, the lines of constant $t'$. Now, let us repeat this exercise to work out the lines of constant $x'$. Suppose we want the line coincident with all events where $x' = 100~{\rm cm}$. What does this line look like in the coordinate system of Brittany? Just substitute $x' = 100~{\rm cm}$  in (\ref{eq:Galileo}) to obtain $100~{\rm cm} = x-vt$, or $t = x/v-100~{\rm cm}/v$. For $v = 50~{\rm cm/s}$, this becomes $t = (x/50 - 2)~{\rm s}$. This is a straight line with gradient $1/(50~{\rm cm/s})$ and $t$-axis intercept $-2$. We can repeat this exercise for all other lines of constant $x'$, and you get the grid lines representing constant $x'$ that we require. Note that they are not perpendicular to the lines of constant $t'$ that we already found. The entire grid of the coordinate system of Vinnie, overlaid on the coordinate system of Brittany is shown in Fig.~\ref{fig:f3}. Note that Vinnie's coordinate grid is skewed with respect to Brittany's coordinate system. The worldline of the bottle of champagne coincides with the line of constant $x' = 0$, so we have found out the hard way that in the coordinate system of Vinnie where the bottle is next to him, it stays fixed in space. 

Now suppose two events are separated by a distance $dx$ and a time interval
$dt$ as measured by the unprimed observer. It follows from (\ref{eq:Galileo}) that the corresponding displacement $dx'$ measured by the prime observer is given by $dx' = dx - v dt$. Because $dt = dt'$ it follows that
\begin{equation}
\frac{dx'}{dt'} = \frac{dx}{dt} - v , 
\end{equation}
or 
\begin{equation}
u'_x = u_x - v \,,
\end{equation}
where $u_x$ and $u'_x$ are the instantaneous velocities of the object relative to Brittany and Vinnie, respectively. This result, which is called the Galilean addition law for velocities (or Galilean velocity transformation), is used in everyday observations and is consistent with our intuitive notions of time and space. \\

{\bf EXERCISE 1.1}~A brilliant aid in understanding the bottom line of rotational invariance is the {\it surveyors parable} introduced by Taylor and Wheeler~\cite{Taylor}. Suppose a town has daytime surveyors, who determine north and east with a compass, and nightime surveyors, who use the North Star. These notions differ, of course, since the magnetic north is not the direction to the North Pole. Suppose, further, that both groups measure north/south distances in miles and east/west distances in meters, with both being measured from the town center. How does one go about comparing the measurements of the two groups?\\

{\bf EXERCISE 1.2} Vinnie, whose speed in still water is 50~cm/s, must cross a 50~m wide canal and arrive at a point 20~m upstream from where it starts to pick up Brittany. To do so, he must head the gondola $45^\circ$ upstream angle. What is the speed of the canal's current?

\section{Principle of Least Time}

We begin our study of modern physics by examining the phenomena associated with light. Although the phenomena of light are among the oldest examined and there are theories of light that must go back to the first humans, light is particularly interesting from the modern perspective because of the central role that it has played in the development of our ideas, particularly quantum mechanics and relativity.

In the 1660's, Fermat proposed that light travels between two points over the path that is the least travel time of all the possible paths~\cite{Mahoney}. This seems a prescription that anyone can follow. Indeed, you have followed it many times when you pick a travel route between two cities. What is the best way to go between Boston and New York? You take a map with all the roads indicated on it. You classify all routes. On any route, you divide the trip into segments and then estimate your speed in each segment. From the speed and the length of the segment, you can calculate the time for that segment and then you add up the time for each segment to get a total:
\begin{equation}
T( {\rm route}) = \sum_{\rm segments} \Delta t_i = \sum_{\rm segments} \frac{\Delta r_i}{v_i} \,,
\end{equation}
where $\Delta t_i$ is the time in each segment labeled $i$, $\Delta r_i$ is the length of that segment $i$, and $v_i$ is the speed in that segment. You somehow make an ordered list of routes and repeat this process for all routes. Once you have $T({\rm route})$ for all routes, you look down the list of travel times and select the one with the least time. That is the route that you take if you want the least time. Likewise, if you know the speed of light at every place when light goes between two points, you can apply this same procedure to find the routes in space through which the light travels. Is it really this simple? Note that in contrast to our highway problem, there is an infinite number of paths. The problem of making sure that you have all paths is a complex one, and we will postpone a detail discussion for later.

The path that nature chooses for the light is based on a global measure -- the total time of travel of the path. Generally, the idea is that, if we can assume that the extremum is reached smoothly in the very rich path space, then paths which differ slightly have roughly the same value. In particular, the requirement that the global measure has almost the same value for two paths which are the same everywhere except at an isolated point for which the deviation of the path is small implies a condition that constrains the effects at that point. This constraint is a local statement on the path development. This result is intuitive from our experience in finding least time paths for travel. The least time path for a trip is always made up of segments that are themselves the least time path between the points at the ends of that segment.

Another interesting observation is that, although the word time is an important part of the formulation of Fermat's principle, there is no real evolution of the system.  The time in this approach is just some global measure on path space.  This observation is especially relevant when we realize that, at the time of Fermat's formulation, the speed of light had not been measured. Actually, at the time it was not clear whether or not light even had a velocity. However, it is tempted to speculate that Fermat choose time as the measure, because he knew that there were circumstances in which length does not work.  Instead he formulated a global measure which weights each path segment with the inverse of velocity, and then predicted that, if it could be measured, we would find that light travels slower (i.e., with higher inverse velocity) in dense media.

We can now ask ourselves: what are the paths of light in a homogeneous medium? A homogeneous medium is one in which every point is the same. In particular, the speed of light must be the same at every point
\begin{equation}
T ({\rm path})  =  \sum_{\rm segments}^{\rm path} \frac{\Delta r_i}{v_i} 
 =  \frac{1}{v} \sum_{\rm segments}^{\rm path} \Delta r_i \, .
\end{equation}
Here $v = c/n$, where $c$ is the speed of light in vacuum and $n$ the index of refraction of the medium.  Since $\sum_{\rm segments}^{\rm path} \Delta r_i$ is the definition of the length of the path, we see that the time for any path is proportional to the length of the path. Thus the least time path is the shortest-length path which, of course, is the straight line path.

\begin{figure}[tbp]
\postscript{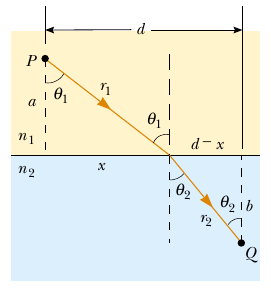}{0.85}
\caption{Geometry for deriving Snell's law of refraction using Fermat's principle~\cite{Serway:14}.}
\label{fig:fermat}
\end{figure}

Fermat's principle successfully accounted for all the known phenomena related to light in the 1660's, basically reflection and refraction. Refraction is the phenomenon that occurs when light passes through a medium that has a varying speed for light. In this case, the ray bends. As the simplest case, chose a system of two media that are themselves homogeneous, separated by a planar interface, and place the two end-points in the different media. Both media are homogeneous, but they have a different speed for light called $v_1$ in media 1 and $v_2$ in media 2.  Our first problem is to determine how to discuss the paths that connect the two points. There are an infinity of them. Physical intuition tells us though that the least time paths in a homogeneous medium must be straight lines and thus the path with the least time overall must be among the paths that are straight within either of the two media and kinked at the interface. A path that is curved in one of the media would clearly be a longer time path than the one with the same start point and hitting the other media at the same point and then traveling in the second media. This is an example of how a global rule does have some local content. This ability to reduce the path space to kinked straight line segments is an important reduction in the nature of the problem. With this reduction in the size of the path space, we can label the paths with the distance of the kink position from the place at which the path would meet the interface. Two things have been accomplished. We now have an ordering for the family of paths that we wish to investigate. Even more significantly, we have reduced the path space to one that can be mapped onto the real line. In this case, we are labeling the paths with the parameter $x$, see Fig.~\ref{fig:fermat}. Remember that functions are mappings of the real line onto the real line. This then gives us access to all the usual tools of mathematics.

Once the path has been reduced to two straight line segments, it is easy to find the least time path. In this example for simplicity of analysis, suppose that a light ray is to travel from point $P$ in medium 1 to point $Q$ in medium 2, where $P$ and $Q$ are at perpendicular distances $a$ and $b$, respectively, from the interface. The speed of light is $c/n_1$ in medium 1 and $c/n_2$ in medium 2. Using the geometry of Fig.~\ref{fig:fermat} we see that the time at which the ray arrives at $Q$ is
\begin{equation}
T (x) = \frac{r_1}{v_1} + \frac{r_2}{v_2} 
 =  \frac{\sqrt{a^2 + x^2}}{c / n_1} + \frac{\sqrt{b^2 + (d-x)^2}}{c/n_2} \, .
\end{equation}
The least time path is the one that has the minimum value for $T (x)$ for all $x$.  To obtain the value of $x$ for which $T$ has its minimum value, we take the derivative of $T$ with respect to $x$ and set the derivative equal to zero
\begin{equation}
\frac{dT}{dx} = \frac{n_1 x}{c (a^2 +x^2)^{1/2}} - \frac{n_2 (d-x)}{c [ b^2 + (d-x)^2]^{1/2} } = 0 \, .
\end{equation}
This yields
\begin{equation}
\frac{n_1 x}{(a^2 + x^2)^{1/2}} = \frac{n_2 (d-x)}{[b^2 + (d-x)^2]^{1/2}} \, .
\label{fer1}
\end{equation}
From Fig.~\ref{fig:fermat}
\begin{equation}
\sin \theta_1 = \frac{x}{(a^2 + x^2)^{1/2}}
\label{fer2}
\end{equation}
and
\begin{equation}
\sin \theta_2 = \frac{d-x}{[b^2 + (d-x)^2]^{1/2}} \, .
\label{fer3}
\end{equation}
Substituting (\ref{fer2}) and (\ref{fer3}) into (\ref{fer1}) we find that
\begin{equation}
n_1 \sin \theta_1 = n_2 \sin \theta_2 \,,
\end{equation}
which is Snell's law of refraction~\cite{Snell}.

In our articulation of Fermat's Principle, we casually assumed that it made sense to use the phrase ``all possible paths'' between two points. In a normal space, that's a lot of paths. To start with does it even make sense to identify ``all paths.'' If you think about it, it means that somehow you produce an ordering so that you can go through the lists to examine all possible cases. An ordering is mapping of the paths onto an ordered set. Without much thought, it should be clear that there are a lot of paths -- an infinity. Are there too many paths to order them like the integers? Two common examples of large sets are the integers for a discrete but infinite set or the points on a line for an infinite but continuous set. The counting of infinite sets is a subtle issue. There are as many integers as there are odd numbers. That’s because they can be ordered together -- put into a one to one correspondence.

How do you determine the number of paths? Do you count them, or do you order them? Counting is a process of matching the elements of two sets. For the case at hand we have the set of paths and a given set whose properties are better understood. The smallest of the standard sets of choice is the discrete infinite set of integers numbers $\mathbb{Z}$. Sets that have the same number of elements than $\mathbb{Z}$ are relatively nice to deal with and once an identification with the numbers is established the elements can be manipulated like numbers. Sets of this size are said to be in the class $\aleph_0$. Anytime that you make a table, you are making a mapping between the set of integers and your set of objects that enter the table.  In order to use the tools of analysis you need to deal with a system that has the right number of members. Functions are mappings of the real line onto the real line. The real line is, in fact, on example of the next larger infinite set, $\aleph_1$. It is bigger than the number of integers.  It is relatively straightforward to convince yourself that the number of paths is larger than the number of points on the real line. This makes for a problem. Most of what we can do in analysis is dealt with through functions. Thus, our manipulations with paths cannot be considered functions and all the things that we learned about the manipulation of functions do not longer hold. Mappings of path space onto the real line are called functionals and thus our ambition of finding the least time as a function of path is a functional.  In our first example, refraction, we used our intuition to label the paths as the same as the point of intersection of the path with the interface of the media. This is clearly only a small sample of all the paths. The important point about our selection of the point of intersection was not only for convenience, it was a reduction in the size of the path space  which allows us to write the time $T$ as a function of $x$. Thus, although it is nice to think of $x$ as the distance along the interface, its real role is as a label in path space. All in all, the simple looking rule
\begin{equation}
T = \sum_{{\rm path}, (x_0,y_0)}^{(x_{f}, y_{f})} \frac{\Delta r_i}{v_i}
\end{equation}
is actually a complicated mathematic structure. All through will ignore most of these complications and go ahead finding a family that is $\aleph_1$ when we are operating in path space.\\

\begin{figure}[tbp]
\postscript{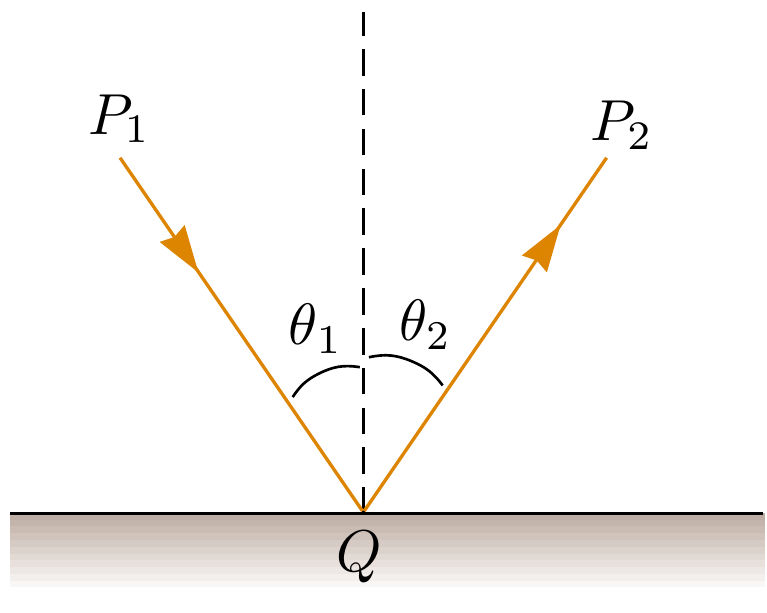}{0.85}
\caption{Reflection law. The incident ray, the reflected ray, and the normal all lie in the same plane; $\theta_1 = \theta_2$~\cite{Serway:14}.}
\label{fig:reflection}
\end{figure}

{\bf EXERCISE 2.1} Consider a ray of light traveling in vacuum from point $P_1$ to $P_2$ by way of the point $Q$ on a plane miror as shown in Fig.~\ref{fig:reflection}. Show that Fermat's principle implies that, on the actual path followed, $Q$ lies in the same vertical plane as $P_1$ and $P_2$ and obeys the law of reflection, that is $\theta_1 = \theta_2$. [Hints: Let the miror lie in the $x$-$z$ plane, and let $P_1$ lie on the $y$ axis at $(0,y_1,0)$ and $P_2$ in the $x$-$y$ plane at $(x_2,y_2,0)$. Finally, let $Q = (x,0,z)$. Calculate the time for the light to traverse the path $P_1 QP_2$ and show that it is minimum when $Q$ has $z=0$ and satisfies the law of reflection.]\\

\begin{figure}[tbp] \postscript{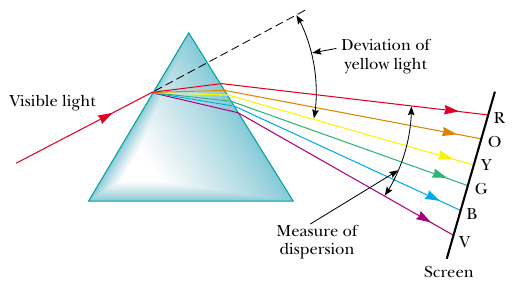}{0.85} 
\caption{Newton's experiment showing that light is composed of colored components. A narrow beam of light is incident on a prism and produces a broadened and colored band which can be reconstituted back into a narrow white beam of light with a second prism~\cite{Serway:14}.}
\label{fig:prisma}
\end{figure}

It is a common experience to use a piece of shaped glass, a triangular cut of glass called a prism, to produce a rainbow of color from sun light. This is basically a refractive phenomenon and a simple extension of Fermat's least time principle can be used to describe it. A narrow beam of white light incident at a non-normal angle on one surface of the glass is refracted; the beam changes direction. The spread of color appears because the different colors in the light have different speeds in the glass, with the red being faster than the blue and all colors slower than for light in air. Hence, the red is bent less than the blue. The separated rays then emerge from the other interface of the glass spread in a familiar rainbow pattern. This spread of color can be seen by placing a piece of paper after the second interface as shown in Fig.~\ref{fig:prisma}. It was Newton who introduce the idea that white light was a complex phenomenon composed of an internal structure -- the colors~\cite{Newton:1704}.  Prior to Newton's interpretataion, the understanding was that the different colors in the prism came from the glass and was not an intrinsic property of the light. To show otherwise Newton placed a prism in the path of a narrow beam of sunlight.  As expected, the beam was spread over a band of angles. He then inserted a second prism and allowed the spread beam to enter it. When arranged carefully, he found that the second prism was able to reconstitute the original beam in the original direction. He labeled the different colors with a continuously varying parameter that had the units of a time, now identified with the reciprocal of the frequency. The length $\lambda$ and time $T$ characterizing a given color are connected by the speed of light in the medium according to $\lambda/T = c/n$ where $n$ is the index of refraction.\\

\begin{figure}[tbp] \postscript{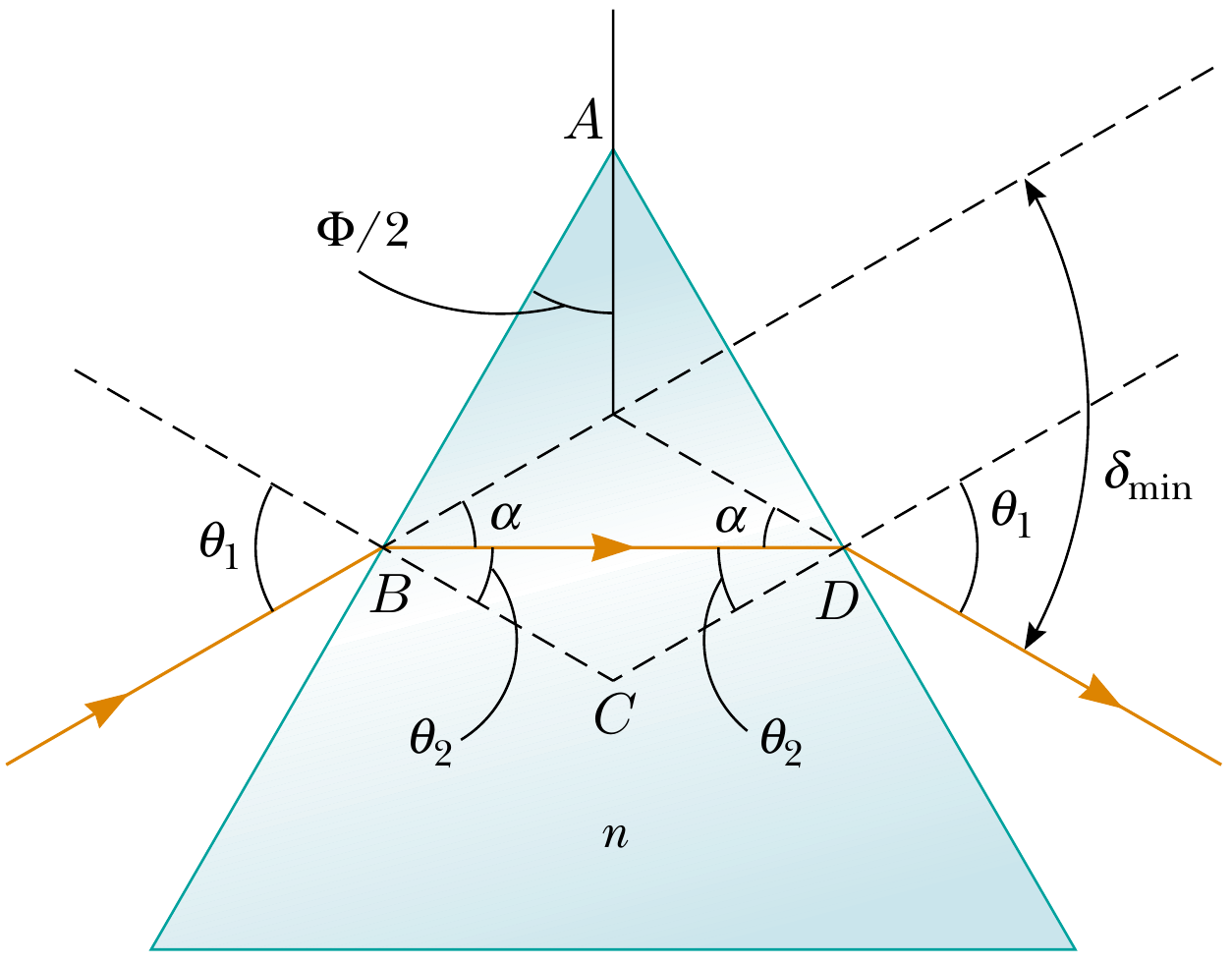}{0.85} 
\caption{Geometry of a dispersing prism with a light ray passing through the prism at the minimum angle of deviation~\cite{Serway:14}.}
\label{fig:prisma-delta-min}
\end{figure}

{\bf EXERCISE 2.2} Nowadays dispersing prisms come in a great variety of sizes and shapes. Typically, a ray entering a dispersing prism will emerge having been deflected from its original direction by an angle $\delta$, known as the angular deviation.  Show that the minimum angle of deviation, $\delta_{\rm min}$, for a prism (with apex angle $\Phi$and index of refraction $n$) occurs when the angle of incidence $\theta_1$ is such that the refracted ray inside the prism makes the same angle with the normal to the two prism faces, as shown in Fig.~\ref{fig:prisma-delta-min}.\\

\begin{figure}[tbp] \postscript{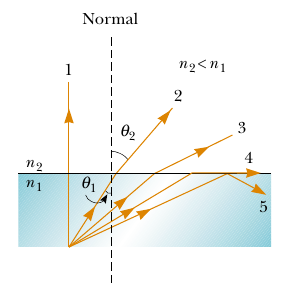}{0.85} 
\caption{Rays travel from a medium of index of refraction $n_1$ into a medium of index of refraction $n_2$, where $n_2 < n_1$. As the angle of incidence $\theta_1$ increases, the angle of refraction $\theta_2$ increases until $\theta_2 = \pi/2$ (ray 4). For even larger angles of incidence, total internal reflection occurs (ray 5)~\cite{Serway:14}.}
\label{fig:prisma-theta-critico}
\end{figure}

{\bf EXERCISE 2.3} An interesting effect called total internal reflection can occur when light is directed from a medium having a given index of refraction toward one having a lower index of refraction. Consider a light beam traveling in medium 1 and meeting the boundary between medium 1 and medium 2, where $n_1$ is greater than $n_2$. Various possible directions of the beam are indicated by rays 1 through 5 in Fig.~\ref{fig:prisma-theta-critico}. The refracted rays are bent away from the normal because $n_1$ is greater than $n_2$. At some particular angle of incidence $\theta_c$, called the critical angle, the refracted light ray moves parallel to the boundary so that $\theta_2 = \pi/2$. For angles of incidence greater than 􏰨$\theta_c$ the beam is entirely reflected at the boundary.  Consider a triangular glass prism with apex angle $\Phi$ and index of refraction $n$.  What is the smallest angle of incidence $\theta_1$ for which a light ray can emerge from the other side?\\

{\bf EXERCISE 2.4} The index of refraction for violet light in silica flint glass is 1.66, and that for red light is 1.62; see Fig.~\ref{fig:prisma}. What is the angular dispersion of visible light passing through a prism of apex angle $60.0^\circ$ if the angle of incidence is $50.0^\circ$?

\section{Interference of Light Waves}

\begin{figure}[tbp] \postscript{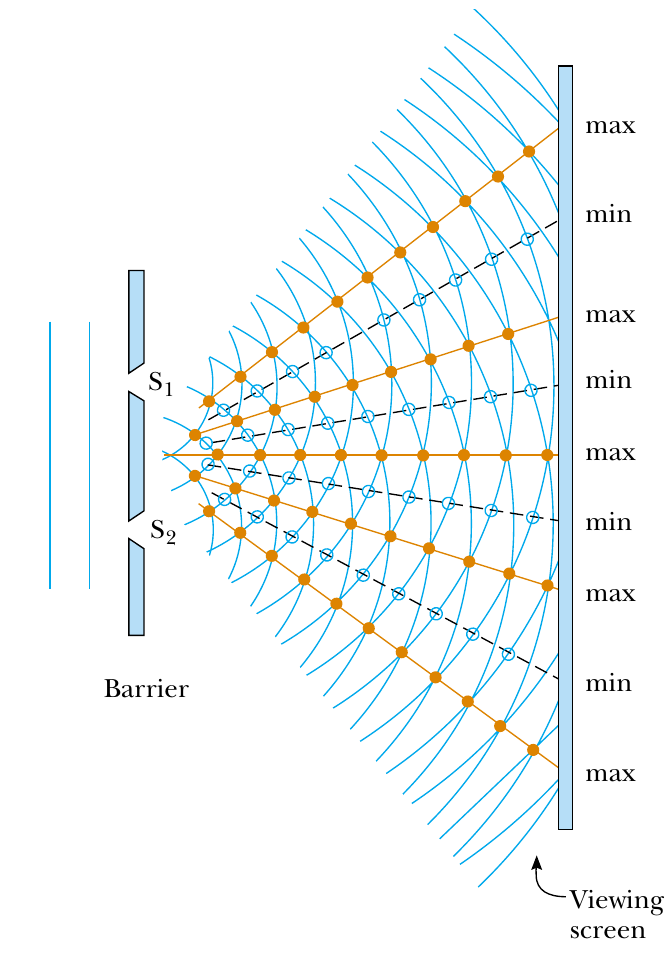}{0.85} 
\caption{Young's double slit interferometer. Light of a single color,
called monochromatic light, shines on an opaque screen with two narrow
slits. The light subsequently passes through to a distant screen on which
the brightness can be observed. The light from $S_1$ and $S_2$ produces a visible pattern of bright and dark parallel bands called fringes. When the light from $S_1$ and that from $S_2$ both arrive at a point on the screen such that constructive interference occurs at that location (max), a bright fringe appears. When the light from the two slits combines destructively at any location on the screen (min), a dark fringe results~\cite{Serway:14}.}
\label{fig:Yds}
\end{figure}

Fermat's least time is a wondrous principle for many applications, e.g., lens design. However, there are situations in which you actually do find light between points that, according to Fermat's principle, should be dark. It is instructive  to consider one simple example. We have seen that according to Fermat's principle in a homogeneous medium light travels in a straight line between two points. Hence, if we place a barrier between these two points the light would be blocked and so no light should be seen at the second point. Consider the experimental set up shown in Fig.~\ref{fig:Yds}, known as Young's double slit interferometer~\cite{Young:1802}. Monochromatic light from a single concentrated source illuminates a barrier containing two small openings.  The light emerging from the two slits is projected onto a distant viewing screen. Distinctly, it is observed that the light deviates from a straight-line path and enters the region that would otherwise be shadowed. The variation in brightness of the projected image as you move across the screen is very eye-catching.

To understand Young's double slit experiment we must conjecture that light of a given color is intrinsically an oscillating system. We have seen that the different components of light identified as colors can be associated with different frequencies, $f$. 
In this sense, each color of light
is identified with a certain time period $T = f^{-1}$, or a wavelength 
$\lambda = cT$.  Light is then described by an amplitude that varies harmonically with time, 
\begin{equation}
A(t) = A_0 \cos ( \omega t), 
\end{equation}
where $A_0$ is the maximum value of the amplitude, $\omega = 2\pi/T$ is the radian frquency, and $T$ is the period for light of that color. Unfortunately, the coefficient $A_0$ of the harmonic factor is also often called the ``amplitude'' of the harmonic signal. It should be clear from the context which amplitude is which. The light propagates between two points in space by having its amplitude travel over all available paths,  and, as it travels, it oscillates with frquency $f$. What you see and can measure is the square of that amplitude.

For the double slit experiment, there is a constant level of brightness at each of the slits.  Yet, we have conjectured that light is oscillatory phenomenon. To clarify the situation we first note that the rate of energy flow per unit area is given by
\begin{equation}
S(t) = A^2(t) = A^2_0 \cos^2(\omega t) .
\end{equation} 
At optical frequencies $S$ is an extremely rapidly varying function of time (indeed twice as rapind as the amplitude, since cosine-squared has double the fequency of cosine), so its instantaneous value would be an impractical quantity to measure. This suggests that we employ an average procedure.\footnote{For visible light,  $\lambda \sim 6 \times 10^{-7}~{\rm m}$, $f \sim 5 \times 10^{14}~{\rm s^{-1}}$ and $T \sim 2 \times 10^{-15}~{\rm s}$. If the time resolution of the eye is milliseconds, what we see is the average of tens of millions of cycles.}  Indeed, the intensity we see is the long time average of many periods,
\begin{equation}
I = \langle S(t) \rangle_t,
\end{equation}
where $\langle \rangle_t$ indicates the time average of the quantity inside the brackets.\footnote{The time average over the interval 0 to $T$ for a function of time is defined as $\langle f  \rangle_t= \frac{1}{T} \sum_0^T f(t) \, \Delta t$ or $\langle f \rangle_t = \frac{1}{T} \int_0^T f(t) d t$.} Using the trigonometric relation $\cos^2(\omega t) = [1+ \cos (2 \omega t)]/2$ it is easily seen that the time average of 
$\cos^2(\omega t)$ for several periods is 1/2, an therefore
\begin{equation}
I = \langle A^2(t) \rangle_t= \frac{A^2_0}{2}  \, .
\label{A-steadyI}
\end{equation}

Consider first the case in which only slit 1 is open. If we arrange the apparatus so that the amplitude at slit is
\begin{equation}
A_1(t) = A_0 \cos(\omega t) \,,
\end{equation}
using (\ref{A-steadyI}) we can determine $A_0$ from the intensity of the light at slit one, $I_1$. The amplitude
at the screen at a given time $t$ is the original amplitude at slit 1 delayed
by the time it takes the light to go from the slit to the screen. In other
words, the amplitude of the light at the screen at time $t$ is the same as the
amplitude of the light at the slit at time $t - r_1/c$ where $r_1$ is the distance 
between the slit and the screen and $c$ is the speed of light. 
This means that the amplitude on the screen from slit 1 alone is
\begin{equation}
{\cal A}_1 = A_0 \cos[\omega (t - r_1/c)] \, .
\end{equation}
This result is not as trivial as it seems. Let us cast it in slightly different
form:
\begin{eqnarray}
{\cal A}_1(t) & = & A_0 \cos \left(\omega t - \frac{2\pi r_1/c}{T} \right) \nonumber \\
& = & A_0 \cos \left(\omega t - \frac{2 \pi r_1}{\lambda} \right) \,,
\end{eqnarray}
{}From the $\omega t$ term,
we see that this is an amplitude that oscillates with a period $T$ so  that the color of the light at the screen is the same as the color of the light at slit 1. The only difference is that there is an extra time independent term in the argument of the $\cos$ function. All this does is shift the the argument that goes in at the start. Again, since this signal varies so rapidly that our sensors can only see the time average over many many periods, this starting angle (called the phase) is not detectable. Since this shift is the only factor that changes as you move to different parts of the screen, the intensity at the screen is uniform. 

If you choose to have only slit 2 open, you would have a similar situation. Since the two slits are located symmetrically relative to the source, the amplitude at slit 2 is the same as that of slit 1 and thus the amplitude at the screen from slit 2 alone would be
\begin{equation}
{\cal A}_2  (t) = A_0 \cos [\omega (t - r_2/c)]
\end{equation}
where we have used the fact that, at a general point on the screen, the two distances, $r_1$ and $r_2$, will be different. Again, this by itself produces an illumination that is uniform and the same color as the original light. Note that if you have just one of the slits open, say slit one, the intensity on the screen is
\begin{equation}
{\cal I}_1  = \langle {\cal A}_1^2 (t) \rangle_t= \frac{A_0^2}{2} = I_1\,  
\end{equation}
As before, the intensity is the 
time average of the amplitude squared. 

\begin{figure*}[tbp] \postscript{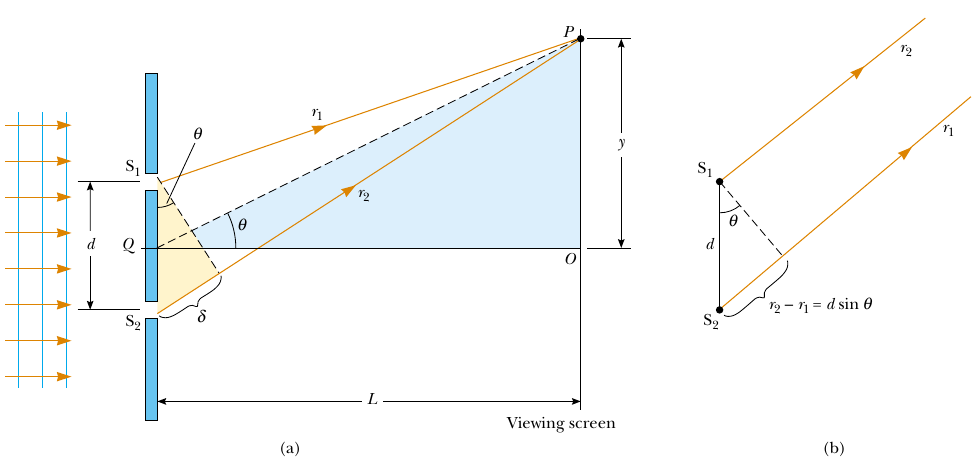}{0.85} 
\caption{{\it (a)}~Geometric construction for describing Young's double-slit experiment (not to scale). {\it (b)}~When we assume that $r_1$ is parallel to $r_2$, the path difference between the two rays is $r_2 - r_1 = d \sin \theta$. For this approximation to be valid, it is essential that $L \gg d$~\cite{Serway:14}.}
\label{fig:double-slit}
\end{figure*}

What happens when both slits are open? The net amplitude at the screen is sum of the two amplitudes from the slits as if they operated independently. This is the point of the fact that the amplitudes of independent sources ``add.'' The amplitudes are the fundamental causal agents. They carry the information about the slits to the screen. This process of adding independent sources as if the other was not present is called superposition. 

The amplitude at the screen is the superposition of the amplitudes from the two slits
\begin{eqnarray}
\!\!\!\!\!{\cal A}_{\rm tot} \!\!\! & = & \!\! {\cal A}_1 + {\cal A}_2 \nonumber \\
& = & \!\! A_0\{ \cos [\omega (t-r_1/c)] + \cos [\omega (t- r_2/c)]\} \nonumber \\
& = & \!\! 2 A_0  \cos \left(\omega \frac{r_1 - r_2}{2c} \right)  \cos \left[\omega \left( t - \frac{r_1 - r_2}{2c} \right) \!\right],
\label{la23}
\end{eqnarray}
where we have  used the trigonometric identity $\cos \alpha + \cos \beta = 2 \cos [(\alpha + \beta)/2] \cos [(\alpha - \beta)/2]$.  Note that $r_1$ and $r_2$ depend on the position on the screen.

In this case there is an oscillating signal at the screen. Again, this is light of the same color with
a position dependent phase that is not observable for fast frequencies with
slow detectors like our eyes. The
important feature of this superposed amplitude is that the amplitude at the
screen now has a position dependent amplitude 
$2A_0 \cos \left[ \omega \left( \frac{r_1 - r_2}{2c} \right) \right]$. 
As you move to different positions on the screen, there will be different intensities and even zero intensities at places where $\cos \left[ \omega \left( \frac{r_1 - r_2}{2c} \right) \right] = 0$ or equivalewntly if $\omega \left( \frac{r_1 - r_2}{2c} \right)$ is an odd multiple of $\pi/2$.

It is useful to obtain expressions for the positions along the screen of the bright and dark fringes measured vertically from $O$ to $P$.  The total rate of energy flow per unit area at the screen is the amplitude (\ref{la23}) squared
\begin{equation}
{\cal S}_{\rm tot}   =  |{\cal A}_1 + {\cal A}_2|^2 \ .
\end{equation}
The difference in the distances for rays from slit one and slit two is $\delta (y) =  r_1  - r_2 = d \sin \theta$. 
For the geometry of Fig.~\ref{fig:double-slit}, in which the slit separation $d$ is very small compared to the distance from the slits to the screen $L$ and $d \gg \lambda$, we can use the small angle approximation $\theta \approx \sin \theta \approx \tan \theta$. These are  justified assumptions because in practice $L$ is often on the order of 1~m,   $d$ a fraction of a millimeter, and $\lambda$ a fraction of a micrometer for visible light.  Putting  this all together we find that $\delta (y)/d \approx y/L$, yielding
\begin{equation}
{\cal I}_{\rm tot} = 4 {\cal I}_1 \cos^2 \left( \frac{\omega \, \delta (y)}{2c} \right)
 = 4 {\cal I}_1 \cos^2 \left( \frac{y \omega d}{2cL} \right),
\label{venticinco}
\end{equation}
where $I_1$  is the intensity at the screen if you only have slit one open. Equation~(\ref{venticinco}) describes the intensity pattern that is observed as you move up or down a distance $y$ measured from the central position on the screen. It predicts a rapidly changing pattern of bright and dark spots. We see that the positions of the bright fringes measured from $O$ are given by
\begin{equation}
y_{\rm bright} = \frac{\lambda L}{d} m \,,
\end{equation}
with $m = 0, \pm 1, \pm 2, \cdots$. The number $m$ is called the order number. For constructive interference, the order number is the same as the number of wavelengths that represents the path difference between the waves from the two slits, i.e. 
\begin{equation}
\delta (y) = d \sin \theta=  m \lambda. 
\label{TonyTratoria}
\end{equation}
The central bright fringe at $\theta = 0$ is called the zeroth-order maximum. The first maximum on either side, where $m = \pm 1$, is called the first-order maximum, and so forth.
Using (\ref{venticinco}) we find that the dark fringes are located at
\begin{equation}
y_{\rm dark} = \frac{\lambda L}{d} (m + \tfrac{1}{2}) \,,
\end{equation}
that is when $\delta$ is an odd multiple of $\lambda/2$, the two waves arriving at point $P$ are out of phase by $\pi$ and give rise to destructive interference. It is important to once again emphasize that it is the amplitudes that add. The amplitude carries the causal information. The  intensity, at a point is derived as the square of the amplitude. What you see is the intensity.

We now know how to construct the amplitude for light with a given frequency. What do you do if you do not have monochromatic light? For any form of the light, you can treat it as a superposition of several frequencies or different colors. Evaluate what happens for each frequency, add the amplitudes, and then squared them. Note that when you take the long time average the mixed frequency terms in the square drop out, $\langle {\cal A}_{\omega_i} {\cal A}_{\omega_j} \rangle_t= 0 \ \forall \ \omega_i \neq \omega_j$. Therefore, 
\begin{eqnarray}
{\cal I}_{\rm tot} & = & \langle ({\cal A}_{\omega_1} + {\cal A}_{\omega_2} + \cdots + {\cal A}_{\omega_n})^2 \rangle_t \nonumber \\
& = & \langle {\cal A}_{\omega_1}^2 \rangle_t + \langle {\cal A}_{\omega_2}^2\rangle_t + \cdots + \langle {\cal A}_{\omega_n}^2 \rangle_t \nonumber \\
& = & {\cal I}_{\omega_1} + {\cal I}_{\omega_2} + \cdots + {\cal I}_{\omega_n} \, .
\end{eqnarray}
This translates into the statement that you have heard since childhood:{\it  light is made up of individual colors}. For a thorough discussion on interference of light waves see e.g.~\cite{Hecht}.\\

{\bf EXERCISE 3.1}~In Fig.~\ref{fig:double-slit}, let  $L= 120~{\rm cm}$ and $d= 0.25~{\rm cm}$. The slits are illuminated with coherent 600~nm light. Calculate the distance $y$ above the central maximum for which the total intensity on the screen is 75\% of the maximum.\\

Young's double slit experiment brings with it the need for a new physical construct: the amplitude. This entity fills all the space. A physical quantity that is defined at all points in space is given the general title of a field. The development of the ideas and techniques of field theory took place in the later half of the 19th century and were applied to optical phenomena by Maxwell. Although this is not modern physics, it is so basic to our understanding of modern physics that we will now spend a few   paragraphs reviewing them. Because of Maxwell, we now understand what the amplitude for light is and in a sense is no longer thought to be unmeasurable. It is a special combination of the electric and magnetic fields. These fields can be and are regularly measured although to do so at optical frequencies is still too difficult.

Maxwell developed a local field theory to describe the phenomena associated with what is called electricity and magnetism~\cite{Maxwell:1865zz}. He reduced all the known laws of electricity and magnetism into four reasonably simple equations,
\begin{equation}
\vec \nabla \cdot \vec E(\vec r , t)  =  \frac{1}{\epsilon_0} \rho (\vec r,t) \,,
\label{Maxwell1}
\end{equation}
\begin{equation}
\vec \nabla \times \vec E  = - \frac{\partial \vec B(\vec r,t)}{\partial t} \,
\label{Maxwell2}
\end{equation}
\begin{equation}
\vec \nabla \cdot \vec B(\vec r , t)  =  0 ,
\label{Maxwell3}
\end{equation}
\begin{equation}
 \vec \nabla \times \vec B(\vec r,t)  =  \mu_0 \vec \jmath (\vec r,t) + \mu_0 \epsilon_0 \frac{\partial \vec E(\vec r,t)}{\partial t} \,,
\label{Maxwell4}
\end{equation}
and the associated force law
\begin{equation}
\vec F = q \vec E + q \vec v \times \vec B \,,
\label{em-force}
\end{equation}
where $\rho (\vec r,t)$ is the charge per unit volume, $\vec \jmath (\vec r,t)$  is the current density or charge per unit area per unit time, $\vec E (\vec r, t)$ is the electric field or force per unit charge, and $\vec B(\vec r, t)$ is the magnetic field or force per unit charge times speed. The force, $\vec F$ is the force on a charged particle with charge $q$ and velocity $\vec v$.
In so doing, he unified the electric and magnetic forces and predicted the fundamental nature of light. These are considerable accomplishments in their own right but also he somewhat inadvertently clarified the idea of the field and the idea of causality. Note, however, that Maxwell's theory is not the first field theory; it was the first field theory of a fundamental force system. The first local field theory and the easiest to appreciate is the description of fluid flow. It was the success of a field theory of fluid flow that motivated Maxwell to attempt to write the rules of the electricity and magnetism in this field theory form.

Like any system of forces, the set of rules articulated by Maxwell must obey Galilean invariance, or we would be able to use electromagnetic phenomena to determine a velocity in space. If you do a careful analysis of the dimensional content of Maxwell's equations you will find  that $\epsilon_0$ and $\mu_0$ have dimensions, and that the combination $(\mu_0 \epsilon_0)^{-1/2}$ has the dimensions of a speed. Actually, this speed is the characteristic speed of travel for changes in the fields, and this is the speed at which light travels. This prediction presented quite a quandary to
19th Century scientists after they realized that Maxwell's equations are not Galilean invariant. This implies that a velocity could be measured and light could be used to do it. In other words, there should be some preferred state of uniform motion in which  Maxwell's equations are true as written, and in this frame the measured speed of light would be $(\mu_0 \epsilon_0)^{-1/2}$ .\\

{\bf EXERCISE 3.2}~Convince yourself that (\ref{Maxwell2}) and (\ref{Maxwell3}) are invariant under Galilean transformations (\ref{eq:Galileo}), but (\ref{Maxwell1}) and (\ref{Maxwell4}) are not.\\

\section{Luminiferious \AE ther}

Scientists from the eighteen hundreds believed in all notions of classical physics. Thus, it was only normal for them to assume that all waves traveled through mediums. Shortly after Maxwell showed that light is indeed an electromagnetic wave it became evident that air was most definitely not the required medium for the propagation of light. This is because electromagnetic waves traveled through space to get to Earth. To solve the problem it was assumed that there is an \ae ther which propagates light waves. This \ae ther was assumed to be everywhere and unaffected by matter. Therefore, it could be used to determine an absolute reference frame (with the help of observing how light propagates through the \ae ther).

\begin{figure}[tbp] \postscript{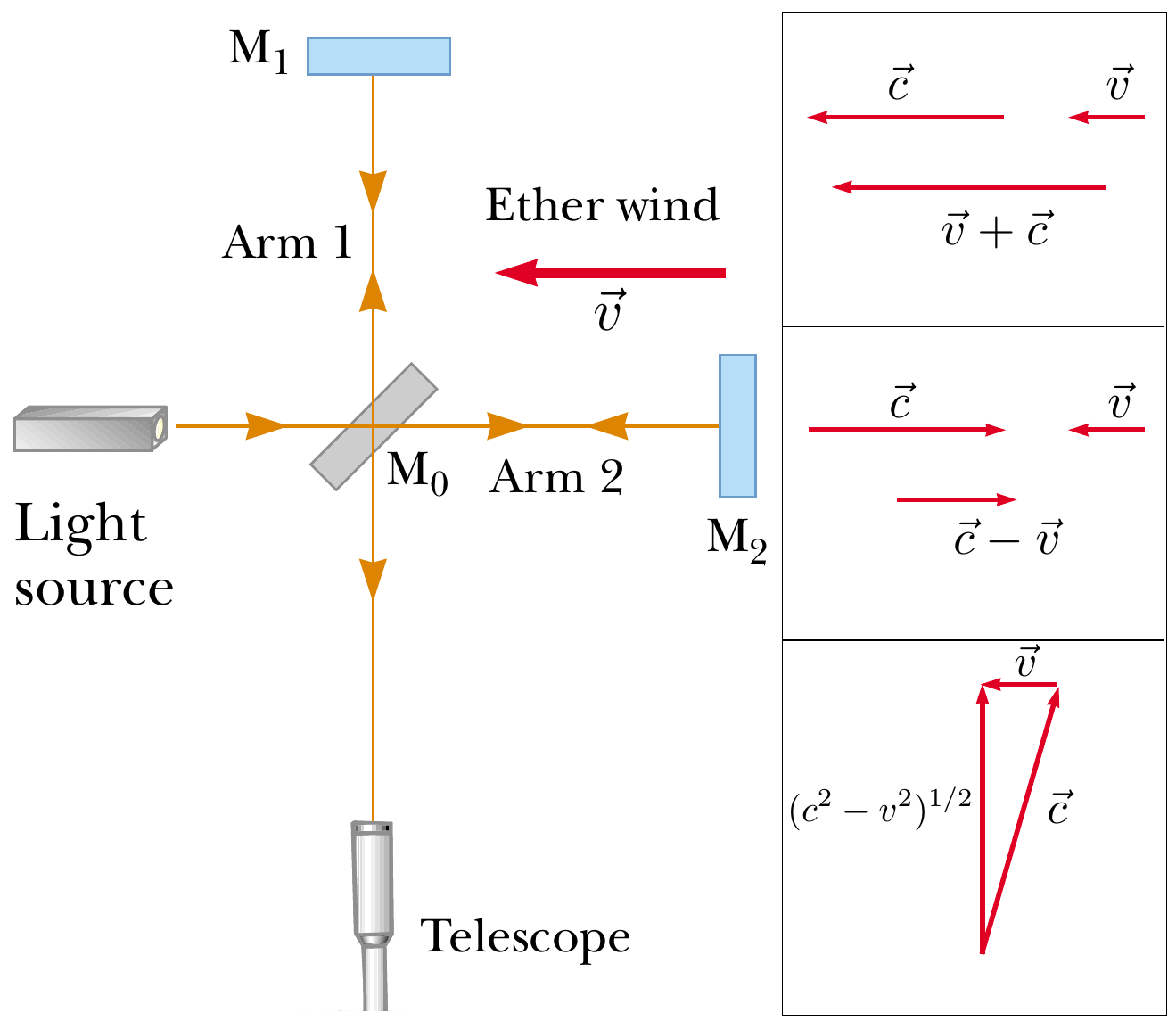}{0.85} 
\caption{Schematic diagram of the Michelson-Morley interferometer. The embedded box shows the velocity addition. If the velocity of the \ae ther wind relative to the Earth
is $\bm{v}$ and the velocity of light relative to the \ae ther is $c$, then the speed
of light relative to the Earth is
$c + v$ in the downwind direction (top),  $c-v$ in the upwind direction (middle), and  $(c^2- v^2)^{1/2}$
in the direction perpendicular to the wind (bottom)~\cite{Serway:14}.}
\label{fig:michelson-morley}
\end{figure}

The famous experiment designed to detect small changes in the speed of light with motion of an observer through the \ae ther was performed by Michelson and Morley~\cite{Michelson:1887zz}. Consider the interferometer shown in Fig.~\ref{fig:michelson-morley}. The viewer will see two beams of light which have traveled along different arms display some interference pattern. If the system is rotated, then the influence of the ``ether wind'' should change the time the beams of light take to travel along the arms and therefore should change the interference pattern. The experiment was performed at different times of the day and of the year. No change in the interference pattern was observed!

\begin{figure}[tbp] \postscript{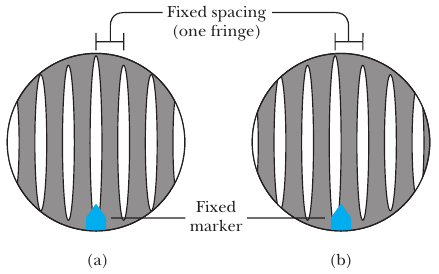}{0.85} 
\caption{Interference fringe schematic showing (a) fringes before rotation and  (b) expected fringe shift after a rotation of the interferometer by $\pi/2$~\cite{Serway:2005}.}
\label{MMfringe}
\end{figure}

To understand the outcome of the Michelson–Morley experiment, let us assume that the interferometer shown in Fig.~\ref{fig:michelson-morley} has two arms of lehgth $L_1$  and $L_2$. The speed of a photon (relative to the source) on the trip ``over'' to the mirror is $c - v$ and so takes a time of $L_1/(c - v)$. On the return trip, the photon has speed of $c + v$ and so takes a time of $L_1/(c + v)$. Therefore the round trip time is
\begin{eqnarray}
t_1  & = & \frac{L_1}{c-v} + \frac{L_1}{c+v}  = 
  \frac{2cL_1}{c^2 - v^2} \nonumber \\ &= &  \frac{2 L_1}{c} \frac{1}{1 - v^2/c^2} 
 =  \frac{2L_1}{c} \left(1 - \frac{v^2}{c^2} \right)^{-1} \, .
\label{treintaytres}
\end{eqnarray}
The photon traveling along the arm perpendicular to the wind must travel with a component of velocity ``upstream'' to compensate for the wind. Because the speed of the beam relative to the Earth is $(c^2 - v^2)^{1/2}$, see Fig.~\ref{fig:michelson-morley}, the time of travel for each half of this trip is $L_2/(c^2 -v^2)^{1/2}$, and the total time of travel for the round-trip is
\begin{equation}
  t_2   =  \frac{2 L_2}{\sqrt{c^2 - v^2}} 
  =  \frac{2 L_2}{c} \frac{1}{\sqrt{1 - v^2/c^2}} \, .
\label{treintaycuatro}
\end{equation}
Now, since $1/(1-x) = \sum_{n=0}^\infty x^n$, we have 
\begin{equation}
t_1 \approx \frac{2L_1}{c} \left( 1 + \frac{v^2}{c^2} \right) \, .
\end{equation}
In addition, 
\begin{eqnarray}
(1+x)^m  & = & 1 + mx + \frac{m (m-1)}{2!} x^2 \nonumber \\ &+ & \frac{m (m-1) (m-2)}{3!} x^3 + \cdots \, .
\end{eqnarray}
Taking $m=-1/2$ and $x = -v^2 /c^2$, we obtain
\begin{equation}
t_2 \approx  \frac{2L_2}{c} \left( 1 + \frac{v^2}{2 c^2} \right) 
 =  \frac{2L_2}{c} \left(1 + \frac{v^2}{2c^2} \right) \, .
\end{equation}
For the Earth's orbit around the sun, $v/c \approx 10^{-4}$ so the approximation
is appropriate. Now the rays recombine at the viewer separated by 
\begin{equation}
\Delta t = t_1 - t_2 \approx \frac{2}{c} \left(L_1 - L_2 + \frac{L_1 v^2}{c^2} - \frac{L_2 v^2}{2 c^2} \right) \, .
\end{equation}
The two light beams start out in phase and return to form an interference pattern. Let us assume that the interferometer is adjusted for parallel fringes and that a telescope is focused on one of these fringes. The time difference between the two light beams gives rise to a phase difference between the beams, producing the interference fringe pattern when they combine at the position of the telescope. As illustrated in Fig.~\ref{MMfringe}, a  difference in the pattern should be detected by rotating the interferometer through $\pi/2$ in a horizontal plane, such that the two beams exchange roles.  Let $t'_1$ and $t'_2$ denote the new round trip light travel times. Then (as above, replacing $L_1$ with $L_2$ in $t_1$ to determine $t'_2$ and replacing $L_2$ with $L_1$ in $t_2$ to determine $t'_1$):
\begin{equation}
t'_1 = \frac{2L_1}{c} \left( 1 + \frac{v^2}{2c^2} \right)\ {\rm and} \ t'_2 = \frac{2L_2}{c} \left( 1 + \frac{v^2}{c^2} \right) \, .
\end{equation}
Then
\begin{equation}
\Delta t' = t'_1 - t'_2 = \frac{2}{c} (L_1 - L_2) + \frac{v^2}{c^3} (L_1 - 2 L_2) 
\end{equation}
and 
\begin{eqnarray}
\Delta t - \Delta t' & = & \frac{2}{c} (L_1-L_2) + \frac{2 v^2}{c^3} \left(L_1 - \frac{L_2}{2} \right) \nonumber \\
& - & \left[ \frac{2}{c} (L_1 - L_2) + \frac{v^2}{c^3} (L_1 - 2 L_2) \right] \nonumber \\
& = & \frac{v^2}{c^3} (L_1 +L_2) \, .
\label{time-changeMM}
\end{eqnarray}
This is the time change produced by rotating the apparatus.  The path difference corresponding to this time difference is
\begin{equation}
\delta =  \frac{v^2}{c^2} (L_1 +L_2) \, .
\end{equation}
The corresponding fringe shift is equal to this path difference divided by the wavelength of light,
\begin{equation}
{\rm Shift} = \frac{v^2}{\lambda c^2} (L_1 +L_2)
\end{equation}
because a change in path of 1 wavelength corresponds to a shift of 1 fringe. In the experiments by Michelson and Morley, each light beam was reflected by mirrors many times to give an increased effective path length $L = L_1 = L_2$ of about 11~m. Using this value, and taking $v$ to be equal to the speed of the Earth about the Sun, gives a path difference of $2.2 \times 10^{-7}~{\rm m}$. This extra distance of travel should produce a noticeable shift in the fringe pattern. Specifically, using light of wavelength 
500~nm, we find a fringe shift for rotation through $\pi/2$ of
\begin{equation}
{\rm Shift} = \frac{\delta}{\lambda} \approx 0.40 \,.
\end{equation}
The precision instrument designed by Michelson and Morley had the capability of detecting a shift in the fringe pattern as small as 0.01 fringe. However, they detected no shift in the fringe pattern. Since then, the experiment has been repeated many times  under various conditions, and no fringe shift has ever been detected. Hence, it was concluded that one cannot detect the motion of the Earth with respect to the \ae ther.\\

{\bf EXERCISE 4.1}~A shift of one fringe in the Michelson-Morley experiment corresponds to a change in the round-trip travel time along one arm of the interferometer by one period of vibration of light (about $2 \times 10^{-15}~{\rm  s}$) when the apparatus is rotated by $\pi/2$. What velocity through the \ae ther would be deduced from a shift of one fringe? (Take the length of the interferometer arm to be 11~m.)\\

Many efforts were made to explain the null results of the Michelson-Morley experiment and to save the \ae ther concept and the Galilean addition law for the velocity of light. The two most prevalent are described in the following exercises.\\

{\bf EXERCISE 4.2}~In 1889, FitzGerald proposed that an object moving through the \ae ther wind with velocity $v$ experiences a contraction in the direction of the \ae ther wind of $\sqrt{1 - v^2/c^2}$~\cite{FitzGerald}. That is, in Fig.~\ref{fig:michelson-morley}, $L_1$ is contracted to $L_1 \sqrt{1-v^2/c^2}$ and then we get $t_1 = t_2$ when $L_1 = L_2$, potentially explaining the results of the Michelson-Morley experiment. Show that for an interferometer with unequal arms Fitzgerald proposal implies
\begin{equation}
\Delta t = \frac{2}{c} \Delta L \left(1 + \frac{v^2}{2c^2} \right) \ .
\end{equation}
Now, since $\Delta t$ is only a function of $v$, we expect $\Delta t$ to vary with a period of 6 months as the Earth changes direction in its orbit around the Sun. In other words, a Michelson-Morley apparatus with unequal arms will exhibit a pattern shift over a 6 month period. In 1932, Kennedy and Thorndike performed such an experiment and detected no such shift~\cite{Kennedy:1932zz}.\\

{\bf EXERCISE 4.3}~Another suggestion to explain the negative result of the Michelson-Morley experiment was the idea that the Earth ``drags the \ae ther along with it'' as it orbits the sun. This idea is rejected because of stellar aberration. Consider light rays from a star directly overhead entering a telescope. Assume that the Earth, in its orbit around the Sun, is moving at a right angle to the incoming rays. In the time it takes a ray to travel down the barrel to the eyepiece, the telescope will have moved slightly to the right. Therefore, in order to prevent the light rays from falling on the side of the barrel rather than on the eyepiece lens, we must tilt the telescope slightly from the vertical, if we are to see the star. Consequently, the apparent position of the star is displaced forward somewhat from the actual position. Show that the angle of displacement  (in radians) is $\theta = \tan^{-1} (v/c) \approx v/c$,  where $v$ is the Earth's orbital velocity. Verify that $\theta = 20.6''$ (roughly the angle subtended by an object 0.1~mm in diameter held at arm's length). As the Earth revolves around the Sun in its nearly circular annual orbit, the apparent position of the star will trace a circle with angular radius $20.6''$. This is indeed observed. If the Earth dragged a layer of \ae ther along with it, the light rays, upon entering this layer, would aquire a horizontal velocity component matching the forward velocity $v$ of the telescope. There would then be no aberration effect.\\

\section{Foundations of Special Relativity}

\subsection{Einstein postulates}

In order to explain nature's apparent conspiracy to hide the \ae ther drift, Lorentz  developed a theory  that was eventually based on two {\it ad hoc} hypotheses: the longitudinal contraction of rigid bodies and the slowing down of clocks (time dilation) when moving through the \ae ther at speed $v$, both by a factor $(1-v^2/c^2)^{1/2}$, where $c$ is the speed of light~\cite{Lorentz:1892,Lorentz:1899,Lorentz:1904,Goldberg:1969}. This would so affect every aparatus designed to measure the \ae ther drift as to neutralize all expected effects. In 1905, in the middle of this development Einstein advanced the principle of relativity based on the following two axioms~\cite{Einstein:1905ve}: 
\begin{enumerate}
\item~{\it The laws of physics are identical in all inertial frames, or, equivalently, the outcome of any experiment is the same when performed with identical initial conditions relative to any inertial frame.}
\item{\it There exists an inertial frame in which light signals in vacuum always travel rectilinearly at constant speed $c$, in all directions, independently of the motion of the source.}
\end{enumerate}
Actually Poincar\'e  discussed essentially the same principle  in~\cite{Poincare:1905}, but it was Einstein who first recognized its full significance and put it to brillant use. In it, he elevated the complete equivalence of all inertial reference frames to the status of an axiom or principle, for which no proof or explanation is to be sought. On the contrary, it explains the failure of all the \ae ther-drift experiments.  At first sight Einstein's relativity principle seems to be no more than a whole-hearted acceptance of the null results of all the \ae ther-drift experiments. {\it However}, by cesing to look for special explanations of those results, and using them rather as the empirical evidence for a new principle of nature, Einstein had turned the tables: {\it predictions could be made}.

\subsection{Relativity of simultaneity}

Before we discuss the predictions of special relativity, we must first understand how an observer in an inertial reference frame describes an event. We define an event as an occurrence characterized by three space coordinates and one time coordinate.  Events are described by observers who do belong to particular inertial frames of reference.  In general, different observers in different inertial frames would describe the same event with different spacetime coordinates. The observer's rest frame is also known as the proper frame.

\begin{figure}[tbp] \postscript{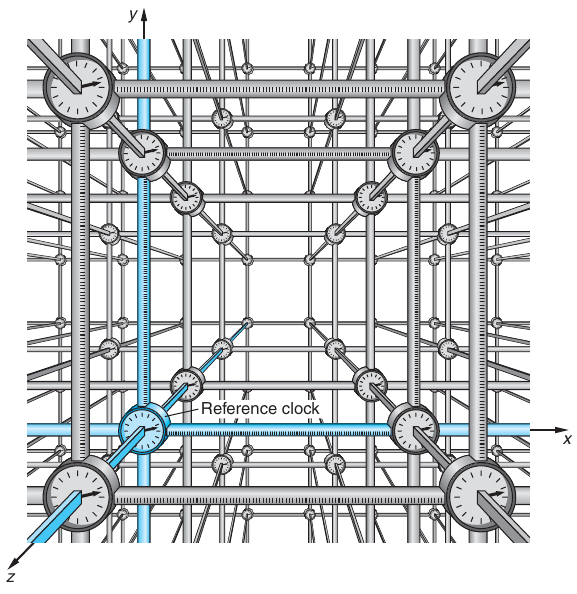}{0.85} 
\caption{Inertial reference frame consisting of a coordinate grid and a set of of synchronized clocks~\cite{Tipler}.}
\label{fig:time-lattice}
\end{figure}

We now turn to discuss the standard construction of a coordinate system.  There are two ordinary methods: the use of confederates at each place and the single observer method. We will first discuss in detail the confederate method and then demonstrate its equivalence to the single observer method, which is the one that we will use throughout.  
A cautionary note is worth taking on into consideration at this juncture. The use of words like confederate or  observer may imply a humanity that is not really intended. Strictly speaking, an observer is a measuring system -- a clock and recording devices -- not necessarily a person. We start by defining the spatial coordinates. 
Because space is the same in all directions at any point we can make a measure of distance that is independent of how we chose the directions of the coordinate system. For our distance measure we adopt 
\begin{equation}
\Delta r =  \sqrt{(\Delta x)^2 + (\Delta y)^2 + (\Delta z)^2}, 
\label{length}
\end{equation}
where $\Delta x$ is the displacement in the one direction and $\Delta y$  and $\Delta z$ are the displacements in the other directions. This distance has the advantage of being independent of the orientation of the axis system. We assume that we can fill space with a confederate at every place and that the distance between the origin observer and each of the confederates remains fixed. Since the alliance of observers are fixed in space, we will label each confederate by how far away he/she is in each of the three coordinate directions. We stress again that because space is homogeneous and isotropic the location of the origin and the directions of the coordinate axis are arbitrary. For the definition of the distance between confederates, we will use the length defined in (\ref{length}), a defined speed of light, and a time to label all distances. This speed is universal for any observer establishing a coordinate system. This means that we need a standard clock. We adopt 
the duration of $9,192,631,770$ oscillations of the radiation corresponding to the transition between the two hyperfine levels of the ground state of $^{133}$Cs to define a second. To find the distance to any confederate, we send a light ray to that confederate who reflects it back and, with the standard clock, the observer at the origin can determine how far away that confederate is, e.g. $\Delta x = c \Delta t/2$,  where $\Delta t$ is the time interval for the round trip of the light.  For labeling the time, we need some ordered system at each place.  By endowing each confederate with a clock, we will have at each place a reference set of events to compare with the events being labeled. We use our standard clock. We tell each confederate to make a standard clock. Since the space is assumed to be homogeneous, all the clocks must run at the same rate for each confederate. This is the first step in getting the time of an event that we want to label, to coordinatize. Since we have now endowed each confederate with a clock, we can use as the space and time label for any event as the time recorded on the nearest confederate's clock and the location of the nearest confederate. Of course the confederates must synchronize their clocks at some time agreeing on the time. It must also be consistent with our understanding that the speed of light is the same in all directions regardless of the velocity of the observer. Of course, this leads to the problem of the relativity of simultaneity and makes it important that we understand the process by which any observer synchronizes clocks. For the moment, since we are dealing with only one frame, we do not need to worry about the relativity of simultaneity but it will cause some concern when we compare the coordinate systems constructed by two relatively moving observers.  Meanwhile, we can accomplish the synchronization by having a burst of light at some very early time released from the origin and, since we know the speed of light and that it is isotropic and we know the location of each confederate, we will know when it passes each confederate and they can set their clocks appropriately.  The confederate scheme for coordinatizing any event is illustrated in Fig.~\ref{fig:time-lattice} and summarized as follows: an observer establishes a lattice of confederates with identical synchronized clocks and the label of any event in spacetime, for that observer, is the reading of the clock and the location of the nearest confederate to that event.

There is a scheme that is equivalent to the confederate scheme that can be accomplished in a less elaborate way by the simple mechanism of having a single clock at the spatial origin and requiring that the observer continuously send out light rays in all directions keeping track of the time of emission. At any event, the incoming light ray is reflected back to the observer. Therefore, the observer has two times and a direction that are associated with any event: the time the reflected ray left and the time of return of the reflected ray and the direction of the reflected light. To yield a spatial coordinatizing that is consistent with the confederate scheme, the spatial distance to the event is 
\begin{equation}
|\vec r \,| = c \, (\tau_2-\tau_1)
\label{distance-def}
\end{equation}
where $\tau_2$ is the later time and $\tau_1$ is the earlier time. The distance is resolved along the coordinate directions according to the direction of the incoming light ray. To be consistent with the time labeling of the confederate scheme, the time coordinate is
\begin{equation}
t = \frac{\tau_1+ \tau_2}{2} \, .
\label{time-def}
\end{equation}

\begin{figure*}[tbp] \postscript{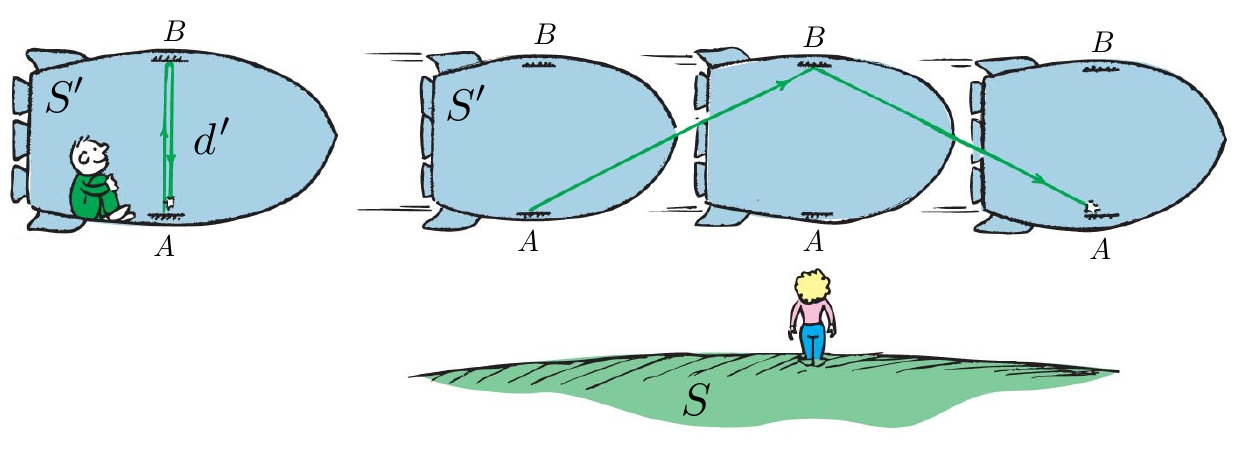}{0.85} 
\caption{Vinnie and the mirror are in a spaceship at rest in the $S'$ frame. The time it takes for the light pulse to reach the mirror and return is measured by Vinnie to be $2d'/c$.  In the frame $S$, the spaceship is moving to the right with speed $v$. For Brittany, the time it takes for the light to reach the mirror and return is longer than $2d'/c$~\cite{Hewitt}.}
\label{fig:Einstein-clock}
\end{figure*}

\begin{figure}[tbp] \postscript{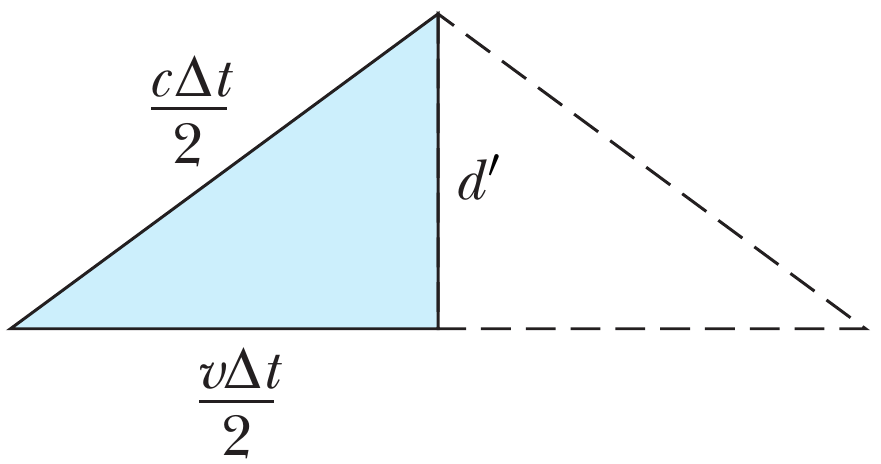}{0.85} 
\caption{A right triangle for computing the time $\Delta t$ in the $S$ frame.}
\label{dilation-triangle}
\end{figure}

Having established our measuring system we now turn to derive predictions of the principle of special relativity using some thought experiments devised by Einstein. Einstein's thought experiments involve an idealized clock in which a light wave is bouncing back and forth between two mirrors. The clock ``ticks'' when the light wave makes a round trip from mirror $A$ to mirror $B$ and back, that is the time that passes as the light travels from one mirror to the other and returns is the unit of time. Assume the mirrors $A$ and $B$ are separated by a distance $d'$ in the rest frame. In that frame a light wave will take a time 
\begin{equation}
\Delta t' = 2 d'/c
\label{qqqqw}
\end{equation}
 for the round trip $A \to B \to A$. This is the proper time interval between two consecutive ticks of the clock. Let $\Delta t$ be the interval between two consecutive ticks of the clock in a frame in which the mirrors move with velocity $v$, as shown in Fig.~\ref{fig:Einstein-clock}. It is noted that when the light wave is bounced back at the mirror $B$, the latter has already move a distance $ v\Delta t/2$, as shown in 
Fig.~\ref{dilation-triangle}. Since light has velocity $c$ in all directions
\begin{equation}
{d'}^2 + \left(v \, \frac{\Delta t }{2} \right)^2 = \left(\frac{c \Delta t}{2} \right)^2\,,
\end{equation}
or
\begin{equation}
\Delta t = \frac{2d'}{\sqrt{c^2 - v^2}} = \frac{\Delta t'}{\sqrt{1 - v^2/c^2}} \, .
\label{t-dilation}
\end{equation}
Hence the ticking of the clock in Vinnie's frame, which moves with velocity $v$ in a direction perpendicular to the separation of the mirrors, is slower by a factor $\gamma = (1- v^2/c^2)^{-1/2}$.

\begin{figure}[tbp] \postscript{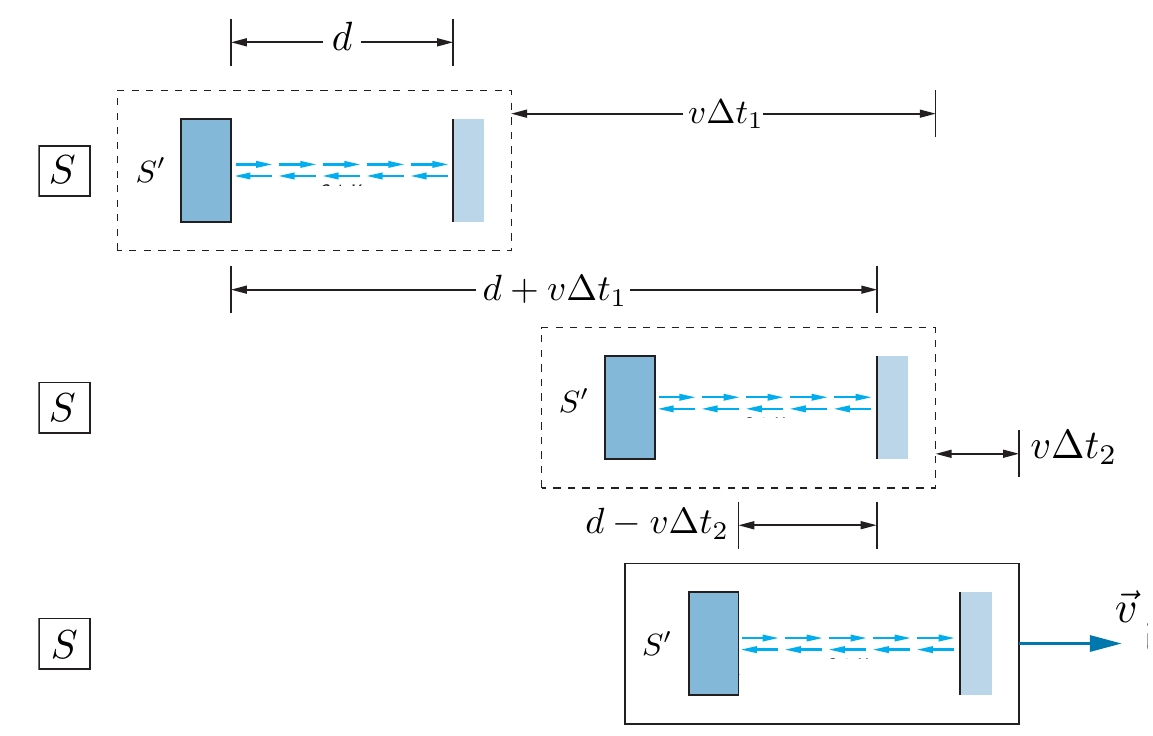}{0.85} 
\caption{The clock carried by Vinnie emits its flash from miror $A$ (on the left of the clock) towards $B$ in the direction of motion. The top, middle, and bottom panels show representative snapshots of the sequence of events.}
\label{length-contarction}
\end{figure}

Now, suppose that the clock is rotated by $90^\circ$ before being set in motion, so that now it has velocity $v$ parallel to the separation between the mirors, see Fig.~\ref{length-contarction}. Suppose the mirrors $A$ and $B$ are separated by $d$ in the moving frame of the clock. After leaving $A$, assume the light wave reaches $B$ at time $\Delta t_1$. As $B$ has moved a distance $v \Delta t_1$, we have
\begin{equation}
d + v \, \Delta t_1 = c \, \Delta t_1
\end{equation}
or
\begin{equation}
\Delta t_1 = \frac{d}{c-v} \, .
\end{equation}
Assuming the light wave, after bouncing at $B$, takes time $\Delta t'_2$ to reach $A$ again, we have by the same reasoning
\begin{equation}
d - v  \Delta t_2 = c \Delta t_2 \,,
\end{equation}
or
\begin{equation}
\Delta t_2 = \frac{d}{c +v} \, .
\end{equation}
Hence the interval between two consecutive ticks in the moving frame is
\begin{eqnarray}
\Delta t  & = & \Delta t_1 +  \Delta t_2 = \frac{2d}{c (1 - v^2/c^2)} \nonumber \\
& = & \left(\frac{d}{d'} \right) \frac{\Delta t'}{1 - v^2/c^2} \, ,
\label{starbuckG}
\end{eqnarray}
where we have used (\ref{qqqqw}). Substituting (\ref{t-dilation}) into (\ref{starbuckG}) we have
\begin{equation}
d = \left( 1 - \frac{v^2}{c^2} \right)^{1/2} d' \, .
\label{l-contraction}
\end{equation}

Note that Einstein postulates retain Galilean invariance, i.e. there is no experiment that can detect a uniform state of motion. However,  the transformation rule (\ref{eq:Galileo})  {\it must be changed}. Indeed, since the way that light travels is determined from Maxwell's equations, we have to find the transformation law between inertial observers that will preserve Maxwell's equations. Another way to say this is that we know the correct transformations of space and time between inertial observers must be such that Maxwell's equations are invariant. Actually, it is even more general than that: we will have a set of transformations that leave a certain velocity, the speed of light, invariant. \\

{\bf EXERCISE 5.1}~{\it (i)}~The escape velocity from Earth is $4 \times 10^4$~km/h. What would be the percent decrease in length of a 95.2~m long spacecraft traveling at that speed? {\it (ii)}~At what speed do the relativistic formulas for length and time intervals differ from the classical values by 1\%? {\it (iii)}~A space explorer $A$ sets off at a steady $0.95c$ to a distant star. After exploring the star for a short time, he returns at the same speed and gets home after a total absence of $80~{\rm yr}$ (as measured by earth-bound observers). How long do $A$'s clocks say he was gone, and by how much has he aged as compared to his twin $B$ who stayed behind on Earth. [Note: This is the famous ``twin paradox.'' It is fairly easy to get the right answer by judicious insertion of a factor of $\gamma$ in the right place, but to understand it, you need to recognize that it involves three inertial frames: the earth-bound frame $S$, the frame $S'$ of the outbound rocket, and the frame $S''$ of the returning rocket. Write down the time dilation formula for the two halves of the journey and then add. Noticed that the experiment is not symmetrical between the two twins: $B$ stays at rest in the single inertial frame $S$, but $A$ occupies at least two different frames. This is what allows the result to be unsymmetrical.]\\

{\bf EXERCISE 5.2}~The muons created by cosmic rays in the upper
atmosphere rain down more-or-less uniformly on the earth's surface,
although some of them decay on the way down, with a half life of about
2.2~$\mu$s (measured in their rest frame). A muon detector is carried in a
balloon to an altitude $h = 2000$~m, and in the course of an hour detects 650
muons travelling at $0.99c$ toward the earth. If an identical detector
remains at sea level, how many muons should it register in one hour?
Calculate the answer taking account of the relativistic time dilation and
classically. (Remember that after $n$ half-lives, $2^{-n}$ of the original
particles survive.) Needless to say, the relativistic answer agrees with
experiment.\\

\subsection{Lorentz transformations}

\begin{figure}[tbp] \postscript{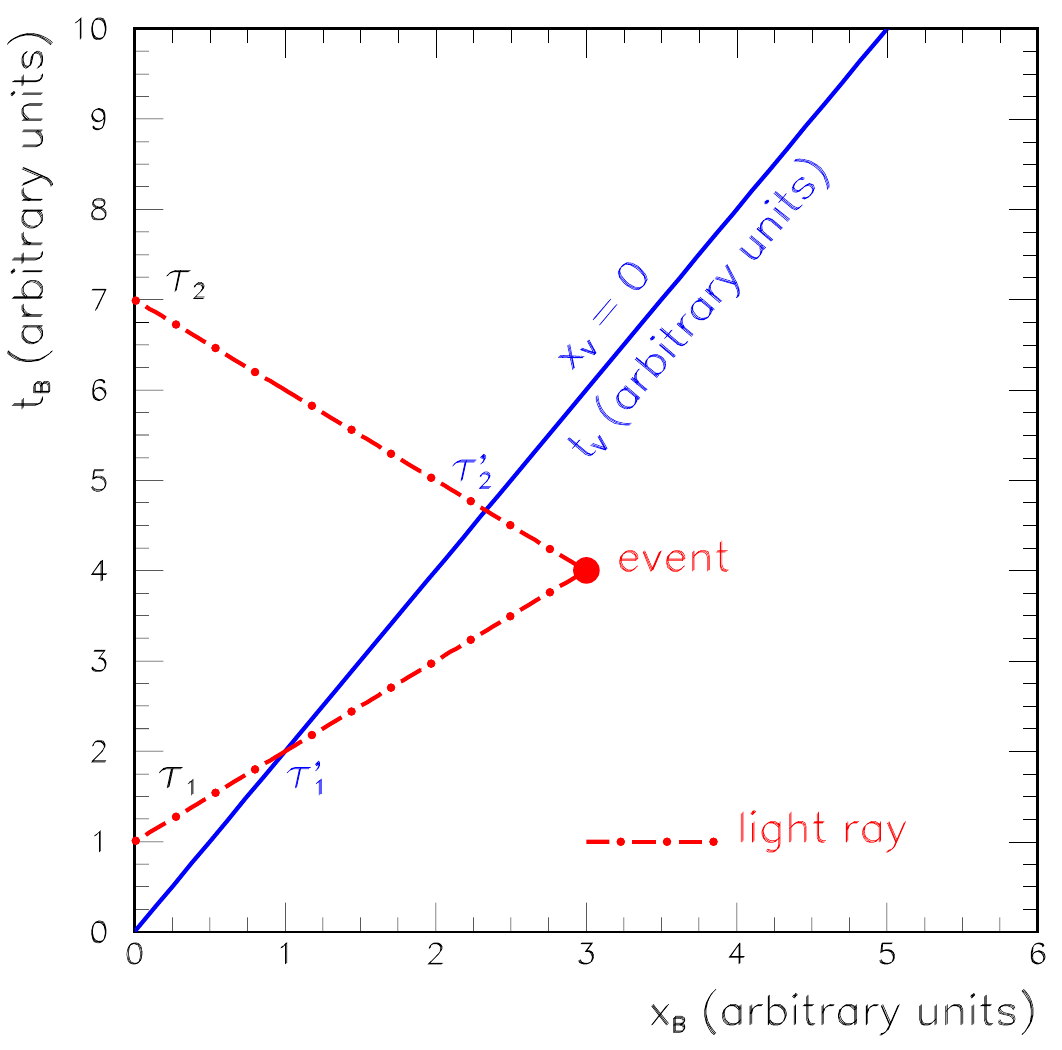}{0.85} \caption{The rules for coordinatizing an event for two relatively moving observers. Note that the times $\tau_1$, $\tau_2$, $\tau'_1$, and $\tau'_2$ are the (proper) times read on each of the observer's clocks. Brittany (the unprimed observer) measures $\tau_1$ (emitted ray) and $\tau_2$ (reflected ray), whereas Vinnie (the prime observer) measures $\tau'_1$ and $\tau'_2$. Recall that Vinnie's coordinate grid is skewed with respect to Brittany's coordinate system. }
\label{red-event}
\end{figure}

We begin by stressing that when relatively moving observers label an event all observers must use the same two light rays, see Fig.~\ref{red-event}. In other words, any event is characterized uniquely by the two light rays that pass through it; all observers that are finding the labels of a particular event use the same transmitted and received rays. This apparent coincidence is actually a reflection of the fact that all observers agree on the speed of light and that the intersection of two light rays is an event and thus a unique label of an event.

Now consider our two observers, Vinnie and Brittany, that share the same origin and want to coordinatize the same red event. The transverse coordinates are the same for Vinnie and Brittany and we used this information to construct our clocks. The red event in Fig.~\ref{red-event}  is coordinatized by Brittany as ($x_{\rm B},t_{\rm B}$). By definition, Vinnie would label it ($x_{\rm V},t_{\rm V}$). The Lorentz transformations are the relationship between ($x_{\rm B}, t_{\rm B}$) and $(x_{\rm V}, t_{\rm V})$.

Start by finding the coordinates of Vinnie's proper times $\tau'_1$ and $\tau'_2$ in terms of the coordinates of the red event in Brittany's reference frame. Event $\tau'_1$ (i.e. the light ray emitted by Vinnie) has the form $(vt_1, t_1)$ in Brittany's coordinates, since it is on Vinnie's time axis and he is moving at a speed $v$ with respect to her. This event is also on a light ray with the red event. The equation of that light ray is
\begin{equation}
x - x_{\rm B} = c (t - t_{\rm B} )\, .
\end{equation}
Putting in the coordinates of the event $\tau'_1$ which is on this line,
\begin{equation}
v t_1 - x_{\rm B}  = c (t_1 - t_{\rm B}) \, .
\end{equation}
Solving for $t_1$ we have
\begin{equation}
t_1 = \frac{c t_{\rm B} - x_{\rm B}}{c -v} \, .
\end{equation}
Because of time dilation,
\begin{equation}
\tau'_1 = t_1 \sqrt{1 - v^2/c^2} \, .
\end{equation}
Combining these
\begin{equation}
\tau'_1 = \frac{ct_{\rm B} - x_{\rm B}}{c-v} \sqrt{1- \frac{v^2}{c^2} }\, .
\label{tauprime1}
\end{equation}
Similarly for event $\tau'_2$
\begin{equation}
\tau'_2 = t_2 \sqrt{1 - \frac{v^2}{c^2}}
\end{equation}
and 
\begin{equation}
\tau'_2 = \frac{ct_{\rm B} + x_{\rm B}}{c-v} \sqrt{1- \frac{v^2}{c^2} }\, .
\label{tauprime2}
\end{equation}
Substituting (\ref{tauprime1}) and (\ref{tauprime2}) into the definitions (\ref{distance-def}) and (\ref{time-def}), 
and doing some straightforward algebra we obtain
\begin{eqnarray}
x_{\rm V}  & = & \frac{x_{\rm B} - v t_{\rm B}}{\sqrt{1-v^2/c^2}} = \gamma (x_{\rm B} - v t_{\rm B}) \,,\nonumber \\
y_{\rm V} & = & y_{\rm B} \,,\nonumber \\
z_{\rm V} & = & z_{\rm B} \,,\nonumber \\
t_{\rm V} &=& \frac{t_{\rm B} - v x_{\rm B}/c^2}{\sqrt{1-v^2/c^2}} = \gamma (t_{\rm B} - v x_{\rm B}/c^2) \,,
\end{eqnarray} 
which are the appropriate Lorentz transformations, for the case in which the transverse directions are unaffected by the velocity transformation~\cite{Lorentz:1904}.

An interesting feature of these relations is that the combination $(ct_{\rm V})^2 - (x_{\rm V})^2 - (y_{\rm V})^2 - (z_{\rm V})^2$ does not involve the velocity and is therefore equal to Brittany's coordinates for the same event,
$(ct_{\rm B})^2 - (x_{\rm B})^2 - (y_{\rm B})^2 - (z_{\rm B})^2$.
This is a special case of the general form for the invariants of the Lorentz transformations.\\

\begin{figure}[tbp] \postscript{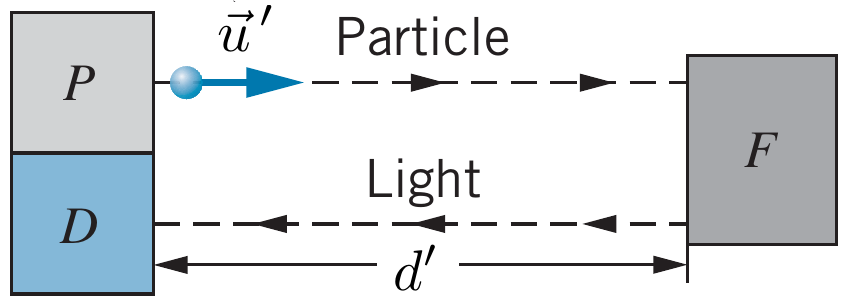}{0.85} \caption{Einstein timing device in which  a particle is emitted by $P$ at a speed $u'$. When the particle reaches $F$, it triggers the emission of a flash of light that travels to the detector $D$~\cite{Krane}.}
\label{Einstein-clock-particle}
\end{figure}

{\bf EXERCISE 5.3}~Suppose that in Einstein's timing device is now modified as shown in Fig.~\ref{Einstein-clock-particle}.  A source $P$ emits particles that travel at speed $u'$ according to Vinnie who is at rest with respect to the device. The flashing bulb $F$ is triggered to flash when a particle reaches it. The flash of light makes the return trip to the detector $D$, and the clock ticks. The time interval $\Delta t'$ between ticks measured by Vinnie is composed of two parts: one for the particle to travel the distance $d'$ at speed $u'$ and another for the light to travel the same distance at speed $c$. Describe the sequence of events as seen by Brittany relative to whom Vinnie moves at speed $v$. Derive the component of the  Lorentz velocity transformation in the direction of Vinnie's motion with respect to Brittany.\\

{\bf EXERCISE 5.4}~Two rockets are leaving their space station along perpendicular paths, as measured by an observer on the space station. Rocket 1 moves at $0.60c$ and rocket 2 moves at $0.80c$, both measured relative to the space station. What is the velocity of rocket 2 as observed by rocket 1? \\

{\bf EXERCISE 5.5}~Coyote notices that the Road Runner is about to
take a train and he decides to look for a good location for
blowing up the train. He notices a bridge that crosses a great
ravine that the train will pass over. Using Acme binoculars it appears to the Coyote,
while he is standing on the side of the bridge and as the train is
coming towards him, that the train takes up the entire length of
the bridge in one moment. He knows the train will then pass by him
to go to the station to pick up the Road Runner among other
passengers, and will make a return trip, thus once again passing
over the same bridge. The Coyote decides to place three Acme bombs
strategically on the bridge, one on each end and one directly in
the middle. He plans on blowing them up all at the same time while
the train is occupying the entire bridge. He lines up the Acme
wiring to the bombs in such a way that they are synchronized with
{\it infinite} precision in his reference frame. The wiring is
connected to an Acme switch placed on his side of the bridge. He
now returns to the train station to ensure his plans are in place
and that the Road Runner has taken his seat on the train. The Road
Runner spots the Coyote just as the train is leaving and leaves
his seat to run onto the track in front of the train.  The Coyote
realizes that to catch the Road Runner he must take the train to
follow him and climbs to the top of the train and is soon
following right behind the Road Runner. As the train approaches
the bridge the Coyote sees the Road Runner is right in front of
the switch and ready to push it with his beak. Since the train is
moving at relativistic speed $v$, the Coyote sees the bridge
contraction and he is sure that most of the train cars (and
especially his car in the front of the train) will make it across
the bridge safely. The Road Runner flips the switch and the three
bombs blow up synchronized in his (the Road Runner's) reference
frame when the train occupies the entire bridge. Describe
precisely how the artist will draw the cartoon's
events from the view of a passenger riding the train. ``Beep! Beep!''

\section{Minkowski Spacetime}

\subsection{Causal structure}

As should have become clear to any attentive reader, the primary unit in special relativity is the event.  The ``arena'' in which all of the events in the universe take place is known as Minkowski spacetime~\cite{Minkowski}.  The basic assumption of special relativity is that  events take place in a four dimensional structure that contains a three dimensional Euclidean space and a time dimension, that is a $(1,3)$ spacetime that has the interval,
\begin{equation}
\Delta s^2 = c^2 \Delta t^2 - \Delta x^2 - \Delta y^2 - \Delta z^2 \,,
\label{deltas}
\end{equation}
as the invariant measure for the Lorentz-Poincar\'e transformations. It is easily seen that the interval is invariant under Lorentz transformations. Since it is defined by differences in coordinates, it is also invariant under translations in spacetime, the so-called ``Poincar\'e invariance.''  This $(1,3)$ spacetime is different from the Euclidean space plus time of Newtonian physics in that the group of transformations that governs it, the Lorentz-Poincar\'e transformations, preserve a different measure. The Galilean transformations (\ref{eq:Galileo}) preserve the usual Pythagorian distance measure (\ref{length}), which is invariant under rotations and spatial translations. The formal consequences of the Lorentz-Poincar\'e group of symmetries are much simpler, but not different in nature, when looking at our textbook example: the $(1, 1)$ world, in which there are no rotations. In fact, it is legitimate to consider the $(1,3)$ spacetime as the $(1, 1)$ spacetime with rotations added.

\begin{figure*}[tbp] 
\begin{minipage}[t]{0.32\textwidth}
\postscript{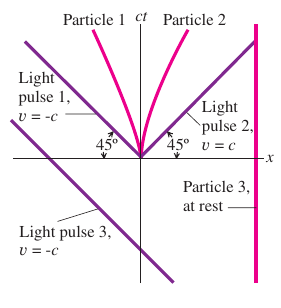}{0.99} 
\end{minipage}
\hfill
\begin{minipage}[t]{0.32\textwidth}
\postscript{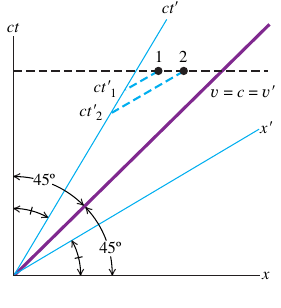}{0.99} 
\end{minipage}
\hfill
\begin{minipage}[t]{0.34\textwidth}
\postscript{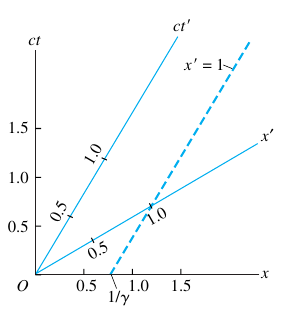}{0.99} 
\end{minipage}
\caption{{\bf Left.} A spacetime diagram showing trajectories of 3 light pulses and 3 particles. Particles 1 and 2 leave $x = 0$ at $t = 0$, accelerating from rest in opposite directions. {\bf Middle.}~The $ct'$- and $x'$-axes drawn on a $ct$-$x$ spacetime diagram. Notice the two sets of equal angles. {\bf Right.}~A spacetime diagram for
$v = 0.60c$. The dashed line $x'= 1$ intercepts the $x$-axis at $x = 1/\gamma =  0.80$.}   
\label{spacetime-diagram} 
\end{figure*}

A trajectory is the connected set of events that represents the places and times through which a particle moves. Trajectories of massive particles and observers are called worldlines. At any event on a trajectory, the slope is the inverse of the velocity relative to the inertial observer that has a time axis  ($x = 0$)  straight up and  a perpendicular set of simultaneity lines of constant $ct$.  We use $ct$ rather than $t$ so that both scales can have the same unit. The path of a particle forms a  worldline as the particle moves in one-dimensional motion. At any point, the slope of the worldline is $d(ct)/dx = (c dt)/(v dt) = c/v$. Thus a light pulse with $\pm c$ speed has a slope of $\pm 1$ on a spacetime diagram, giving angles of $45^\circ$ with the $\pm x$-axes. Since massive particles have speeds less than $c$, all worldlines are steeper than those $45^\circ$ angles. That is, nothing known has a worldline with a slope between $-1$ and $1$. The worldline of a particle at rest is vertical and so has infinite slope. In Fig.~\ref{spacetime-diagram} we show 3 worldlines and 3 light rays. 

How does the $S'$ reference frame appear on a $ct$-$x$ spacetime diagram? Recall that we always set $x' = 0$ at $x = 0$ when $t' = 0 = t$. As in our previous example, let Vinnie move in $S'$ at a speed $v$ in the $+x$-direction. Now, $x' = 0$ all along the $ct'$-axis, thus the $ct'$-axis have a worldline of slope $c/v$. For example, if $v = 0.600c$, the $ct'$-axis is at an angle of $\tan^{-1} (1/0.600) = 59.0^\circ$ from the $x$-axis. Surprisingly enough, the $x'$-axis is not drawn perpendicular to the $ct'$-axis. Since $ct' = 0$ all along the $x'$-axis, the Lorentz transformation for $t'$ gives $(t - vx/c^2) = 0$ or $ct = (v/c)x$ for the $x'$-axis. Thus the $x'$-axis is drawn with a slope of $v/c$. For $v = 0.600c$, the $x'$-axis is at an angle of $\tan^{-1} (0.600) = 31.0^\circ$ from the $x$-axis. That is, the $x'$-axis makes the same angle with the $x$-axis as the $ct'$-axis makes with the $ct$-axis. In Fig.~\ref{spacetime-diagram} we show that the trajectory of a light pulse leaving $x' = 0 = x$ at $t' = 0 = t$ with a velocity $+c$ bisects the angle between either set of axes.

Now, assume Brittany measures events 1 and 2 to occur simultaneously in $S$ at positions $x_1$ and $x_2$, where $x_2 > x_1$. We will use spacetime diagrams to show that Vinnie, who moves in the positive $x$-direction relative to Brittany, measures event 2 to occur before event 1. Events that are simultaneous in $S$ have the same time $t$, so in Fig.~\ref{spacetime-diagram} we draw a dashed line parallel to the $x$-axis (constant $t$). We put a dot on that line for event 1, and farther from the $ct$-axis we put
another dot for event 2 (because $x_2 > x_1$). How do we read a value of a point on a graph? We draw a line through that point parallel to one axis and measure where it intercepts the other axis. Then, to measure the times of the two events in Vinnie's $S'$ frame, in Fig.~\ref{spacetime-diagram} we draw dashed lines parallel to the $x'$-axis that intercept the $ct'$-axis at $ct'_1$ and $ct'_2$. We see that $ct'_2 < ct'_1$. The events are not simultaneous in $S'$, and Vinnie measures event 2 to occur before event 1~\cite{Gleeson}. Let us now discuss a simple example of length contraction. Vinnie, at rest in frame $S'$, holds a stick with its left end at $x' = 0$ and its right end at $x' = 1$. (We have left off the unit for generality.) At any time $t'$ measured in frame $S'$, the left end is at $x' = 0$ (on the $ct'$-axis) and the right end is on the $x' = 1$ dashed line. In frame $S$, at $t = 0$ (on the $x$-axis) we see from Fig.~\ref{spacetime-diagram} that the left end of the stick is at $x = 0$ and the right end is at $x = 1/\gamma$. Thus in $S$ the stick has a contracted length of $1/\gamma$.

\begin{figure}[tbp] \postscript{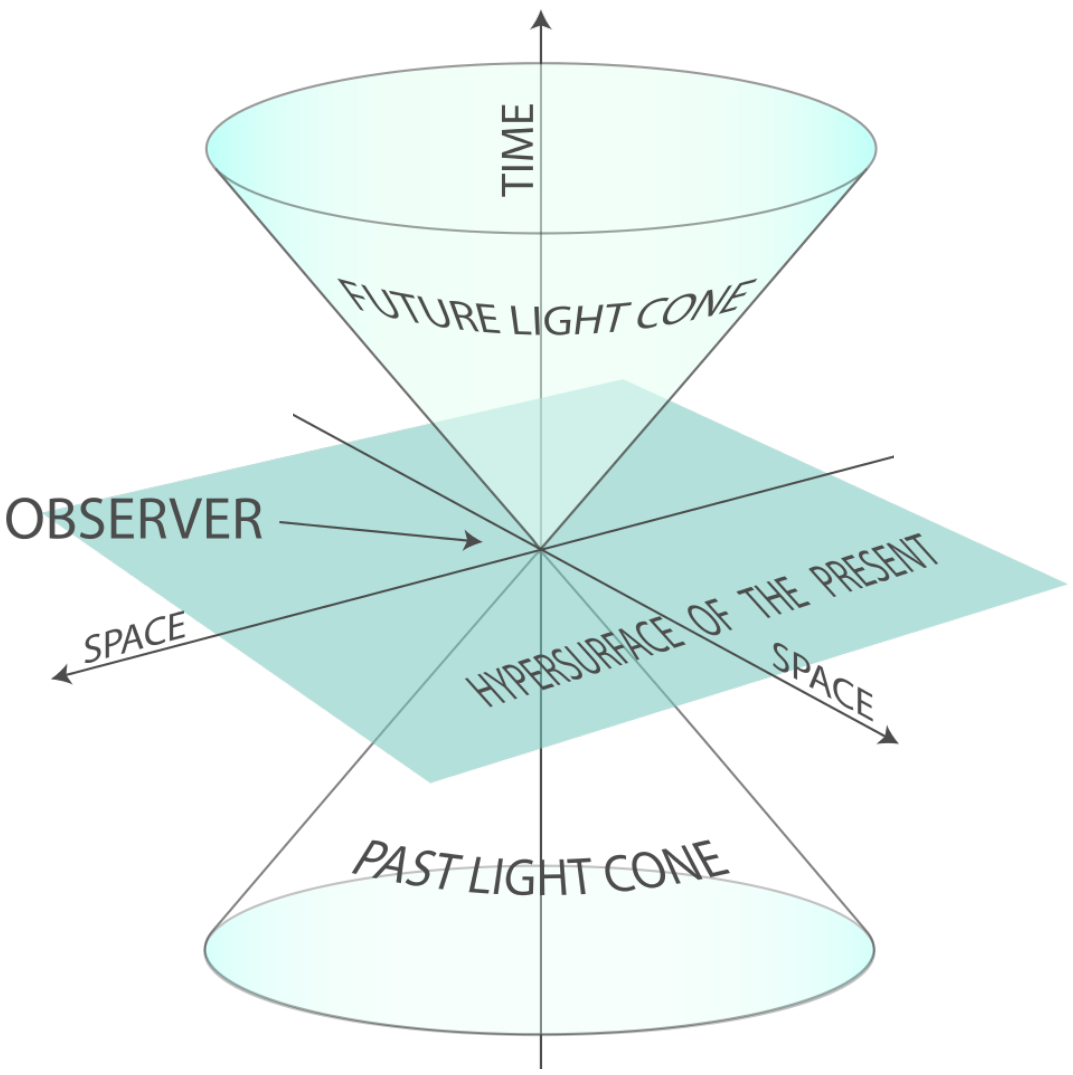}{0.85} \caption{Future, past, and elsewhere. For any event, in this case the event at the vertex of the two cones, all the other events in spacetime can be categorized into a future, a past, and an elsewhere. Since the trajectories of light rays are unchanged by Lorentz transformation, this classification of the relationship between two events is the same for all inertial observers.}  
\label{lightcone} 
\end{figure}

Spacetime around any one event is divided into regions separated by the trajectories of light rays emanating from that event, see Fig.~\ref{lightcone}. This separation of events is the same for all Lorentz observers since the light ray trajectories are unchanged by the Lorentz transformations. All the events in the upper light cone are the future of the event in question. This is in the sense that, from the origin event and any event in the future, there exists an inertial observer for whom the interval between these two events is a pure time, $(\tau, \vec 0 )$, i.e. no spatial separation, and that the time of the other event is after the now of our original event, $\tau > 0$.  Similarly, events in what is called the backward light cone from our original event are in the past. There exists an inertial observer for whom the second event is a pure time, $(\tau ,\vec 0)$, but in this case $\tau < 0$. 
The union of the events in the past and of the events in the future to our origin event is the set of time-like events relative to our origin event, that is all events relative to the original event with intervals in any inertial coordinate system such that $\Delta s^2 >0$. If the event under discussion is in the upper light cone or future of our origin event, it is characterized by the positive sign of the square root of the  interval 
and if the event is in the lower or past light cone by the negative sign. This is called the proper time $\tau$ between the events. 

There are clearly a large number of events that are not time-like relative to our origin event. These are called elsewhere or space-like events relative to our origin event. Similar to our construction of future and past, for any elsewhere event there exists a Lorentz observer for whom the events are separated by a spatial interval, $(0,\vec r)$. \\

{\bf EXERCISE 6.1}~A ``cause'' occurs at point 1, with coordinates $(x_1 , t_1 )$ in the reference frame $S$, and its ``effect''  occurs at point 2, with coordinates $(x_2 , t_2 )$. Use the Lorentz transformation to find $t'_2 - t'_1$, and show that $t'_2 - t'_1 >0$;  that is, $S'$ (moving at velocity $v$ with respect to $S$) can never see the ``effect'' coming before its ``cause.''

\subsection{Lorentz invariance}

In the previous section, we introduced the idea of  Minkowski spacetime. In this section, we will  develop an efficient formalism for expressing ideas in Minkowski spacetime. As in Euclidean space, a vector formalism is possible. Given an origin event and an inertial observer, a coordinate system can be established. An event is a place and a time, a set of four numbers, $(t,\vec x)$, that specifies that event in that coordinate system.\footnote{Note that $\vec r \equiv \vec x$; we will use both notations interchangeably hereafter, rather than settle on one particular type of notation.} We can designate the coordinates with an index $x^\mu$ with $x^0 = ct$, $x^1 = x$, $x^2 = y$, and $x^3 = z$.  In this notation, a Lorentz transformations is expressed by
\begin{equation}
x'^\mu = \sum_{\nu = 0}^3 {\Lambda^\mu}_\nu \ x^\nu \,,
\label{LTsum}
\end{equation}
with 
\begin{eqnarray}
\Lambda^0_{\phantom{0} 0} & = & \gamma, ~~~~~~~~~~~~~~~~~~~~~~~~~~~~~~ \Lambda^i_{\phantom{i} 0} =
\gamma v^i/c, \nonumber \\
\Lambda^i_{\phantom{i} j} & = & \delta^i_{\phantom{i}j} + (\gamma -1)
\frac{v^i\, v_j}{v^2}, ~~~~~~~~~ \Lambda^0_{\phantom{0} j} =
\gamma v_j/c \,, 
\label{Lorentz-matrix}
\end{eqnarray}
where $\gamma = (1 - v^2/c^2)^{-1/2}$.\footnote{In what follows, Greek indices $(\mu, \nu, \cdots)$ run from 0 to 3 and Latin indices $(i,j,\cdots)$ from 1 to 3.} The rotations, which are a subgroup of the Lorentz transformations, are the described by the usual rotation elements operating in the three by three Euclidean subspace of the four by four matrix ${\Lambda^\mu}_\nu$. There is a broadly accepted convention introduced by Einstein, which simplifies the notation considerably by eliminating the summation symbol for cases in which the same index appears up and down in the same equation. In this notation (\ref{LTsum}) appears simply as
\begin{equation}
x'^\mu = {\Lambda^\mu}_\nu \ x^\nu \, .
\label{LT-nosum}
\end{equation}

Given two events we can talk about the interval between them. In this
language, there is a four vector
\begin{equation}
\Delta x^\mu = ( c(t_2 - t_1), (x_2-x_1), (y_2 - y_1) , (z_2 - z_1)) \, ,
\end{equation}
such that the invariant interval (\ref{deltas}) is now expressed by
\begin{eqnarray}
\Delta s^2 & = & g_{\mu \nu} \ \Delta x^\mu \ \Delta x^\nu \nonumber \\
& = & c^2 \Delta t^2 - \Delta x^2 - \Delta y^2 -\Delta z^2 \,,
\label{interval0}
\end{eqnarray}
where
\begin{equation}
g_{\mu \nu} =  \left(
                \begin{array}{cccc}
                1 & 0 & 0 & 0
                \\ 0 & -1 & 0 & 0
                \\ 0 & 0 & -1 & 0
                \\ 0 & 0 & 0 & -1
                \end{array}
        \right)
\end{equation}
is the metric tensor of Minkowski spacetime. The metric tensor arises directly from the physics of spacetime. It 
can be used to lower indices, as in 
\begin{equation}
x_\mu = g_{\mu \nu} \ x^\nu \, . 
\end{equation}
The inverse metric $g^{\mu \nu}$ is defined so that $g^{\mu \nu} g_{\nu \zeta} = {\delta^\mu}_\zeta$, that is $g^{\mu \nu}$ raises indices, 
\begin{equation}
g^{\mu \nu} \, x_\nu = g^{\mu \nu} (g_{\nu \zeta} \, x^\zeta) = (g^{\mu \nu} g_{\nu \zeta}) x^\zeta = {\delta^\mu}_\zeta \, x^\zeta = x^\mu.
\end{equation}  
For details, see Appendix~\ref{appendixA}.

The invariance of  the interval $\Delta s^2 = {\Delta s'}^2$ places a constraint on the form of (\ref{LT-nosum}), 
\begin{equation}
g_{\mu \nu} x'^\mu x'^\nu = g_{\mu \nu} \ {\Lambda^\mu}_\alpha \ {\Lambda^\nu}_\beta \ x^\alpha x^\beta \, =
g_{\alpha \beta} x^\alpha x^\beta \, , \label{1}
\end{equation}
which implies 
\begin{equation}
g_{\alpha \beta} = g_{\mu \nu} {\Lambda^\mu}_\alpha {\Lambda^\nu}_\beta \, .
\label{dummies}
\end{equation}
This condition can be used as the defining equation for a Lorentz transformation. The  sixteen elements of a $4 \times 4$ matrix are the coefficients of a Lorentz transformation if they satisfy (\ref{dummies}). Note that since $g_{\mu\nu}$ is symmetric there are only ten independent equations which leaves six free parameters. Indeed, this is just what is needed -- three parameters to label the boost relative velocity and three parameters to label rotations in a three space.\\

{\bf EXERCISE 6.2}~Show that $\Lambda^\mu_{\phantom{ \mu} \nu}$, with elements given by (\ref{Lorentz-matrix}), is a matrix associated to a Lorentz transformation.\\

Having introduced the idea of 4-vectors, let us now turn to discuss their use for describing the motion of a particle in spacetime terms. Consider a time-like (or light-like) interval as
expressed by (\ref{interval0}) in any Lorentz frame separating points on a trajectory of a particle. The same interval can be expressed in coordinates such that at each moment the particle is at rest. Such a frame is called an instantaneous rest frame. Since in the instantaneous rest frame the particle is at rest, because of the invariance of the interval without loss of generality we may write 
\begin{equation}
\Delta s^2 = c^2 \ \Delta \tau^2 \, .
\label{taurelation}
\end{equation}
Because the interval is assumed time-like (or light-like), we could take the square root of (\ref{taurelation}) to define the  proper time interval
\begin{equation}
\Delta \tau = \frac{1}{c} \, \Delta s \, .
\end{equation}
For a curved
trajectory to be time-like, each segment must be time-like. A cumulative time can be assigned to a time-like trajectory $(t_0,x_0; t_f,x_f)$ by adding the proper
time for each segment of a sensibly rectified approximation to the curve,
\begin{eqnarray}
\tau [{\rm traj}] & = & \sum_{i=0}^{f-1} \left[ (t_{i+1} - t_i)^2 - \frac{1}{c^2} ( x_{i+1} - x_i)^2  \right. \nonumber \\ &- & \left. \frac{1}{c^2} (y_{i+1} - y_i)^2 
- \frac{1}{c^2}  (z_{i+1}- z_i)^2 \right]^{1/2}  \, .
\end{eqnarray}
In the 
limit of small segments, this cumulative time is the proper time over the trajectory
\begin{equation}
 \tau [{\rm traj}] = \int_{(t_0,x_0)}^{(t_f, x_f)} \frac{1}{c} \ ds \, .
\end{equation}

Now, consider a particle which follows a timelike wordline through spacetime. This curve can be specified by giving three spatial coordinates $x^i$ as a function of time in a particular inertial frame. Alternatively, the four dimensional way of describing a worldline is to give all 4 coordinates of the particle $x^\mu$ as a single-valued function of a parameter that varies along the worldline. Many parameters are possible, but a natural one is the proper time that gives the spacetime distance $\tau$ along the worldline.  A worldline is then described by the equations
\begin{equation}
x^\mu = x^\mu (\tau) \, .
\end{equation}
 In other words, a trajectory which is a connected set of events that would be coordinatized by some inertial observer as $(t, \vec x(t))$, can be parametrized by the elapsed proper time of the time-like trajectory, $(t(\tau), \vec x (\tau))$, where
\begin{eqnarray}
 \tau [{\rm traj}] & = & \int_{(t_0,x_0)}^{(t_f, x_f)} \sqrt{1 -\frac{1}{c^2} \frac{d\vec x}{dt} \cdot \frac{d\vec x}{dt}} \ dt \nonumber \\
& = & \int_{(t_0,x_0)}^{(t_f, x_f)} \sqrt{1 - \frac{\vec u \cdot \vec u}{c^2}} \ dt ,
\label{tauintegral}
\end{eqnarray}
and where $\vec u = d\vec x /dt$ is the 3-velocity vector. The elapsed proper time is a functional of the trajectory but is a function of the labels of the events at the end points of the integral. Since it is a function of the time on the trajectory we can derive a differential form of (\ref{tauintegral}),
\begin{equation}
\frac{d\tau}{dt} = \sqrt{1 - \frac{\vec u \cdot \vec u}{c^2}} \, .
\label{taudiff}
\end{equation}
For a time-like trajectory, we can now define a four vector velocity $\bm{U}$, whose components $U^\mu$ are the derivatives of the position along the worldline with respect to the
proper time parameter
\begin{equation}
 U^\mu = \frac{dx^\mu}{d\tau} = \frac{dx^\mu/dt}{d\tau/dt} \, .
\label{umu}
\end{equation}
The 4-velocity $\bm{U}$ is thus tangent to the world line at each point
because a displacement is given by $\Delta x^\mu = U^\mu \Delta \tau$.  The four components of the 4-velocity vector $\bm{U}$ can be expressed in terms of the 3-velocity $\vec u$.  Substituting (\ref{taudiff}) into (\ref{umu}),  
for the 0-th component we have
\begin{equation}
\frac{dx^0}{d\tau} = U^0 = c \frac{dt}{d \tau} = \frac{c}{(1 - u^2/c^2)^{1/2}} \, ,
\end{equation}
whereas for $\mu = 1,2,3$ we get
\begin{equation}
\vec U = \frac{1}{(1 - u^2/c^2)^{1/2}} \frac{d\vec x}{dt} = \frac{1}{(1- u^2/c^2)^{1/2}} \ \vec u \,,
\end{equation}
Therefore, 
\begin{equation}
\bm {U} \equiv U^\mu = (\gamma c, \gamma \vec u) \, .
\label{4velocity}
\end{equation}
An immediate consequence of this result is that
\begin{equation}
\bm{U} \cdot \bm{U}   =  g_{\mu \nu} U^\mu U^\nu = g_{\mu \nu} \frac{(dx^\mu/dt) (dx^\nu/dt)} {(d\tau/ dt)^2} 
  =  c^2
\label{itsecure}
\end{equation}
so that the 4-velocity is always a time-like and future-pointing 4-vector.\\

{\bf EXERCISE 6.3} Show that if you change from a reference frame $S$ to $\overline S$, which moves with velocity $v$ relative to $S$ along a common $x$-axis, the Lorentz transformation of $U$ is given by
\begin{eqnarray}
\overline U^0 & = & \gamma (U^0 - \beta U^1) \,,\nonumber \\
\overline U^1 & = & \gamma (U^1 - \beta U^0) \,, \nonumber \\
\overline U^2 & = & U^2 \,,\nonumber \\
\overline U^3 & = & U^3 \, ,
\end{eqnarray}
where $\beta = v/c$.\\

The relation between the 4-acceleration vector
\begin{equation}
\bm{A} = \frac{d \bm{U}}{d\tau} \equiv \frac{d^2 x^\mu}{d \tau^2}
\label{4accel}
\end{equation}
and the 3-aceleration vector 
\begin{equation}
\vec a = \frac{d^2 x^i}{dt^2} 
\end{equation}
is more complicated
\begin{eqnarray}
\bm{A}  & = &  \frac{d\bm{U}}{d\tau} = \gamma \frac{d\bm{U}}{dt} = \gamma  \frac{d}{dt} (\gamma c , \gamma \vec u) \nonumber \\
 & = &  \gamma \left(\frac{d \gamma}{dt} c, \frac{d\gamma}{dt} \vec u + \gamma \vec a\right) \, .
\label{4acceleration}
\end{eqnarray}
Note that since
\begin{equation}
\frac{d\gamma}{dt} =  \frac{\vec u \cdot d\vec u/dt}{c^2 \ (1 - u^2/c^2)^{3/2}},
\end{equation}
in the instantaneous rest frame of the particle $(\vec u=0)$ and hence (\ref{4acceleration}) simplifies to $\bm{A} = (0, \vec a)$. Thus, $\bm{A} =0$ if and only if the {\it proper acceleration}, i.e., the magnitude of the 3-acceleration vector in the rest frame, vanishes. For $A^2$, being an invariant, has the same value in all frames, so we may as well evaluate it in the rest frame, 
\begin{equation}
\bm{A} \cdot \bm{A} = -\alpha^2
\label{accelsquare}
\end{equation}
where $\alpha$ is the proper acceleration. From  (\ref{accelsquare}) we see that $\bm{A}$ is space-like vector. 
By the same articfice it is easily seen from (\ref{4velocity}) and (\ref{4acceleration})  that
\begin{equation}
\bm{U} \cdot \bm{A} =0 \,,
\label{uaortho}
\end{equation}
i.e., the 4-acceleration vector is always orthogonal to the 4-velocity. \\

{\bf EXERCISE 6.4}~{\it (i)} Suppose that a point $P$ in spacetime with coordinates $x^\mu = (x^0, \vec  x)$ lies inside the backward light cone as seen in frame $S$. This means that $x^\mu  \, x_\mu >0$ and $x^0 < 0$ at least in frame $S$. Prove that these two conditions are satisfied in all frames. Since this means that all observers agree that $t<0$, this justifies calling the inside of the backward light cone the absolute past. {\it (ii)}~Show that the statement that a point $x^\mu$ in spacetime lies on the forward light cone  is Lorentz invariant.\\

{\bf EXERCISE 6.5}~{\it (i)} Show that if $\bm{q}$ is time-like, there is a frame $\overline S$ in which it has the form $\bm{\bar q} = (\bar q_0, 0, 0, 0)$. {\it (ii)}~Show that  if $\bm{q}$ is forward time-like in one frame $S$, then it is forward timelike in all inertial frames.\\

We end with a thought-provoking observation.  Consider a Lorentz transformation in which the new frame (prime coordinates) moves with velocity $v$ along the $z$ axis of the original frame
(unprimed coordinates).  We will leave it to the reader to convince
themselves that
\begin{eqnarray}
ct' & = &   \phantom{-} \cosh (\vartheta) \, \,  ct  -  \sinh (\vartheta) \, \, z  \nonumber \\
z' & = &   -\sinh (\vartheta) \, \,c t +   \cosh (\vartheta) \, \, z 
\end{eqnarray}
with $x$ and $y$ unchanged; here, $\cosh (\vartheta) = (1 - v^2/c^2)^{-1/2}$.  Because $\cos (i \vartheta) = \cosh (\vartheta)$ and $\sin (i \vartheta) = \sinh (\vartheta),$ we see that a Lorentz boost may be regarded as a rotation through an imaginary angle $i\vartheta$ in the $itc$-$z$ plane.\\

For a deeper insight into the properties of Minkowski spacetime see e.g.~\cite{Rindler1,Rindler2}.

\section{Particle Dynamics}

Though there are many approaches to relativistic mechanics, the result is always the same.\footnote{The first development of relativistic mechanics was given by Planck, whose starting point was the relativization of Newton's law of motion~\cite{Planck:1906}. A second soon followed by Lewis and Tolman, who chose as their starting point the relativization of Newton's law of momentum conservation in particle collisions~\cite{Lewis}. A development from energy conservation can be found in~\cite{Ehlers}. 
The academic exposision given herein will build upon the content of the exquisite book by Hartle~\cite{Hartle}.} If Newton's well-tested theory is to hold in the  slow-motion limit, and unnecessary complications are to be avoided, then only one Lorentz-invariant mechanics appears to be possible. Furthermore, it is persuasively elegant, and has been remarkably successful in matching nature perfectly in modern high-speed interactions where Newton's theory is out by many orders of magnitude.

Newton’s first law of motion holds in special relativistic mechanics as well as nonrelativistic mechanics. Note that in the absence of forces, a body is at rest or moves in a
straight line at constant speed. This is summarized by
\begin{equation}
\frac{d\bm{U}}{d\tau} =0 \,,
\label{534}
\end{equation}
since in view of (\ref{4velocity}),  this equation implies that $\vec u$ is constant in any inertial frame.

The objective of relativistic mechanics is to introduce the
analog of Newton’s second law: 
\begin{equation}
\vec F = m \vec a.
\label{Newton2law}
\end{equation}
 There is nothing from which this law can be derived,
but plausibly it must satisfy certain properties:
{\it (i)}~it must satisfy the principle of relativity, i.e., take the same form in every inertial frame;
{\it (ii)}~it must reduce to (\ref{534}) when the force is zero;
and {\it (iii)}~it must reduce to (\ref{Newton2law}) in any inertial frame when the speed of the particle is much less than the speed of light. The choice
\begin{equation}
m \frac{d\bm{U}}{d\tau} = \bm{f}
\label{eqmo}
\end{equation}
naturally suggests itself. The constant $m$, which characterizes the particle's inertial properties, is called the  mass, and $\bm{f}$ is called the 4-force. Requirement {\it (i)}~is satisfied because this is a four vector equation, {\it (ii)}~is evident, and {\it (iii)}~is satisfied with a proper choice of $\bm{f}$. This is the correct law of motion for special relativistic mechanics and the special relativistic generalization of Newton’s second law. Using (\ref{4accel}), the equation of motion (\ref{eqmo}) can be rewritten in the evocative form 
\begin{equation}
\bm{f} = m \bm{A} \, .
\label{eqmo-2}
\end{equation}
Although this represents 4-equations, they are not all independent. The normalization of the 4-velocity (\ref{4velocity}) implies 
\begin{equation}
m \frac{d(\bm{U} \cdot \bm{U})}{d \tau} = 0 \,,
\end{equation}
which from (\ref{uaortho}) implies 
\begin{equation}
\bm{f} \cdot \bm{U} = 0 \, .
\label{fuortho}
\end{equation}
Now, (\ref{fuortho})  shows that there are only three independent equations of motion -- the same number as in Newtonian mechanics.

The equation of motion (\ref{eqmo}) leads naturally to the relativistic ideas of energy and momentum.
If the 4-momentum is defined by
\begin{equation}
\bm{p} = m \, \bm{U} \,,
\end{equation}
then the equation of motion (\ref{eqmo}) can be rewritten as 
\begin{equation}
\frac{d\bm{p}}{d \tau} = \bm{f} \, .
\label{eqmo666}
\end {equation}
An important property of the 4-momentum follows from its definition and the normalization of the 4-velocity
\begin{equation}
p_\mu \, p^\mu = m^2 c^2 \, .
\label{invariant-mass}
\end{equation}
In view of (\ref{4velocity}) the components of the 4-momentum are related to the 3-velocity $\vec u$ in an inertial frame according to
\begin{equation}
p^0 = \frac{mc}{\sqrt{1-u^2/c^2}} \quad {\rm and} \quad \vec p = \frac{m\vec u}{\sqrt{1-u^2/c^2}} \, .
\label{3M}
\end{equation}
For small speeds $u \ll c$,
\begin{equation}
p^0 = m c + \frac{1}{2} m \frac{u^2}{c} + \cdots \quad {\rm and} \quad \vec p = m \vec u + \cdots \, .
\end{equation}
Therefore,  at small velocities $\vec p$ reduces to the usual three-momentum, whereas $p^0$ reduces to the kinetic energy per units of $c$ plus the particle's mass in units of $c$. For this reason $\bm{p}$ is also called energy-momentum four-vector, and its components in an inertial frame are given by
\begin{equation}
p^\mu = (E/c, \vec p) = (m \gamma c, m \gamma \vec u) \,,
\label{4momento}
\end{equation}
where $E$ is the energy. Note that (\ref{invariant-mass}) can be solved for the energy in terms of the 3-momentum to give
\begin{equation}
E = (m^2 c^4 + {\vec p}^{\, 2} c^2)^{1/2} \,,
\label{mcsq}
\end{equation}
which shows how the mass is part of the energy of the relativistic particle. Indeed, for a particle at rest (\ref{mcsq}) reduces to $E = mc^2$.

In a particular inertial frame the connection between the relativistic equation  of motion (\ref{eqmo}) and Newton's laws can be made more explicit by defining the 3-force $\vec F$ as
\begin{equation}
\frac{d\vec p}{dt} \equiv \vec F \, .
\label{Newton2again}
\end{equation}
This has the same form as Newton's law but with the relativistic expression for the three momentum (\ref{3M}). The only difference arises from the different relation of momentum to velocity (\ref{3M}). Evidently, 
\begin{equation}
\vec f = \frac{d\vec p}{d\tau} =\frac{d\vec p/dt}{dt /d\tau} = \gamma \vec F \, .
\end{equation} 
Using (\ref{4velocity}) and (\ref{fuortho}), the 4-force acting on a particle can be written in terms of the 3-force as
\begin{equation}
\bm{f} = (\gamma \vec F \cdot \vec u, \gamma \vec F) \, .
\end{equation}
where $\vec u$ is the particle's 3-velocity. The time component of the equation of motion (\ref{eqmo666}) is
\begin{equation}
\frac{dE}{dt} = \vec F \cdot \vec u
\label{vecfdotu}
\end{equation}
which is a familiar relation from Newtonian mechanics. This time component of the equation of motion (\ref{eqmo666})  is a consequence of the other three. Therefore, in terms of the three force, the equations of motion take the same form as they do in usual Newtonian mechanics, but with the relativistic expressions for energy and momentum. When the velocity is small (\ref{4velocity}) shows that the relativistic version of Newton's second law (\ref{Newton2again}) reduces to the familiar nonrelativistic form. Newtonian mechanics is then the low-velocity approximation of relativistic mechanics.\\

{\bf EXERCISE 7.1}~Starting from the defnition of the force $\vec F$ on an object, prove
that the transformation of the components of $\vec F$ as we pass from a frame $S$ to a second frame $S'$ traveling at speed $v$ in the $x$-direction relative to $S$ is
\begin{equation}
F'_x = \frac{F_x - \beta \vec F  \cdot \vec u/c}{ 1 - \beta u_x/c}, \quad \quad F'_{y,z} = \frac{F_{y,z}}{\gamma ( 1 - \beta u_x/c) } ,
\end{equation}
where $\beta = \beta (v)$ and $\gamma = \gamma(v)$ relate to the relative speed of the two frames and $\vec u$ is the velocity of the object as measured in $S$.\\

A point worth noting at this juncture is that when two particles with respective 4-momenta $\bm{p}_a$ and $\bm{p}_b$ are involved, and $v_{ab}$ is their relative speed (that is, the speed of one in the rest-frame of the other) we have
\begin{equation}
\bm{p}_a \cdot \bm{p}_b = m_a E_b = m_b E_a = c^2 \gamma(v_{ab}) m_a m_b,
\label{R615}
\end{equation}
where, typically, $m_a$ is the mass of the first particle and $E_b$ is the energy of the second particle in the rest-frame of the first. For proof we need only evaluate $\bm{p}_a \cdot \bm{p}_b$ in the rest-frame of either particle.

The discussion so far has concerned particles with nonzero mass, which move at speeds less than the speed of light. Let us now consider massless particles that move at the speed of light along null trajectories. Evidently the proper time can no longer be used as a parameter along the trajectory of a light ray -- the proper time interval between any two points on it is zero. However there are many other parameters that could be used. For example, the curve $x=ct$ could be written parametrically as 
\begin{equation}
x^\mu = U^\mu \lambda,
\end{equation}
where $\lambda$ is the parameter and $U^\mu = (c,c,0,0)$.  Note that here $\bm{U}$ is a null vector and thus in contrast to (\ref{itsecure})
\begin{equation}
\bm{U} \cdot \bm{U} = 0 \, .
\end{equation}
Different choices of parametrization will give different tangent 4-vectors, but all have zero length. With this choice of parametrization,
\begin{equation}
\frac{d \bm{U}}{d\lambda} = 0 \,,
 \end{equation}
so that the equation of motion of a light ray is the same as for a particle (\ref{534}).\\

{\bf EXERCISE 7.2}~The light of a distant galaxy is analized using an spectrometer. Studying the distribution of spectral lines leads to identification of  $7,300~\dot {\rm A}$ line of hydrogen, which in the lab has $\lambda = 4,870~\dot{\rm A}$. If the change in $\lambda$ if due to the Doppler effect, determine at what velocity is the galaxy moving with respect to the Earth. If there is independent evidence that indicates the observed galaxy is at $5 \times 10^9$~light years, determine the time at which the galaxy started receding from the Earth, assuming that the recession velocity did not  change over time? (In 1929, Hubble discovered that such a time, wich is the reciprocal of the Hubble constant, is approximately the same for all galaxies~\cite{Hubble:1929ig}. This motivated the idea of an expanding universe.)

\section{Particle Decay and Collisions}
\subsection{Conservation of 4-momentum and all that...}

The basic law of collision mechanics is the conservation of the 4-momentum vector: {\it The sum of the 4-momenta of all the particles going into a point-collision is the same as the sum of the 4-momenta of all those coming out.} (The collision may or may not be elastic, and there may be more, or fewer, or other particles coming out than going in.\footnote{An elastic collision is an encounter between two bodies in which the total kinetic energy of the two bodies after the encounter is equal to their total kinetic energy before the encounter. Elastic collisions occur only if there is no net conversion of kinetic energy into other forms}) We can write this in the form
\begin{equation}
{\sum}^* \bm{p}_i = 0 \,,
\label{xxxc}
\end{equation}
where a different value of $i = 1, 2, \cdots$ is assigned to each particle going in and to each particle coming out, and $\sum^*$ is a sum that counts pre-collision terms positively and post-collision terms negatively. For a closed system, the conservation of the total 4-momentum can be shown to be the result of the homogeneity of spacetime.\footnote{Indeed, the invariance under translations in the description of physical systems (homogeniety of spacetime) implies through Noether theorem~\cite{Noether:1918zz} the conservation of the 4-momentum. The invariance under rotations (isotropy of spacetime) yields the conservation of the angular momentum.} Whether the law is actually true must, of course, be decided by experiment. As you have no doubt guessed, the verdict is clear. Countless experiments have shown that the total 4-momentum of an isolated system is constant.

If we have a system of particles, with 4-momenta $\bm{p}_i$, subject to no forces except mutual collisions, the total 4-momentum  $\bm{p}_{\rm tot} =\sum \bm{p}_i$ is timelike and future-pointing, so there exists an inertial frame $S$ in which the spatial components of $\bm{p}_{\rm tot}$ vanish. $S$ should be called the center-of-momentum frame, but it is usually called the center-of-mass (CM) frame. \\

{\bf EXERCISE 8.1}~{\it (i)}~Show that the 4-momentum of any material particle
($m>0$) is forward timelike. {\it (ii)}~Show that the sum of any two
forward timelike vectors is itself forward timelike, and hence that
the sum of any number of forward time-like vectors is itself forward
time-like. {\it (iii)}~Use the results in {\it (i)} and {\it (ii)} to
convince yourself that for any number of particles there exists a center-of-mass
frame, that is a frame in which the total 3-momentum is zero. {\it
  (iv)}~Relative to an arbitrary frame $S$, show that the velocity of
the center-of-mass frame is given by $\beta_{\rm CM} = \sum_k \vec p_kc/\sum_k E_k.$, and the Lorentz factor is $\gamma_{\rm CM} = \sum_k E_k/E_{\rm CM}$.\\

{\bf EXERCISE 8.2}~Consider the elastic, head-on collision, in which two particles of
(masses $m_a$ and $m_b$) approach one another traveling along the
$x$-axis, collide and emerge traveling along the same axis. In the
center-of-mass frame (by its definition) $\vec{p}_{a,i} = -
\vec{p}_{b,i} $. Use the conservation of 4-momentum to show 
that $\vec{p}_{a,f} = - \vec{p}_{a,i}$; that is, the momentum of particle $a$ (and likewise $b$) just reverses itself in the
  CM frame.\\

The invariant mass of two particles with 4-momenta $\bm{p}_a$ and $\bm{p}_b$ is defined by $m_{ab}^2 c^2 = (\bm{p}_a + \bm{p}_b)^2$. The invariant mass is particularly useful, for example, in finding the masses of short-live unsatble particles from the momenta of their observed decay products. Consider the decay of a particle $X \to a + b$. Since $\bm{p}_X = \bm{p}_a + \bm{p}_b$ it follows that
\begin{eqnarray}
m_X^2 c^2  \!\!& = & \! (\bm{p}_a + \bm{p}_b)^2 = \bm{p}_a^2 + \bm{p}_b^2 + 2 \bm{p}_a \cdot \bm{p}_b \nonumber \\
& = & m_a^2 c^2 + m_b^2 c^2 + 2 E_a E_b/c^2 - 2 \vec p_a \cdot \vec p_b \, . 
\end{eqnarray}
In a high energy experiment, the 3-momenta and masses of particles $a$ and $b$ must be measured. For charged particles this requires a magnetic field and tracking of the trajectory to measure the bending, as well as some means of particle identification (e.g. through the rate at which they lose energy in passing through matter). One must also identify the vertex and measure the opening angle. We discuss this next.

\subsection{Two-body decay of unstable particles}

The simplest kind of particle reaction is the two-body decay of unstable particles. Well known examples include decays of charged pions or kaons into muons and neutrinos, decays of neutral pions into two photons, and decays of neutral kaons into pairs of pions. The unstable particle is the mother particle and its decay products are the daughter particles. 
Consider the decay process $X \to ab$. In the CM frame for $a$ and $b$ the mother particle $X$ is at rest. Then its 4-momentum is $\bm{p}_X = (M c, 0,0,0)$. Denote the 4-momenta of the two daughter particles by $\bm{p}_a = (E_a/c, \vec p_a)$ and $\bm{p}_b = (E_b/c, \vec p_b)$. Conservation of 4-momentum requires that $\bm{p}_X = \bm{p}_a + \bm{p}_b$ and hence $\vec p_a = -\vec p_b$. We can therefore omit the subscript on the particle momenta and hence energy conservation takes the form
\begin{eqnarray}
E_a + E_b & = & \sqrt{m_a^2  c^4 + p^2 c^2} +  \sqrt{m_b^2  c^4 + p^2 c^2}  \nonumber \\ & = & M c^2 \, ,
\label{bugatuta1}
\end{eqnarray}
where $p^2 = \vec p \cdot \vec p$. Solving (\ref{bugatuta1}) for $p$ we get
\begin{equation}
p = c \frac{ \sqrt{[M^2 - (m_a -m_b)^2] [M^2 - (m_a +m_b)^2]}}{2M} \, .
\label{bugatuta2}
\end{equation} 
An immediate consequence of (\ref{bugatuta2}) is that
\begin{equation}
M \geq m_a + m_b \, ,
\end{equation}
i.e. a particle can decay only if its mass exceeds the sum of the masses of its decay products. Conversely, if some particle has a mass that exceeds the masses of two other particles, then this particle is unstable and decays unless the decay is forbidden by some conservation law, such as conservation of charge, momentum (as in exercise 8.7),  and angular momentum.

Another point to note is that the momenta of the daughter particles and hence also their energies are fixed by the masses of the three particles. Let us complete our calculation by deriving the formulae for the energies of the daughter particles. This is straightforward if we begin from the energy conservation formula (\ref{bugatuta1}) and express $E_b$ in terms of $E_a$, {\it viz}. $E_b = \sqrt{E_a^2 - m_a^2 c^4 + m_b^2 c^4}$, and then solve for $E_a$ to get
\begin{equation}
E_a = \frac{1}{2M} (M^2 + m_a^2 - m_b^2) c^2
\end{equation}
and similarly
\begin{equation}
E_b = \frac{1}{2M} (M^2 + m_b^2 - m_a^2) c^2\, .
\end{equation}
We also note that there is no preferred direction in which the daughter particles travel (the decay is said to be {\it isotropic}), but if the direction of one of the particles is chosen (e.g. by the positioning of a detector) then the direction of the second particle is fixed by momentum conservation: the daughter particles are travelling {\it back-to-back}  in the rest frame of the mother particle. 

\begin{figure}[tbp] \postscript{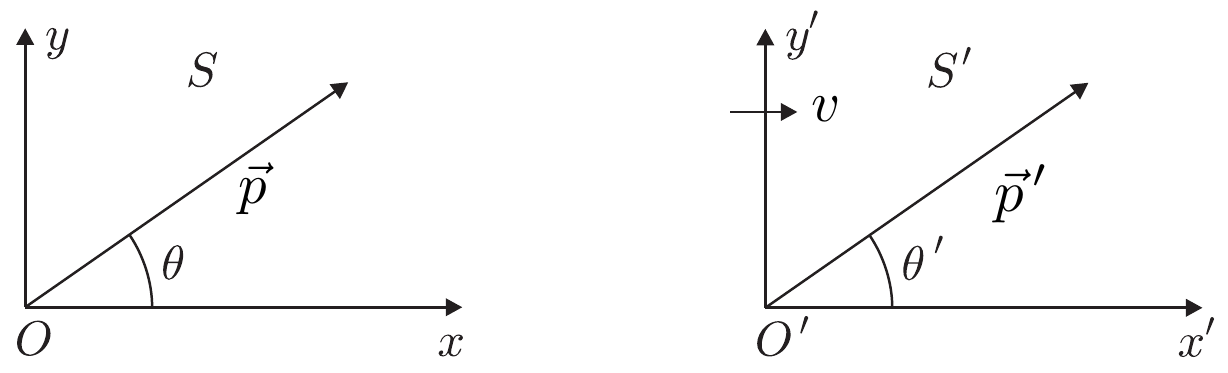}{0.85} \caption{Relations between angles.}
\label{fig:angles}
\end{figure}

Of interest is also the two-body decay of unstable particles in flight. For example, high energy beams of muons are produced in accelerators by letting the internal beam of protons impinge on a target of metal (thin foils or wires are used in practice) to produce pions and kaons which are then steered in a vacuum tube in which they decay into muons and neutrinos􏰛 Other cases of great interest are the decays of very short􏰚lived reaction products of high energy collisions such as for example the decays of $B$ mesons or of $D$ mesons, which are copiously produced in modern high energy colliders􏰛 To illustrate the importance of a proper discussion of such in-flight decays suffice it to say that this is frequently the only way to measure the mass of a neutral particle.

Therefore, we now have  the following 4-momenta of the three particles: for the mother particle we write $\bm{p}_X = (E/c,0,0,p)$ and for the daughter particles we have $\bm{p}_a = (E_a/c, \vec p_{a \perp}, p_{az})$ and $\bm{p}_b = (E_b/c, \vec p_{b \perp}, p_{bz})$. This means that we have chosen the $z$-axis along the direction of flight of the mother particle. The immediate consequence of this is that by momentum conservation the two-dimensional transverse momentum vectors are equal in magnitude and opposite in sign:
\begin{equation}
\vec p_\perp \equiv \vec p_{a\perp} =- \vec p_{b \perp} \, .
\end{equation}
The energies and the $z$ components of the particle momenta are related to those in the CM frame by a Lorentz boost with a boost velocity equal to the speed of the mother particle. We label  the kinematic variables in the CM frame with asterisks, and write the Lorentz transformation of particle in the form of
\begin{eqnarray}
E_a/c & = & \gamma (E_a^*/c + \beta p_{az}^*) \nonumber \\
p_{az} & = & \gamma (p_{az}^* + \beta E_a^*/c) \nonumber \\
\vec p_{a \perp} & = & \vec p_{a\perp}^{\, *} \,,
\end{eqnarray}
and similarly for particle $b$. Here $\beta = p c/E$ and $\gamma = E/(Mc^2)$.

This completely solves
the problem; e.g., we can find the angles which the two daughter particles make with the $z$-axis and with each other as functions of the momentum of the mother particle. To determine the relations between angles consider a particle with momentum $\vec p$ moving in the $x$-$y$ plane and making an angle $\theta$ with the $x$-axis in an inertial frame $S$. What is the corresponding angle $\theta'$ in the frame $S'$, which is moving with velocity $v$ along the $x$ axis? In $S$ we have
\begin{equation}
p^\mu = (E/c, p \cos \theta, p \sin \theta, 0) \,,
\end{equation}
whereas in $S'$ it follows that
\begin{equation}
p'^\mu = (E'/c, p' \cos \theta', p' \sin \theta', 0) \,,
\end{equation}
see Fig.~\ref{fig:angles}. Applying the $S \to S'$ Lorentz transformation we obtain
\begin{eqnarray}
p' \cos \theta' & = & \gamma^* (p \cos \theta - \beta^* E/c) \, , \nonumber \\
p' \sin \theta' & = & p \sin \theta \,,
\end{eqnarray}
so
\begin{equation}
\tan \theta' = \frac{p \sin \theta}{\gamma^* ( p \cos \theta - \beta^* E/c)} \,
\end{equation}
or
\begin{equation}
\tan \theta' = \frac{\sin \theta}{\gamma^* (\cos \theta - \beta^*/\beta)}\,,
\end{equation}
where $\beta^* = v/c$ is the velocity of $S'$ with respect to $S$ and $\beta = pc/E$ is that of the particle in $S$.
The inverse relation is found to be
\begin{equation}
\tan \theta = \frac{\sin \theta'}{\gamma^* (\cos \theta' + \beta^*/\beta')} \,,
\label{problema84}
\end{equation}
where $\beta' = p'c/E'$ is the velocity of the particle in $S'$.

\begin{figure*}[tbp] 
\begin{minipage}[t]{0.49\textwidth}
\postscript{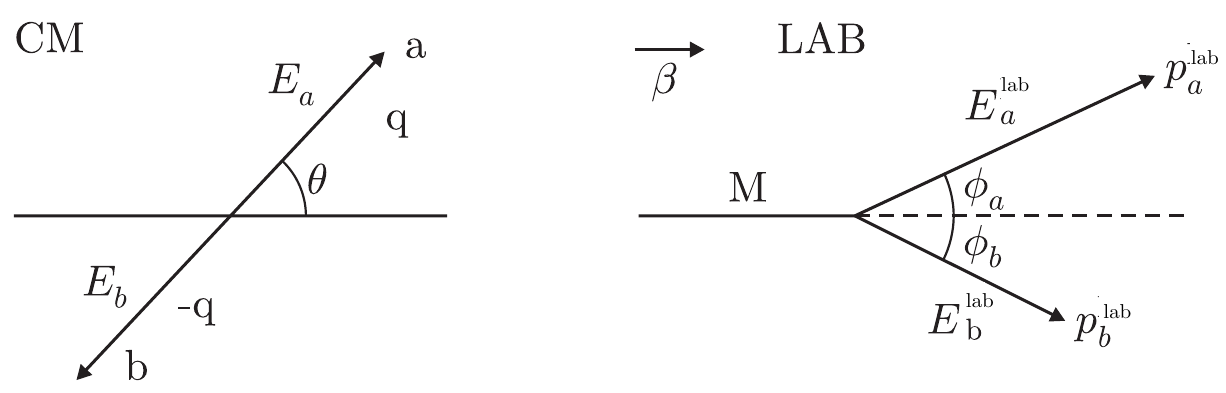}{0.99} 
\end{minipage}
\hfill
\begin{minipage}[t]{0.49\textwidth}
\postscript{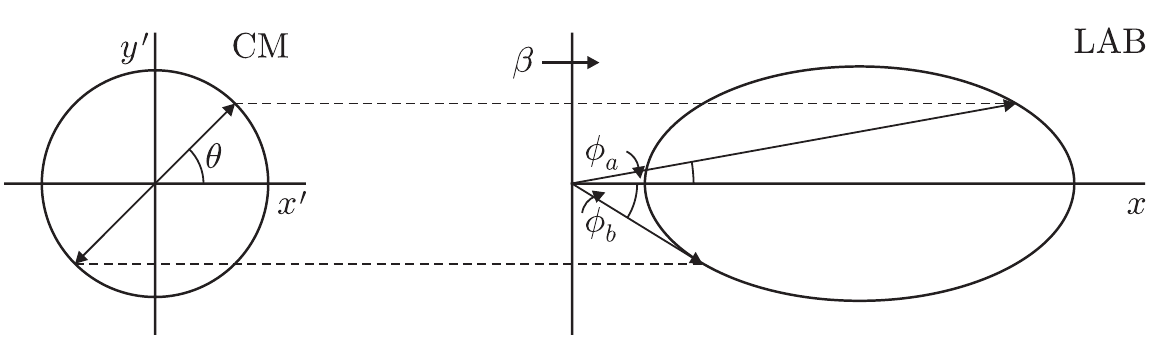}{0.99} 
\end{minipage}
\caption{Schematic representation of the angular distribution of fragments in the lab~\cite{Subir}.}
\label{fig:Subir}
\end{figure*}

Now, consider the situation exhibited in Fig.~\ref{fig:Subir}, in which the mother particle of mass $M$ is traveling with velocity $\beta = |pc|/E$ in the lab system and decays into particles $a$ and $b$. In the CM frame, particle $a$ has energy $E_a$ and momentum $\vec q$ at an angle $\theta$ with respect to the $x'$-axis. The particle momenta in the lab frame are then given by
\begin{eqnarray}
p^{\rm lab}_{a } \cos \phi_a & = & \gamma (q \cos \theta + \beta E_a/c) \,, \nonumber \\
p_{b }^{\rm lab} \cos \phi_b & = & \gamma (-q \cos \theta + \beta E_b/c)
\end{eqnarray}
We can use the inverse Lorentz transformation to obtain the corresponding variables in the CM frame from measured parameters in the lab.\\

{\bf EXERCISE 8.3} What is the opening angle between the two photons for a $\pi^0$ decay in flight?\\

{\bf EXERCISE 8.4} A particle of unknown mass $M$ decays into two particles of known masses $m_a = 0.5~{\rm GeV}/c^2$ and  $m_b = 1.0~{\rm GeV}/c^2$, whose momenta are measured to be $\vec p_a = 2.0~{\rm GeV}/c$ along the $x_2$ axis and $\vec p_b = 1.5~{\rm GeV}/c$ along the $x_1$ axis. Find the unknown mass $M$ an its speed.  \\

{\bf EXERCISE 8.5} A particle $X$ traveling along the positive axis of frame $S$ with
speed $0.5 c$ decays into to identical particles $X \to aa,$ both of
which continue to travel on the $x$ axis. {\it (i)}~Given that $m_X =
2.5 m_a$, find the speed of either $a$ particle in the rest frame of
particle $X$. {\it (ii)} By making the necessary transformation on the
result of part (i), find the velocities of the two $a$ particles in
the original frame $S$. \\

It is also of interest to approach the problem in a different way; namely, without using the Lorentz transformation. We then start from energy and momentum conservation
\begin{equation}
E = E_a + E_b  = \sqrt{m_a^2 c^4 + p_a^2 c^2} + \sqrt{m_b^2 c^4+ p_b^2 c^2},
\label{kevlin}
\end{equation}
and
\begin{equation}
\vec p = \vec p_a + \vec p_b \ ,
\end{equation}
respectively. Substituting in (\ref{kevlin}) $p_b^2$ by $(\vec p - \vec p_a)^2$ we obtain an equation with unknown momentum $p_a >0$ and angle $\theta_a$ between $\vec p_a$ and the $z$ axis. Solving for $p_a$ is a straightforward if lengthy calculation. After a little algebra we obtain
\begin{equation}
p_a  =  \frac{ (M^2 + m_a^2 - m_b^2) c^2 \ p \cos \theta_a \pm 2 E \ \sqrt{{\cal A}}}{2 (M^2 c^2 + p^2 \sin ^2 \theta_a)} \,,
\end{equation}
where
\begin{equation}
{\cal A} = M^2 p^{*2} - m_a^2 p^2 \sin ^2 \theta_a  .
\end{equation}
This result is of interest in the following sense. By demanding $p_a$ to be real we have ${\cal A} \geq 0$. This condition is satisfied for all angles $\theta_a$ if $Mp^*/(m_a p) >1$. In this case 
the negative sign in front of the square root must be rejected since otherwise we would get unphysical negative values of $p_a$ for $\theta_a > \pi/2$. On the other hand, if $Mp^*/(m_a p) <1$, there is a region of the parameter space in which both signs in the square root must be kept: for each value of $\theta_a < \theta_{a, {\rm max}}$ there are two values of $p_a$ and correspondingly also two values of $p_b$.

\subsection{Two-body scattering}
\label{Mandelstamsection}

In high energy physics, cross sections and decay rates are written
using kinematic variables that are relativistic invariants. For any
``two particle to two particle'' process $a b \to cd$, we have at
our disposal the four-momenta associated with each particle. The 
invariant variables are  six scalar products: $\bm{p}_a \cdot \bm{p}_b,$ $\bm{p}_a \cdot
\bm{p}_c$, $\bm{p}_a \cdot \bm{p}_d$, $\bm{p}_b \cdot \bm{p}_c$, $\bm{p_b} \cdot \bm{p_d}$, $\bm{p}_c \cdot \bm{p}_d$.  Rather than these, it is conventional to use the
related (Mandelstam) variables~\cite{Mandelstam:1958xc}
\begin{eqnarray}
s  & = & c^2 (\bm{p}_a + \bm{p}_b)^2 \,, \nonumber \\
t  & = & c^2 (\bm{p}_a - \bm{p}_c)^2 \,, \nonumber \\
u & = & c^2 (\bm{p}_a - \bm{p}_d)^2  \, . 
\label{eq:mandelstam}
\end{eqnarray}
Because $\bm{p}_i^2 = m_i^2 c^2$ (with $i = a,\, b,\, c,\, d$) and 
$\bm{p}_a + \bm{p}_b = \bm{p}_c + \bm{p}_d$ (due to energy momentum conservation) it follows that
\begin{eqnarray}
s + t + u & = & \sum_i m_i^2 c^4 + c^2 \left[2 \bm{p}_a^2 + 2 \bm{p}_a . (\bm{p}_b - \bm{p}_c - \bm{p}_d)\right] \nonumber \\
          & = & \sum_i m_i^2 c^4, 
\end{eqnarray}
i.e., only two of the three variables are independent. 

As an illustration, we take a look at M{\o}ller scattering: $e^-e^- \to e^-e^-$~\cite{Moller:1932}. In the CM frame the  3-momenta of the incoming particles satisfy $\vec p_a = - \vec p_b = \vec p_i$ and therefore $E_a = E_b= E = (p^2 c^2+ m_e^2c^4)^{1/2}$, with \mbox{$|\vec p_i|^2 =p^2$.} Conservation of 4-momentum yields $\vec p_c = - \vec p_d = \vec p_f$, with $|\vec p_f|^2 =p^2.$ Substituting the 4-momenta $\bm{p}_a = (E/c, \vec p_i)$, $\bm{p}_b = ( E/c, -\vec p_i)$, $\bm{p}_c = (E/c, \vec
p_f),$ $\bm{p}_d = (E/c, -\vec p_f),$ into (\ref{eq:mandelstam})  it follows that
\begin{eqnarray}
  s  & = & 4 (p^2 c^2 + m_e^2 c^4), \nonumber \\
  t  & = & -c^2 (\vec p_i - \vec p_f)^2  = -2p^2 c^2 ( 1 - \cos \theta^*), \nonumber \\ 
  u  & = & -c^2 (\vec p_i + \vec p_f)^2 = -2p^2 c^2 (1 + \cos  \theta^*),  
\end{eqnarray} 
where  $m_e$ is the electron mass and $\theta^*$
is the scattering angle, i.e., $\vec p_i \cdot \vec
p_f = p^2 \cos \theta^*$; recall that $\bm{p}_a \cdot \bm{p}_b = g_{\mu \nu} p^\mu \, p^\nu = (E/c)^2 + p^2$. 
Restriction to the physically allowed region yields
\begin{equation}
\left. \begin{array}{c}
-1 \leq \cos \theta^* \leq 1 \\
\vec p^{\, 2} \geq 0 \end{array} \right\} \Leftrightarrow \left\{ \begin{array}{c} 
- 4 c^2|\vec p \,|^2 \leq t \leq 0 \\
s \geq (m_a + m_b)^2 c^4 \end{array} \right. \, .
\end{equation}
In other words, as $p^2 \geq 0,$ we have $s \geq 4m_e^2c^4$; and
since $-1 \leq \cos \theta^* \leq 1$, we have $t\leq 0$ and $u \leq 0$.
Note that $t =0$ ($u=0$) corresponds to forward (backward) scattering.

In the CM frame of the reaction $ab \to cd$, $s$ is
equal to the square CM energy $E_{\rm CM}^2$, where
$E_{\rm CM}$ is the sum of the energies of particles $a$ and $b$, $t$
represents the square of the momentum transfer between particles $a$
and $c$, and $u$ (which is not an independent variable) represents the
square of the momentum transfer between particles $a$ and $d$. This is
called the $s$-channel process.  As we have seen, in the $s$-channel 
$s$ is positive, while $t$ and $u$ are negatives. The process is elastic if $m_a = m_c$ and $m_b=m_d$.\\

{\bf EXERCISE 8.6} Consider the scattering process $ab \to cd$. Determine $s$ in the lab frame in which particle $b$ is at rest and the projectile $a$ (beam) has an energy $E_a^{\rm lab}$. Use the invariance of $s$ to find the relation between $E_{\rm CM}$ and $E_a^{\rm lab}$. Show that for energies $\gg$ masses, $E_{\rm CM} \sim \sqrt{2 m_b c^2 E_a^{\rm lab}}$, so that the useful energy $E_{\rm CM}$ increases only as the square root of $E_a^{\rm lab}$. This is why modern particle accelerators are usually colliding beams rather than fixed-target machines.\\

Let us now take a closer look at more generic $ab \to cd$ processes. The CM frame is defined by
\begin{equation}
\vec p_a^{\,*} + \vec p_b^{\,*} =\vec 0 = \vec p_c^{\,*} + \vec p_d^{\,*} \, .
\label{bugaluni1}
\end{equation}
Note that (\ref{bugaluni1}) leads to the follwoing relations between the 3-momenta: $\vec p_a^{\, *} = - \vec p_b^{\, *} = \vec p_i$, $\vec p^{\, *}_c = - \vec p^{\, *}_d= \vec p_f$. The particle 4-momenta are then given by
$\bm{p}_a = (E^*_a/c, \vec p_i)$, $\bm{p}_b =(E_b^*/c, - \vec p_i)$, $\bm{p}_c = (E_c^*/c, \vec p_f)$, $\bm{p}_d = (E_d^*/c, - \vec p_f)$, where $E_a^* = \sqrt{\vec p_i^{\, 2} c^2 + m_a^2 c^4}$ and $E_b^* = \sqrt{\vec p_i^{\, 2} c^2 + m_b^2 c^4}$. After some algebra, we can express $E_{a,b}^*$, $|\vec p_i|$ and $|\vec p_f|$ in terms of $s = c^2 (\bm{p}_a + \bm{p}_b)^2 = (E_a^* + E_b^*)^2$ as follows:
\begin{equation}
E_{a,c}^* = \frac{1}{2 \sqrt{s}} \left( s + m_{a,c}^2 c^4 - m_{b,d}^2 c^4 \right)
\label{bugaluni2}
\end{equation}
and 
\begin{equation}
p_i^2  c^2 = E_a^{*2} - m^2_a c^4  = \frac{1}{4s} \lambda (s, m_a^2 c^4, m_b^2 c^4) \,,
\label{bugaluni3}
\end{equation}
where we have used the K\"allen (triangle) function which is defined by
\begin{eqnarray}
\lambda(a,b,c) & = & a^2 + b^2 + c^2 - 2ab -2ac -2bc \nonumber \\
& = & \left[a - (\sqrt{b} + \sqrt{c})^2 \right] \left[ a - (\sqrt{b} - \sqrt{c})^2 \right] \nonumber \\
& = & a^2 - 2 a(b+c) + (b-c)^2 \, .
\end{eqnarray}
Note that $\lambda$ is symmetric under $a\leftrightarrow b \leftrightarrow c$, and $\lambda (a,b,c) \to a^2$, for $a \gg b,c$. This enables us to determine some properties of scattering processes.  In the high energy limit, $s \gg m_i^2 c^4$,  (\ref{bugaluni2}) and (\ref{bugaluni3}) simplify because of the asymptotic behavior of $\lambda$ and we obtain
\begin{equation}
E_a^* = E_b^*=E_c^*=E_d^* = c |\vec p_i| = c |\vec p_f| = \sqrt{s}/2 \, .
\end{equation}

\begin{figure}[tbp] \postscript{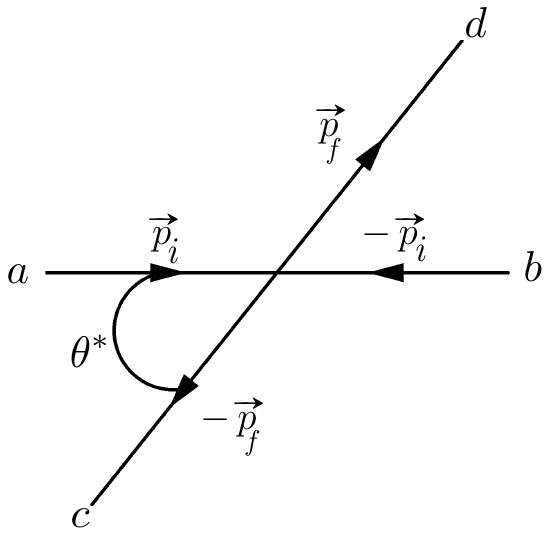}{0.85} \caption{Scattering angle.}
\label{fig:scatteringangle}
\end{figure}

In the CM frame, the scattering angle is defined by
\begin{equation}
\vec p_i \cdot \vec p_f = |\vec p_i| \cdot |\vec p_f| \cos \theta^* \, ,
\end{equation}
see Fig.~\ref{fig:scatteringangle}. Now, using
\begin{equation}
\bm{p}_a \cdot \bm{p_c} = E_a^* E_c^*/c^2 - |\vec p_a^{\, *}| |\vec p_c^{\, *}| \cos \theta^*
\end{equation}
and 
\begin{eqnarray}
t  & = & c^2 (\bm{p}_a - \bm{p}_c)^2 = (m_a^2 + m_c^2) c^4 - 2 c^2 \bm{p}_a \cdot  \bm{p}_c \nonumber \\
& = & c^2 (\bm{p}_b - \bm{p}_d)^2
\end{eqnarray}
we can write the scattering angle as a function of $s,$ $t,$ and  $m_{a,b,c,d}^2$,
\begin{equation}
\cos \theta^* = \frac{s (t-u) + (m_a^2 - m_b^2) (m_c^2 - m_d^2) c^8}{\sqrt{\lambda(s,m_a^2c^4, m_b^2c^4)} \sqrt{\lambda(s, m_c^2c^4 , m_d ^2c^4)} } \, .
\end{equation}
This means that $2 \to 2$ scattering is described by two independent variables: $\sqrt{s}$ and $\theta^*$, or $\sqrt{s}$ and $t$.\\

\begin{figure*}[tbp] \postscript{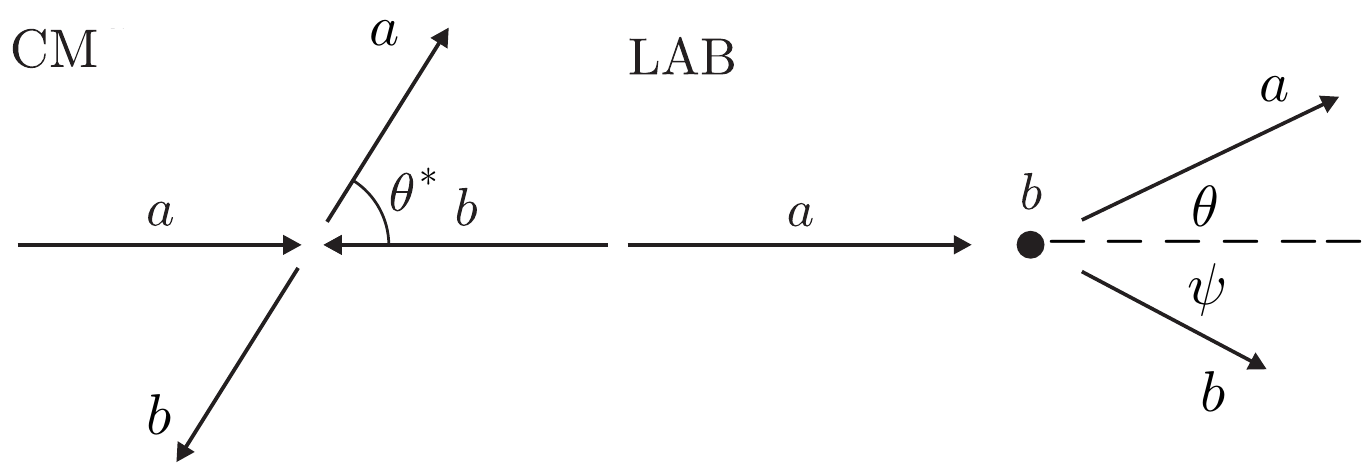}{0.85} \caption{Two-body scattering.}
\label{fig:2-2-scattering}
\end{figure*}

{\bf EXERCISE 8.7}~Consider the elastic collision shown in the Fig.~\ref{fig:2-2-scattering}. In the lab frame $S$, particle $b$ is initially at rest; particle $a$ enters with 4-momentum $\bm{p}_a$ and scatters through an angle $\theta$; particle $b$ recoils at an angle $\psi$. In the CM frame $S'$, the two particles approach and emerge with equal and opposite momenta, and particle $a$ scatters through an angle $\theta^*$. {\it (i)}~Show that the velocity of the CM frame relative to the lab frame is $\bm{v} = \vec p_a c^2/(E_a + m_b c^2).$ {\it (ii)}~By transforming the final moemntum $a$ back from the CM to the lab frame, show that
\begin{equation}
\tan \theta = \frac{\sin \theta^*}{\gamma  (\cos \theta^* + v/u^*_a)},
\label{tantheta}
\end{equation}
where $u^*_a$ is the speed of $a$ in the CM frame. {\it (iii)}~Show that in the limit that all speed are much smaller than $c$, this result agrees with the non-relativistic result
\begin{equation}
\tan \theta = \frac{\sin \theta^*}{\lambda + \cos \theta^*},
\end{equation}
where $\lambda = m_a/m_b.$ {\it (iv)}~Specialize now to the case $m_a = m_b$. Show that in this case, $v/u^*_a = 1$, and find a formula like (\ref{tantheta}) for tan~$\psi.$ {\it (v)}~Show that the angle between the two outgoing momenta is given by $\tan (\theta + \psi) = 2/ \beta^2 \gamma \sin \theta^*.$ Show that in the limit that $v \ll c$, you recover the well known nonrelativistic result that $\theta + \psi = 90^\circ$. 

\subsection{Threshold energy}

One way to create exotic heavy particles is to arrange a collision between two lighter particles
\begin{equation}
a + b \to d+ e + \dots + g
\end{equation}
where $d$ is a heavy particle of interest and $e, \dots, g$ are other possible particles produced in the reaction. 
In all such cases the theoretical minimum expenditure of energy occurs when all the end-products are mutually at rest. 
For consider, quite generally, two colliding particles: a projectile $a$ and a stationary  target $b$, with respective 4-momenta $\bm{p}_a$ and $\bm{p}_b$. If the emergent particles have 4-momenta $\bm{p}_i$ $(i = 1, 2,  \cdots)$, then
\begin{equation}
\bm{p}_a + \bm{p}_b =  \bm{p}_d + \bm{p}_e \cdots + \bm{p}_g = \sum_{i} \bm{p}_i \, .
\label{Rmuda}
\end{equation}
Squaring (\ref{Rmuda}), using (\ref{invariant-mass})  and (\ref{R615}), and adopting a self explanatory notation we find
\begin{equation} 
m_a^2 + m_b^2 + \frac{2 m_b E_a}{c^2} = \sum_i m_i^2 + 2 \sum_{(i<j)} m_i m_j \gamma(v_{ij}) \, .
\end{equation}
All the masses in this equation are fixed by the problem. The only variable on the left-hand-side is $E_a$, the energy of the projectile  relative to the rest-frame of the target, and therefore relative to the lab. The minimum of the right-hand-side evidently occurs when all the Lorentz factors are unity; that is, when there is no relative motion between any of the outgoing particles. So the threshold energy of the projectile is given by
\begin{equation}
E_a = \frac{c^2}{2 m_b} \left[\left(\sum_i m_i \right)^2 - m_a^2 -m_b^2 \right] \, .
\label{cientocuarentaydos}
\end{equation}
This formula applies even when the projectile is a photon that gets absorbed in the collision. 

As an illustration, we consider  the case of a free stationary proton (of mass $m_p$) being struck by a moving proton, whereupon not only the two protons but also a pion (of mass $m_\pi$) emerge. (Such reactions are often written in the form $p p \to p p \pi^0$.) The question is: what is the minimum (threshold) energy of the incident proton for this reaction to be possible? It is not simply $(m_p  + m_\pi) c^2$; that is, it is not enough for the proton's kinetic energy to equal the rest energy of the newly created particle. For, by momentum conservation, the post-collision particles cannot be at rest, and so a part of the incident kinetic energy must remain kinetic energy and thus be wasted. Particularizing (\ref{cientocuarentaydos}) to the case at hand,
\begin{equation}
E_p - m_p c^2 = c^2 \left(2m_\pi + \frac{m_\pi^2}{2 m_p}  \right)\, .
\label{boardgear1}
\end{equation}
The {\it kinematic} efficiency of this and all analogous processes can be defined as the ratio $k$ of the rest energy $m_\pi c^2$ of the new particle to the kinetic energy of the proton. When (\ref{boardgear1}) applies, we thus have
\begin{equation}
k = m_\pi \left( 2m_\pi + \frac{m_\pi^2}{2m_p} \right) = \frac{2}{4 + 
(m_\pi/m_p)} \, .
\label{boardgear2}
\end{equation}
The kinematic efficiency is thus always less than 50\%. In our particular example, $m_\pi/m_p \approx 0.14$ and $k \approx 48\%$. But when $m_d$ greatly exceeds the masses of all other particles involved, the details of the collision do not matter and (\ref{cientocuarentaydos}) yields the very unfavorable kinematic efficiency
\begin{equation}
k \approx 2 m_b/m_d \,  . 
\end{equation}
For example, when Richter~\cite{Augustin:1974xw} and Ting~\cite{Aubert:1974js} created the $J/\psi$  by colliding electrons with positrons, $k$ would have been $\sim 1/1850$. The way out of the difficulty was to use a method that is almost 100\% efficient: the method of colliding beams. Here both target and bullet particles are first accelerated to high energy (for example, electrons and positrons can be accelerated in the same sychrotron, in opposite senses), then accumulated in magnetic ``storage rings'', before being loosed at each other. No ``waste'' kinetic energy need be present after the collision, since there was no net momentum going in. For the process $e^+ e^- \to J/\psi$, with $m_b = m_e \approx 0.5~{\rm MeV}/c^2$ and $m_{J/\psi} \approx 3100~{\rm MeV}/c$, we find for the two threshold energies
\begin{equation}
E_{\rm CM} \approx m_{J/\psi} c^2 \approx 3100~{\rm MeV}
\end{equation}
whereas
\begin{equation}
E_{\rm lab} \approx \frac{m_{J/\psi}^2 c^2}{2 m_e} = 9600000~{\rm MeV} \, .
\end{equation}  

\vspace{0.25cm}

{\bf EXERCISE 8.8}~In 1928 Dirac predicted the existence of negative energy states of electrons when he developed his famous relativistic wave equation for massive fermions~\cite{Dirac:1928hu}. The antimatter character of these states became clear in 1933 after the discovery of the positron (the antielectron)  by Anderson~\cite{Anderson:1933mb}. The first positrons to be observed were created in electron-positron pairs by high energy cosmic ray photons in the upper atmosphere.
Show that an isolated photon cannot convert to an electron-positron pair in the process $\gamma \to e^+e^-$, because this process inevitably violates conservation of momentum. What actually occurs is that a photon collides with a stationary nucleus with the result $\gamma + \, {\rm nucleus} \, \to e^+ e^- + \, {\rm nucleus} \, .$ The nucleus is just a ``catalyst'' that can absorb some 3-momentum.\\

{\bf EXERCISE 8.9}~The fastest particles in the universe moving below the speed of light with respect to Earth are the highest energy cosmic rays. {\it Cosmic ray} is a general term for an elementary particle or an atomic nucleus propagating through the intergalactic medium. Protons are an abudant example. Cosmic rays are detected through the cascades they produce when they enter our atmosphere, and energies up to $E_{\rm CR} \sim 10^{11.5}~{\rm GeV}$ have been observed~\cite{Bird:1994uy}. For a proton, this corresponds to $\gamma \sim 10^{11}$ and a velocity of only a few parts in $10^{22}$ less than the speed of light. Shortly after the CMB was discovered~\cite{Penzias:1965wn}, Greisen~\cite{Greisen:1966jv}, Zatsepin, and Kuz'min~\cite{Zatsepin:1966jv}
 (GZK) noted that the relic photons would make the universe opaque to cosmic rays of sufficiently high energy. The microwave radiation field is isotropic with a blackbody spectrum in a frame called the CMB frame. The galaxies are moving only slowly compared to the speed of light with respect to this frame.  At a temperature of 2.726~K the characteristic energy of a CMB photon is $6 \times 10^{-4}~{\rm eV}$, and there are an average of 410 CMB photons per ${\rm cm}^3$. What happens when an ultrahigh energy cosmic ray proton collides with a CMB photon? If the proton is moving fast enough, the collision can initiate reactions like the photoproduction  of pions, $p\gamma_{\rm CMB}  \to n \pi^+$ and $p\gamma_{\rm CMB} \to p \pi^0$, that will degrade the proton's energy. Therefore, we would not expect to see cosmic ray protons above the photopion production energy threshold  if their source is distant enough that they would almost surely have collided with a CMB photon {\it en route} to Earth, the so-called ``GZK cutoff.'' The discovery of a suppression at the high energy end  of the cosmic ray spectrum was first reported by the HiRes~\cite{Abbasi:2007sv} and Pierre Auger~\cite{Abraham:2008ru} collaborations; by now the significance is well in excess of $20\sigma$ compared to a continuous power law extrapolation from lower energies~\cite{Abraham:2010mj}. This suppression is consistent with the GZK prediction. The GZK cutoff is a remarkable example of the profound links between different regimes of physics, connecting as it does the behavior of the rarest, highest-energy particles in the cosmos to the existence of Nature's most abundant particles -- the low energy photons in the relic microwave radiation of the Big Bang -- while simultaneously demanding the validity of special relativity over a mind-boggling range of scales. How energetic need a proton be to be above threshold for the GZK reaction? Show that the proton mean free path on the CMB is approximately 10 million light years. This is only a few times the size of the local group of galaxies.

\subsection{Transverse mass, rapidity, and pseudorapidity}

Let us now introduce some variables that are of common use in collider physics, which derive from the fact that in accelerators the incident velocities of the particles taking part in a collision are along the beam axis. This leads to the definition of invariants with respect to boosts to the rest frames of observers moving at different velocities parallel to the beam axis, or others that although they are not invariant have transformation properties that are easy to handle and useful for analysis. You may find a little strange that we are interested in quantities who are only invariant with respect to a set of observers whose velocities are all parallel to a single $z$-axis. What is special about these observers? In accelerators, one is often colliding particles whose momentum is not equal and opposite, but whose directions are down a common beam $z$-axis. In this case, the CM frame is moving at some velocity down the $z$-axis, so you will often wish to study physics in this frame. However, if you are stuck in the lab frame, you are boosted with some velocity $v_z$ with respect to this frame, and the direction of the boost is parallel to the beam axis.

The rapidity of a particle is defined as
\begin{equation}
y = \frac{1}{2} \ln \left(\frac{E + p_z c}{E- p_z c} \right) \, .
\label{rapidity}
\end{equation}
Why would you want to define such a quantity? Suppose we are dealing with a very high energy product of a collision, in the highly relativistic regime. Suppose now this particle is directed essentially in the $x$-$y$ plane, perpendicular to the beam direction. Then $p_z$ will be small, and the rapidity will be close to zero. Now, let the same highly relativistic particle be directed predominantly down the beam axis, say in the $+z$ direction. In this case, $E \simeq  p_zc$ and $y \to + \infty$. Similarly, if the particle is travelling down the beam axis, $E \simeq - p_zc$  and $y \to - \infty$. Therefore, the rapidity is zero when a particle is close to transverse to the beam axis, but tends to $\pm \infty$ when a particle is moving close to the beam axis in either direction. The rapidity is then related to the angle between the $x$-$y$ plane and the direction of emission of a product of the collision.

It turns out  that $E$ and $p_z$ can separately be expressed as functions of rapidity. To demonstarte this we  rewrite
the energy-momentum-mass relation (\ref{mcsq}) as
\begin{equation}
E^2 = M_T^2 c^4 + p_z^2 c^2 \,,
\label{TMasso}
\end{equation}
where 
\begin{equation}
M_T^2 c^4 = p_x^2 c^2 + p_y^2 c^2 + m^2 c^4 \, .
\end{equation}
is the ``transverse mass.''  Since the $x$ and $y$ components of momentum of a particle and its mass $m$ are all invariant with respect to boosts parallel to the $z$-axis, $M_T$ is a Lorentz invariant quantity. The quantity $(p_x^2 + p_y^2)^{1/2}$ is the invariant transverse momentum variable, usually denoted as $p_\perp$. Now, we rewrite (\ref{TMasso}) as
\begin{equation}
\left(\frac{E}{M_T c^2}\right)^2 - \left(\frac{p_z}{M_T c} \right)^2 = 1
\end{equation}
By comparing with the familiar relation between hyperbolic functions $\cosh^2 y - \sinh^2 y =1$ we can write the particle 4-momentum as~\cite{Beringer:1900zz}
\begin{equation}
 (E/c = M_T c \cosh y, \, p_x, \, p_y, \, p_z = M_T c \sinh y) \, .
\end{equation}

\vspace{0.25cm}

{\bf EXERCISE 8.10} {\it (i)}~There are various neat ways of writing the rapidity. Show that 
\begin{equation}
y = \ln \left(\frac{E + p_z c}{M_T c^2}\right)
\end{equation}
and 
\begin{equation}
y = \tanh^{-1} \left(\frac{p_z c}{E} \right).
\end{equation}
{\it (ii)}~Show that upon Lorentz transforming parallel to the beam axis with velocity $v = \beta c$, the equation for the transformation on rapidity is a particularly simple one,
\begin{equation}
y' = y - \tanh^{-1} \beta \, .
\label{rapLT}
\end{equation}

\vspace{0.25cm}

The particularly simple transformation law (\ref{rapLT}) has an important consequence. Suppose we have two particles ejected after a collision, and they have rapidities $y_1$ and $y_2$ when measured by some observer. Now, let some other observer measure these same rapidities, and obtain $y'_1$ and $y'_2$. The difference between the rapidities in the unprimed frame is $y_1 - y_2$, and in the primed frame it becomes
\begin{eqnarray}
y'_1 - y'_2 & = & y_1 - \tanh^{-1} \beta - y_2 + \tanh^{-1} \beta \nonumber \\
& = & y_1 - y_2 \, .
\end{eqnarray}
Thus, the difference between the rapidities of two particles is invariant with respect to Lorentz boosts along the $z$-axis. This is the key reason why rapidities are so crucial in accelerator physics.  Rapidity differences are invariant with respect to Lorentz boosts along the beam axis. Rapidity is often paired with the azimuthal angle $\phi$ at which a particle is emitted, so that the angle of emission of a particle from an interaction point is often given as the coordinate pair $(y, \phi)$. This way, the angular separation of two events, $(y_2 - y_1, \phi_2 - \phi_1)$ is invariant with respect to boosts along the beam axis. Histograms binned in either the angular separation of events or the rapidity separation of events can be contributed to by events whose CM frames are boosted by arbitrary velocities with respect to the rest frame of the detector, the lab frame. The resulting histograms are undistorted by these CM frame boosts parallel to the beam axis, as the dependent variable is invariant with respect to this sub-class of Lorentz boosts.

The only problem with rapidity is that it can be hard to measure for highly relativistic particles. You need both the energy and the total momentum, and in reality it is often difficult to get the total momentum vector of a particle, especially at high values of the rapidity where the $z$ component of the momentum is large, and the beam pipe can be in the way of measuring it precisely. However, there is a way of defining a quantity that is almost the same  as the rapidity, but it is much easier to measure than $y$ for highly energetic particles. This leads to the concept of pseudo–rapidity $\eta$.\\

{\bf EXERCISE 8.11}~Show that in the highly relativistic limit
\begin{equation}
y \simeq \eta = - \ln \left(\tan \frac{\theta}{2} \right) \, ,
\end{equation}
where $\theta$ is the angle made by the particle trajectory with
the beam pipe.\\

Pseudorapidity is particularly useful in hadron colliders, where the composite nature of the colliding protons means that interactions rarely have their CM frame coincident with the detector rest frame, and where the complexity of the physics means that $\eta$ is far quicker and easier to estimate than $y$. Furthermore, the high energy nature of the collisions mean that the two quantities may in fact be almost identical.\\

\acknowledgments{L.A.A. is supported by U.S. National Science Foundation (NSF) CAREER Award PHY1053663 and by the National Aeronautics and Space Administration (NASA) Grant No. NNX13AH52G; he thanks the Center for Cosmology and Particle Physics at New York University for its hospitality. Any opinions, findings, and conclusions or recommendations expressed in this material are those of the author and do not necessarily reflect the views of the NSF or NASA.}

%%%%%%%%%%%%%%%%%%%%%%%%%%%%%%%%%%%%

\onecolumngrid

\section*{Answers and Comments on the Exercises}

1.1~With our knowledge of Euclidean geometry a comparison of the measurements is straightforward: {\it (i)}~convert miles to meters (or vice versa); {\it (ii)}~distances computed with the Pythagorean theorem do not depend on which group does the surveying; {\it (iii)}~it is easily seen that ``daytime'' coordinates  can be obtained from nightime coordinates by a simple rotation. The moral of this parable is therefore: {\it (i)}~the same units should be used for all distances; {\it (ii)}~the (squared) distance is invariant; {\it (iii)}~different frames are related by rotations.\footnote{1 mile = 1,609.344 meters. The angle on the horizontal plane between magnetic north (the direction the north end of a compass needle points, corresponding to the direction of the Earth's magnetic field lines) and true north (the direction along a meridian towards the geographic North Pole)  varies depending on position on the Earth's surface, and changes over time.}\\

1.2~Call the direction of the flow of the canal the $x$ direction, and the direction straight across the canal the $y$ direction. Using the given distances it is easily seen that as the gondola cross the canal the angle with the vertical is $\theta = 21.8^\circ$. Equate the vertical component of the velocities to find the speed of the gondola relative to the shore, $v_{\rm gs} \cos \theta = v_{\rm gw} \sin 45^\circ$, where $v_{\rm gs}$ is the velocity of the gondola with respect to the shore and $v_{\rm gw}$ is the velocity of the gondola with respect to the water. We have $v_{\rm gs} = 38~{\rm cm/s}$. Equate the horizontal components of the velocities to find $v_{\rm gs} \sin \theta = v_{\rm gw} \cos 45^\circ - v_{\rm ws}$, where $v_{\rm ws}$ is the velocity of the water relative to the shore. The speed of the canal current is then $21~{\rm cm/s}$.\\

2.1~We already know that the actual path is a straight within one medium. Thus, the segments from $P_1$ to $Q$ and from $Q$ to $P_2$ are straight lines and the corresponding distances are $P_1Q = \sqrt{x^2 + y_1^2 + z^2}$ and $QP_2 = \sqrt{(x-x_1)^2 + y_2^2 + z^2}$. Then, the total time for the journay $P_1QP_2$ is $T=\left(\sqrt{x^2 +y_1^2 + z^2} + \sqrt{(x-x_1)^2 +y_2^2 +z^2}\right)/c$. To find the position of $Q = (x,0,z)$ for which this is a minimum we must differentiate with respect to $z$ and $x$ and set the derivatives equal to zero: $\frac{\partial T}{\partial z} = \frac{z}{c \sqrt{\cdots}} + \frac{z}{c \sqrt{\cdots}} = 0 \Rightarrow z = 0$, which says that $Q$ must lie in the same vertical plane as $P_1$ and $P_2$, and $\frac{\partial T}{\partial x} = \frac{x}{c \sqrt{\cdots}} + \frac{x-x_1}{c \sqrt{\cdots}} = 0 \Rightarrow \sin \theta_1 = \sin \theta_2$ or $\theta_1 = \theta_2$.\\

2.2~At the first refraction the ray is deviated through an angle $\theta_1 - \theta_2$, and at the second refraction it is further deflected through $\theta_4 - \theta_3$. The total deviation is then $\delta = (\theta_1- \theta_2) + ( \theta_4 - \theta_3)$. We need to show that $\theta_3 = \theta_2$ and so $\theta_4 = \theta_1$.
 Since the polygon $ABCD$ contains two right angles, the angle $BCD$ must be the supplement of the appex angle $\Phi$, see Fig.~\ref{fig:prisma-delta-min}. As the exterior angle to triangle $BCD$, $\Phi$ is also the sum of the alternate interior angles, that is $\Phi = \theta_2 + \theta_3$. Hence $\delta = \theta_1 + \theta_4 - \Phi$. What we would like to do now is write $\delta$ as a function of both the angle of incidence for the ray (i.e. $\theta_1$) and the prism angle $\Phi$. If the prism index is $n$ and it is immersed in air ($n_{\rm air} \approx 1$) it follows from Snell's law that $\theta_4 = \sin^{-1} (n \sin \theta_3) = \sin^{-1} [ \sin (\Phi - \theta_2)$. Upon expanding this expression, replacing $\cos \theta_2$ by $(1 - \sin \theta_2)^{1/2}$, and using Snell's law we have 
$\theta_4 = \sin^{-1} [(\sin \Phi) (n^2 - \sin^2 \theta_1)^{1/2} - \sin \theta_1 \cos \Phi]$. The deviation is then 
$\delta = \theta_1 + \sin^{-1} [(\sin \Phi) (n^2 - \sin^2 \theta_1)^{1/2} - \sin \theta_1 \cos \Phi] - \Phi$.
 It is evident that the deviation suffered by a monochromatic beam on traversing a given prism (i.e. $n$ and $\Phi$ are fixed) is a function only of the incident angle at the first face $\theta_1$. The smallest deviation can be determined analytically by differentianting the expression for $\delta(\theta_1)$ and then setting $d\delta/d\theta_1 =0$, but a more indirect route will certainly be simpler. Differentiating  $\delta(\theta_1,\theta_2)$ and setting it equal to zero we get $\frac{d \delta}{d \theta_1} = 1 + \frac{d \delta} {d\theta_2} =0$, or  $d \theta_2/d \theta_1 = -1$. Taking the derivative of Snell's law at each interface, we have $\cos \theta_1 d \theta_1 = n \cos \theta_2 d \theta_2$ and $\cos \theta_4 d \theta_4 = n \cos \theta_3 d \theta_3$. Note as well that, since $\Phi$ is the sum of alternate interior angles, $d\theta_2 = - d\theta_3$, because $d\Phi=0$. Dividing the expressions obtained for Snell's law and substituting for the derivatives it follows that $\cos \theta_1/\cos \theta_4 = \cos \theta_2 /\cos \theta_3$.  Making use of Snell's law once again, we can rewrite this as $\frac{1- \sin^2 \theta_1}{1-\sin^2 \theta_4} = \frac{n^2 - \sin^2 \theta_1}{n^2 - \sin^2 \theta4}$. The value of $\theta_1$ for which this is true is the one for which $d\delta/ d \theta_1 =0$. Provided $n \neq 1$, it follows that $\theta_1 = \theta_4$ and therefore $\theta_2 = \theta_3$. This means that the ray for which the deviation is a minimum traverses the prism symmetrically, i.e. parallel to the base.\\

2.3~At the first refraction, $1.00 \sin \theta_1 = n \sin \theta_2$. The critical angle at the second surface is given by 􏱳$n \sin \theta_3 = 1.00$, or $\theta_3 = \sin^{-1} (1.00/n)$; but $(\pi/2 - \theta_2) + (\pi/2 - \theta_3) + \Phi = \pi$, which gives $\theta_2 = \Phi - \theta_3$.
Thus, to have $\theta_3 < \sin^{-1} (1.00/n)$ and avoid total internal reflection at the second surface, it is necessary that $\theta_2 > \Phi - \sin^{-1} (1.00/n)$. Since $\sin \theta_1 = n \sin \theta_2$, this requirement becomes $\sin \theta_1 > n \sin [ \Phi - \sin^{-1} (1.00/n)]$, or $\theta_1 > \sin^{-1} \{ n \sin [ \Phi - \sin^{-1} (1.00/n)]\}$. Through the application of trigonometric identities, $\theta_1 > \sin^{-1} ( \sqrt{n^2 -1} \sin \Phi - \cos \Phi)$.\\

2.4~For the incoming ray, $n \sin \theta_2 = \sin \theta_1$. Then $(\theta_2)_{\rm violet} = \sin^{-1}(\sin 50.0^\circ/1.66) = 27.48^\circ$ and $(\theta_2)_{\rm red} = \sin^{-1}(\sin 50.0^\circ/1.62) = 28.22^\circ$. For the outgoing ray, $\theta_3 = 60.0^\circ - \theta_2$ and $\sin \theta_4 = n \sin \theta_3$. This leads to $(\theta_4)_{\rm violet} = \sin^{-1}( 1.66 \, \sin 32.52^\circ) = 63.17^\circ$ and $(\theta_4)_{\rm red} = \sin^{-1}( 1.62 \, \sin 31.78^\circ) = 58.56^\circ$. The angular dispersion is the difference $\Delta \theta_4 = (\theta_4)_{\rm violet} - (\theta_4)_{\rm red} = 63.17^\circ - 58.56^\circ = 4.61^\circ$.\\

3.1~From (\ref{venticinco}) we have ${\cal I}_{\rm tot} =  {\cal I}_{\rm max} \cos^2 \left(\frac{\pi d y}{\lambda L} \right)$, which leads to $y = \frac{\lambda L}{\pi d} \cos^{-1} \sqrt{{\cal I}_{\rm tot}/{\cal I}_{\rm max}}$. For the condition ${\cal I}_{\rm tot} = 0.75 \, {\cal I}_{\rm max}$, we obtain $y = 48~\mu{\rm m}$.\\

3.2~Let us first determine how time derivatives transform under a Galilean transformation. Since the position vector transforms according to (\ref{eq:Galileo}) we have ${\vec r}^{\,\prime} = \vec r - \vec u t = \vec r - \vec u t' \Rightarrow \vec r = {\vec r}^{\,\prime} + \vec u t'$, yielding $\frac{\partial}{\partial t'} = \frac{\partial \vec r}{\partial t'} \frac{\partial}{\partial \vec r} + \frac{\partial t}{\partial t'} \frac{\partial }{\partial t} = \vec u \cdot \vec \nabla + \frac{\partial}{\partial t}$, where $\vec u \cdot \vec \nabla = \sum_{i = 1}^3 u^i \frac{\partial}{\partial x_i}$. Since ${\vec \nabla}' = \vec \nabla$, in the absence of sources (\ref{Maxwell2}) and (\ref{Maxwell3}) transform to $\vec \nabla \times {\vec E}'  +  \frac{\partial}{\partial t } {\vec B}' + ( \vec u \cdot \vec \nabla) {\vec B}' =0$ and $\vec \nabla \cdot \vec B' = 0$, respectively.  Note that since (\ref{em-force}) does not depend on the acceleration, the statement that ${\vec F}^{\prime} = \vec F$ under Galilean transformation should hold. This means that $\vec E +   \vec v \times \vec B = {\vec E}' +  {\vec v}^{\, \prime} \times {\vec B}'$. By using the Galilean transformation for speed, ${\vec v}^{\, \prime} = \vec v - \vec u$, the previous equation can be rewritten as $\vec E +  \vec v \times \vec B = {\vec E}' +   \vec v \times {\vec B}' -  \vec u \times \vec {B}'$. By factorizing the second term in the l.h.s. and the second term in the r.h.s. we can further simplify the preceding equation to $\vec E + \vec v \times (\vec B - {\vec B}') = {\vec E}' - \vec u \times {\vec B}'$. Note that for any given $\vec u$ this equation has the solution ${\vec B}' = \vec B$ and ${\vec E}' = \vec E + \vec u \times \vec B$. By substituting the expressions for ${\vec B}'$ and ${\vec E}'$ in the transformed Maxwell equations in the absence of sources we get $\vec \nabla \cdot B = 0$ and $\vec \nabla \times \vec E +  \vec \nabla \times (\vec u \times \vec B) +  \frac{\partial}{\partial t} \vec B +  (\vec u \cdot \vec \nabla) \vec B = \vec \nabla \times \vec E +  \frac{\partial}{\partial t} \vec B =0$. This is because for $\vec u =$~constant, $\vec \nabla \times (\vec u \times \vec B) = \vec u (\vec \nabla \cdot \vec B) - (\vec u \cdot \vec \nabla ) \vec B = - (\vec u \cdot \vec \nabla) \vec B$. Hence, (\ref{Maxwell2}) and (\ref{Maxwell3}) are clearly invariant under Galilean transformations. 
Let us look now to the other pair of Maxwell's equations. By using the expressions which we derived above, we can write (\ref{Maxwell4}) as $\vec \nabla \times \vec B - \mu_0 \epsilon_0  \frac{\partial}{\partial t} {\vec E}' - \mu_0 \epsilon_0 (\vec u \cdot \vec \nabla) {\vec E}' = \vec \nabla \times \vec B - \mu_0 \epsilon_0 \frac{\partial}{\partial t} \vec E - \mu_0 \epsilon_0 (\vec u \cdot \vec \nabla) \vec E - \mu_0 \epsilon_0 \frac{\partial}{\partial t} (\vec u \times \vec B) - \mu_0 \epsilon_0 (\vec u \cdot \vec \nabla) (\vec u \times \vec B) =0$ and (\ref{Maxwell1}) as $\vec \nabla \cdot {\vec E}' = \vec \nabla \cdot \vec E +  \vec \nabla \cdot (\vec u \times \vec B) = \vec \nabla \cdot \vec E -  \vec u \cdot (\vec \nabla \times \vec B) =0$. Thus, for (\ref{Maxwell1}) the requirement of Galilean invariance leads to the condition that $\vec u \cdot (\vec \nabla \times \vec B) =0$
which is not true in the general case. Analogically reasoning can be used to show that (\ref{Maxwell4}) is not invariant under Galilean transformations.\\

4.1~For $L_1 = L_2= L$ and $\Delta t - \Delta t' = T = 2 \times 10^{-15}~{\rm s}$, from (\ref{time-changeMM}) it follows that $v = \sqrt{c^3 T/ (2L)} = 5 \times 10^4~{\rm m/s}$.\\

4.2~If there is a contraction by a factor of $(1 - v^2/c^2)^{1/2}$ in the direction of the \ae ther wind, from (\ref{treintaytres}) it follows that $t_1 = 2L_1  (1 - v^2/c^2)^{-1/2}/c$. Combining the round trip time for arm 1 with (\ref{treintaycuatro}) leads to the desired result $\Delta t = t_1 - t_2 = 2 (L_1-L_2) (1 - v^2/c^2)^{-1/2} / c\approx 2 (L_1 - L_2) [ 1 + (-1/2) (-v^2/c^2)] /c = 2 \Delta L [1 + v^2/(2c^2)]/c$. \\

4.3~Clearly, $\theta = \tan^{-1} \left(\frac{v\Delta t}{c \Delta t}\right) = \tan^{-1} \left(\frac{v}{c}\right)$. Now, for $|x| < 1$ we have $\tan^{-1} x = \sum_{n=0}^\infty (-1)^{n} \frac{x^{2n+1}}{2n+1}$, and so $\tan^{-1} (v/c) \approx v/c$. The Earth's orbital velocity is $v/c \approx 10^{-4}$, so $\theta \approx 10^{-4} (360^\circ/2\pi) (60'/1^\circ) (60''/1') \approx 20.6''$.\\

5.1~{\it (i)}~For a spacecraft speed of $1.1 \times 10^4~{\rm m/s}$, we have $1 - d/d' = 1 - \sqrt{1-v^2/c^2} = 7 \times 10^{-10}$, or $7 \times 10^{-8}\%$. {\it (ii)}~For a 1\% change, $\sqrt{1-v^2/c^2} =  0.99$, which gives $v = 0.14 c$. {\it (iii)}~For $v/c = 0.95$,  the Lorentz factor for both the outward and return trip is $\gamma = (1 - v^2/c^2)^{-1/2} = 3.20$. The times for the two halves of the journey satisfy $\Delta t_{\rm B}^{\rm out} = \gamma \Delta t_{\rm A}^{\rm out}$ and $\Delta t_{\rm B}^{\rm back} = \gamma \Delta t_{\rm A}^{\rm back}$, so by addition, the times for the whole journey satifty the same relation. Thus $\Delta t_{\rm A} = \Delta t_{\rm B}/\gamma = 25~{\rm yr}$, which is the amount by which twin $A$ has aged.\\ 

5.2~The half-life (measured in the muon rest frame) is $t_{1/2}^{\rm proper} = 2.2~\mu$s, and that measured in the Earth frame is $t_{1/2}^{\rm earth} = \gamma t_{1/2} ^{\rm proper}.$ The time of flight measured in the earth frame is $T^{\rm earth} = h/v$. For $v=0.99c$, we have $\gamma = 7.09$ and so the number of half-lives that have elapsed is $ n^{\rm rel} = \frac{T^{\rm earth}}{t_{1/2}^{\rm earth}} = \frac{h}{v \gamma t_{1/2}^{\rm proper}} =0.43$.  Hence, the number of muons that survive to the ground should be about $N^{\rm rel} = N_0 /2^n = 650/2^{0.43} = 482.$ To find the classical answer, we take $\gamma \to 1$, or equivalently $n^{\rm clas} = \gamma n^{\rm rel} = 3.04$ half-lives, and $N^{\rm clas} = 650/2^{3.04}
\approx 76.$ \\

5.3~The time interval $\Delta t'$ between ticks measured by Vinnie is $\Delta t' = d'/u' + d'/c$. The emitted particle, which travels at speed $u$ according to Brittany, reaches $F$ in an interval $\Delta t_1$ after traveling the distance $u \Delta t_1$ equal to the contracted length $d$ plus the additional distance $u \Delta t_1$ moved by the clock in the interval, i.e. $u \Delta t_1 = d + v \Delta t_1$.
In the interval $\Delta t_2$, the light beam travels a distance $c \Delta t_2$ equal to the length $d$
minus the distance $v \Delta t_2$ moved by the clock in that interval, i.e. $
c \Delta t_2 = d - v \Delta t_2$.
We now solve for $\Delta t_1$ and $\Delta t_2$, add to find the total interval $\Delta t$ between ticks according to Brittany, use the time dilation formula (\ref{t-dilation})  to relate this result to $\Delta t'$, and finally use the length contraction formula (\ref{l-contraction}) to relate $d$ to $d'$. After doing the algebra, we obtain the relativistic velocity addition law for velocity components that are in the direction of $v$,
\begin{equation}
u = \frac{u' + v}{1 + u' v /c^2} \, .
\label{217}
\end{equation}
We can also regard (\ref{217}) as a velocity transformation, enabling us to convert a velocity $u'$ measured by Vinnie to a velocity $u$ measured by Brittany.\\

\begin{figure}[tbp] \postscript{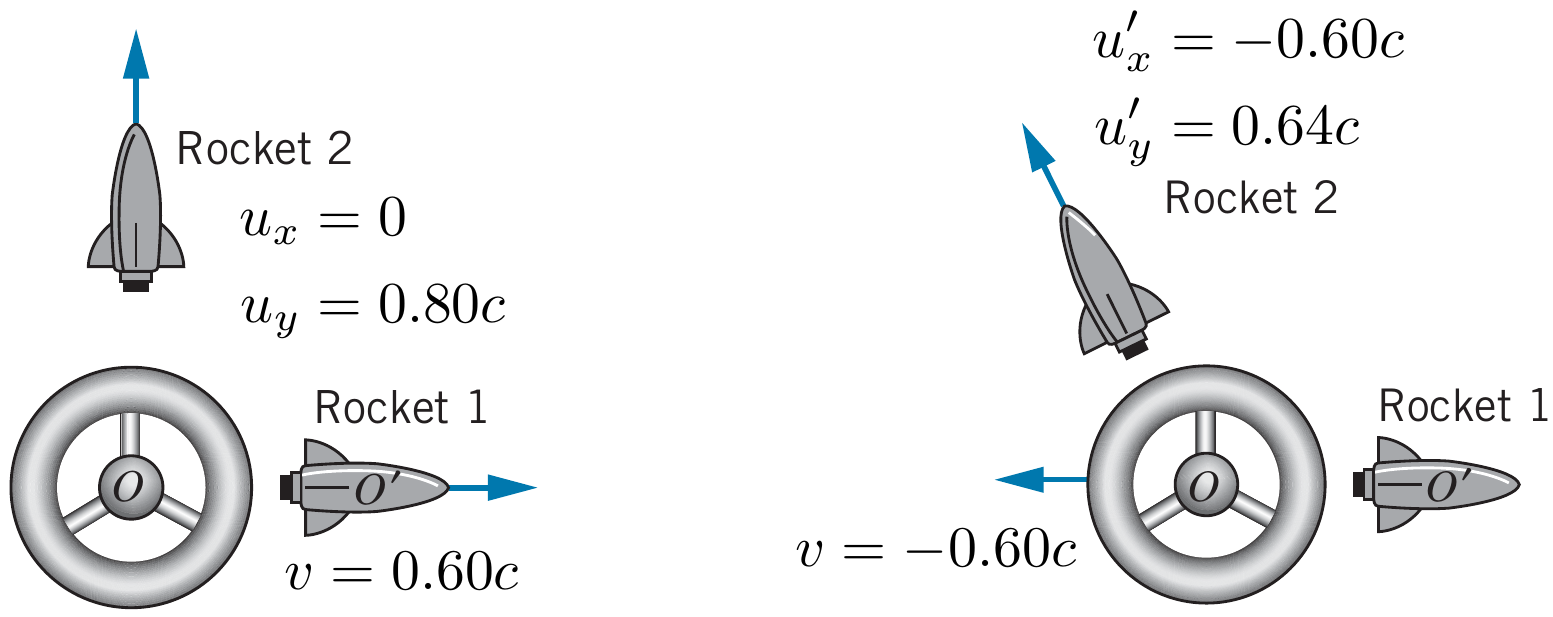}{0.85} \caption{On the left we show the situation in exercise 5.4. as viewed from reference frame of $O$ and on the right as viewed from the reference frame $O'$~\cite{Krane}.}
\label{Rocket-SS}
\end{figure}

5.4~Observer $O$ is the space station, observer $O'$ is rocket 1 (moving at $v = 0.60c$), and each observes rocket 2, moving (according to $O$) in a direction perpendicular to rocket 1. We take this to be the $y$ direction of the reference frame of $O$. Thus, $O$ observes rocket 2 to have velocity components $u_x = 0$, $u_y = 0.80c$, as shown in Fig.~\ref{Rocket-SS}.
The relationship between the velocities measured by $O$ and $O'$ is given by the Lorentz velocity transformation. In (\ref{217}) we derived the velocity transformation in the direction of $v$. Substituting each rocket's velocity we have
\begin{equation}
u'_x = \frac{u_x - v}{1- u_x v/c^2} = - 0.6c \, .
\end{equation}
We can derive the velocity transformations in the $y$-direction  from the Lorentz coordinate transformation. To this end we first differentiate the coordinate transformation $y' = y$ and obtain $dy' = dy$. Then we differentiate the time coordinate transformation, $dt' = [dt - (v/c^2) dx]/\sqrt{1- v^2/c^2}$.
Combining these two expressions we find
\begin{eqnarray}
u'_y & = & \frac{dy'}{dt'} = \frac{dy}{[dt - (v/c^2) dx]/ \sqrt{1-v^2/c^2}} = \sqrt{1-v^2/c^2} \frac{dy}{dt - (v/c^2) dx} 
  =  \sqrt{1 - v^2/c^2} \frac{dy/dt}{1 - (v/c^2) dx/dt} \nonumber \\
& =& \frac{u_y \sqrt{1 - v^2/c^2}}{1 - v u_x/c^2} = 0.64c\, .
\end{eqnarray}
Hence, according to $O'$, the situation looks like the cartoon in Fig.~\ref{Rocket-SS}. The speed of rocket 2 according to $O'$ is $\sqrt{0.60c)^2 + (0.64 c)^2 } = 0.88c$. According to the Galilean transformation $v'_y$ would be identical to $v_y$, and thus the speed would be $\sqrt{(0.6c)^2 + (0.8c)^2 } = c$. The Lorentz transformation  prevents relative speeds from reaching or exceeding the speed of light. \\

5.5~\underline{\it Setting up reference frames.}  We refer to two reference
frames: Road Runner's reference frame, which has velocity $= 0$
relative to the ground, and Wile E. Coyote's reference frame, which
has relativistic velocity $v$ relative to the ground.  Coyote's frame
has velocity $v$ relative to Road Runner's frame, as well.  Define the
origin of Road Runner's reference frame to be the left side of the
bridge and the origin of Wile Coyote's reference frame to be the left
side (back) of the train.  With these reference frames, we can find
position coordinates corresponding to where each bomb explodes in each
reference frame.

\begin{figure*}[tbp]
\begin{minipage}[t]{0.49\textwidth}
\postscript{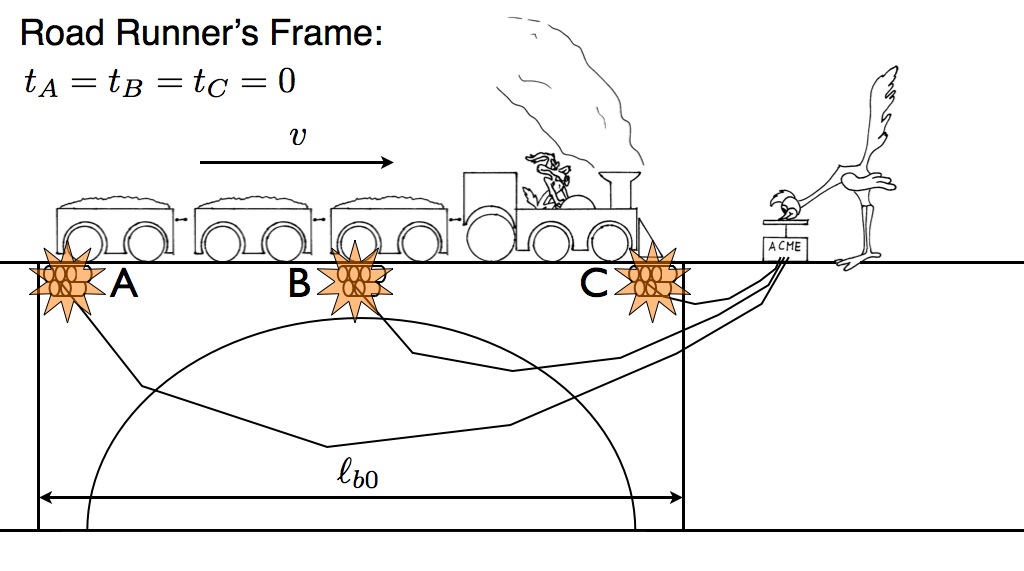}{0.99}
\end{minipage}
\hfill
\begin{minipage}[t]{0.49\textwidth}
\postscript{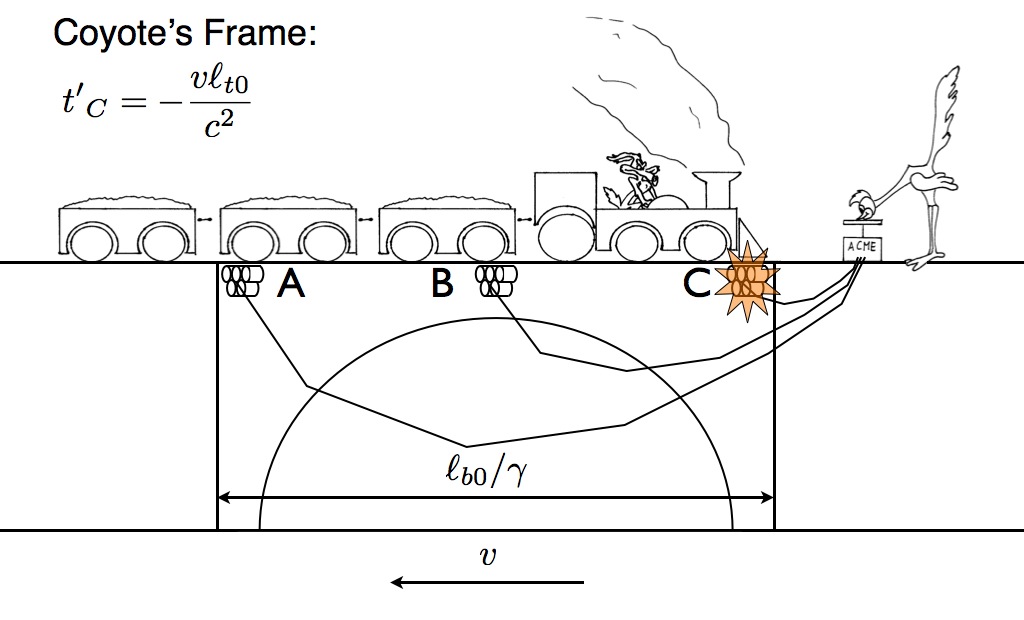}{0.99}
\end{minipage}
\begin{minipage}[t]{0.49\textwidth}
\postscript{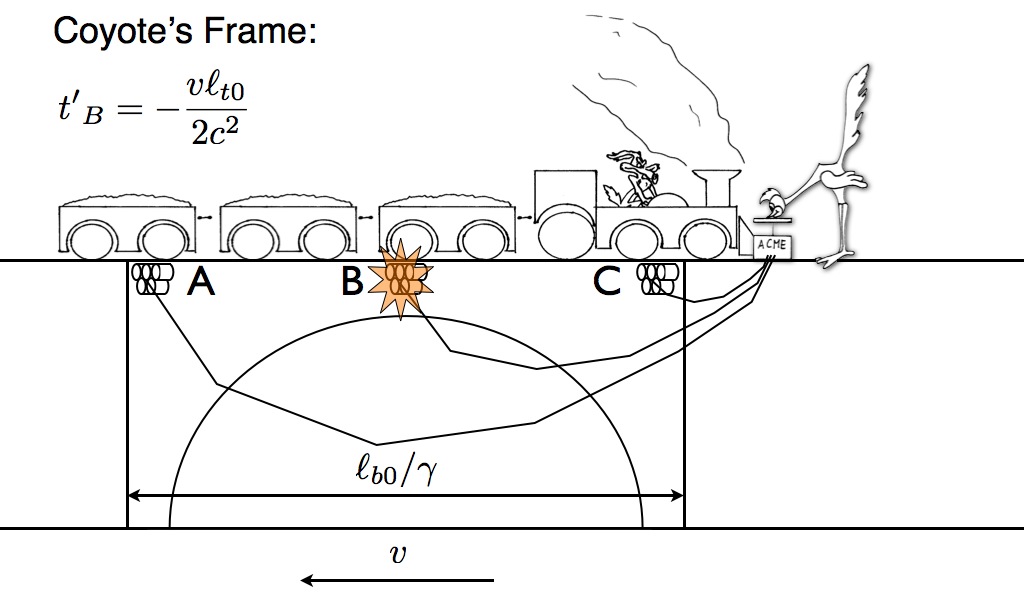}{0.99}
\end{minipage}
\hfill
\begin{minipage}[t]{0.49\textwidth}
\postscript{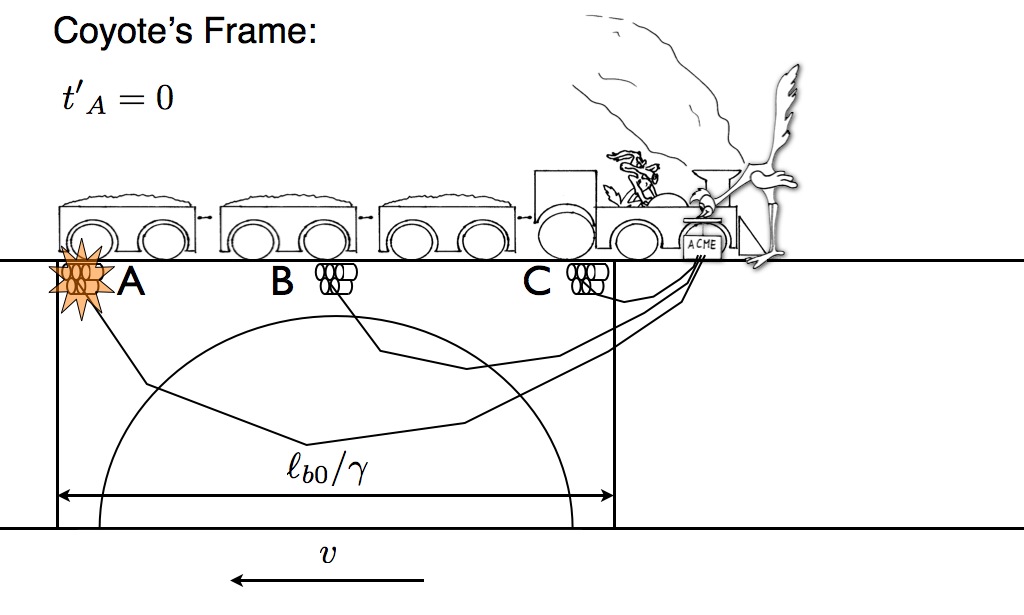}{0.99}
\end{minipage}
\caption{\label{RRC}{\bf Top left.} All three bombs explode simultaneously and
    destroy the train in the Road Runner's reference frame. {\bf Top right.} Snapshot of train when bomb $C$ explodes
    in the Coyote's reference frame. {\bf Bottom left.} Snapshot of train when bomb $B$ explodes in the Coyote's reference frame. {\bf Bottom right.} Snapshot of train when bomb $A$ explodes in
    the Coyote's reference frame. This figure is courtesy of Les Wade.}
\end{figure*}

\underline{\it Where bombs explode in each frame.} In Road Runner's frame, the three bombs (which we label $A,$ $B,$ and $C$) explode at positions: $x_A=0$, $x_B=\frac{\ell_{b0}}{2}$, and $x_C=\ell_{b0}$, where $\ell_{b0}$ is the proper length of the bridge. Since $\ell_{b0}$ is also the contracted length of the train $\ell_t$ in the Road Runner's frame (see Fig.~\ref{RRC}), we can relate the proper lengths of the bridge and the train, $\ell_{b0}$ and $\ell_{t0}$, via the formula for length contraction: $\ell_{t0} = \gamma\ell_t = \gamma\ell_{b0}$.

To find the position of each bomb upon exploding in Wile E. Coyote's frame, apply the following Lorentz transformation to each position-coordinate: $
x^\prime=\gamma\left(x-vt\right)$.
The Road Runner sets off all three bombs at once in his frame.  Let us define this as time $t=0$.  Thus, in Wile Coyote's frame, each bomb explodes at:
\begin{eqnarray}
x_A^\prime&=&\gamma\left(x_A-v\cdot t_A\right)=\gamma\left(0-v\cdot0\right)=0 \nonumber \\
x_B^\prime&=&\gamma\left(\frac{\ell_{b0}}{2}-v\cdot0\right)=\gamma\left(\frac{\ell_{b0}}{2}\right)=\frac{\ell_{t0}}{2} \nonumber \\
x_C^\prime&=&\gamma\left(\ell_{b0}-v\cdot0\right)=\gamma\ell_{b0}=\ell_{t0}  .
\end{eqnarray}
One might think that knowing {\it where} each bomb explodes on top of
knowing that Road Runner explodes each bomb simultaneously paints the
whole picture.  However, as we will soon find out, it is essential to
determine {\it when} each explodes in Wile Coyote's frame, as well.

\underline{\it When bombs explode in each frame.}  Again, the Road Runner
sets off all three bombs at time $t=0$ in his frame (see Fig.~\ref{RRC}).
Therefore, the time that each bomb explodes in Road Runner's frame is:
$t_A= t_B=t_C=0.$ To find the time that each bomb explodes in Wile
E. Coyote's frame, apply the following Lorentz transformation to each
time-coordinate: $t^\prime=\gamma\left(t-\frac{v}{c^2}x\right)$.
Thus, in Wile E. Coyote's frame, each bomb explodes at:
\begin{eqnarray}
t_A^\prime&=&\gamma\left(t_A-\frac{v}{c^2}x_A\right)=\gamma\left(0-\frac{v}{c^2}\cdot0\right)=0 \nonumber \\
t_B^\prime&=&\gamma\left(0-\frac{v}{c^2}\cdot\frac{\ell_{b0}}{2}\right)=-\gamma\left(\frac{v\ell_{b0}}{2c^2}\right)=-\frac{v\ell_{t0}}{2c^2} \nonumber \\
t_C^\prime&=&\gamma\left(0-\frac{v}{c^2}\cdot\ell_{b0}\right)=-\gamma\frac{v\ell_{b0}}{c^2}=-\frac{v\ell_{t0}}{c^2} \, .
\end{eqnarray}
Therefore, bomb $C$ explodes first, bomb $B$ explodes second, and bomb $A$ explodes last.
To summarize, in Road Runner's frame the components of $(ct,x)$ are given by $(0,0)_A$,  $(0,\ell_{b0}/2)_B$,  $(0,\ell_{b0})_C$, whereas  in Wile E. Coyote's frame we have $(0,0)_A$, $(- v\ell_{t0}/(2c), \ell_{t0}/2)_B$, $(-v\ell_{t0}/c,\ell_{t0})_C$. Armed with position and time coordinates, it is time to take on the
task of qualitatively explaining what is seen in Wile Coyote's frame.

\underline{\it Life in the fast lane.} In Fig.~\ref{RRC}
we outline what is seen in Wile Coyote's frame. Everything except for the train itself appears to be smaller to
someone in Wile Coyote's frame.  At time $t_C^\prime$, bomb $C$ explodes
at $x_C=\ell_{t0},$ which blows up the front of the train (see
Fig.~\ref{RRC}). At time $t_B^\prime$, bomb $B$ explodes at
$x_B=\ell_{t0}/2,$ which blows up the middle of the train
(see Fig.~\ref{RRC}).  At time $t_A^\prime$ bomb $A$ explodes at
$x_A^\prime=0,$ which blows up the back of the train (see
Fig.~\ref{RRC}).  Thus, no matter how fast the train is travelling, if
the train occupies the entire length of the bridge at one moment and
the Road Runner blows up the bridge at that moment in his frame, the train will
most definitely be blown up at the same spots on the train in both
reference frames.

\underline{\it Summary.} It turns out that the train gets blown up no
matter which reference frame you are observing the events in.  In
fact, if the Road Runner sees the train blow up in his reference
frame, it must be blown up in the Coyote's frame as well.  The only
difference is the simultaneity of the explosions.  In the Road
Runner's frame, bombs $A,$ $B,$ and $C$ blow up at exactly the same
time ($t=0$).  In Coyote's frame, however, these three events actually
happen at different times.\\

6.1~The $S$-$S'$ Lorentz transformation yields $t'_1 = [t_1 - (v/c^2) x_1]/(1-v^2/c^2)^{1/2}$ and $t'_2 = [t_2 - (v/c^2) x_2]/(1-v^2/c^2)^{1/2}$, hence $t'_2 - t'_1 = [(t_2 - t_1) - (v/c^2) (x_2 - x_1)]/(1 - v^2/c^2)^{1/2}$. In the  frame of reference $S$, the ``cause'' can travel to the ``effect'' with a speed that can be no greater than $c$; that is, $(x_2 -x_1)/(t_2 -t_1) \leq c$, or $(x_2 -x_1) \leq c (t_2 - t_1)$. Substituting for $(x_2 - x_1 )$ in the second term of the $\Delta t'$ relation, we obtain $t'_2 - t'_1 \geq (t_2 - t_1) (1-v/c)/(1- v^2/c^2)^{1/2} \geq 0$.\\

6.2~Lorentz transformations obey the linear transformation rule
and preserve the norm given by the metric tensor $g_{\mu \nu}$. From Eq.~(\ref{1}) we
see that $g_{00} = g_{\mu\nu}\Lambda^{\mu}_{0} \Lambda^{\nu}_{0} = 1 $.  Expanding $g_{\mu\nu}$ into its time and spatial components, we find that: $g_{00}\Lambda^{0}_{0} \Lambda^{0}_{0} + \displaystyle\sum_{i, j=1}^{3} g_{ij} \; \Lambda^{i}_{0} \Lambda^{j}_{0} = 1$.  This leads to 
\begin{equation}
 \left(\Lambda^0_0 \right)^2 = 1+ \displaystyle\sum_{i = 1}^{3} \left(\Lambda^i_0 \right)^2 .
\label{salt0}
\end{equation}
Now, consider a reference frame $O$ which is at rest (i.e.,
$d\vec{x}/dt = 0$) and a reference frame $O',$ which is moving with
velocity $v^i$ with respect to the frame $O.$ The Lorentz relation
between the two systems is $d{x'}^{\alpha} = \Lambda^{\alpha}_{\beta}
dx^{\beta}$.  For $\alpha = 0$, we have
$d{x'}^0 = \Lambda^0_0\, c\, dt + \Lambda^0_i\, dx^i$. 
Therefore, dividing the previous equation by $dt$ and using $d\vec{x}/dt = 0$, we obtain $dx'^{0} = \Lambda^0_0 \, c\, dt$,
or equivalently $dt' = \Lambda^0_0 dt$.
For $\alpha = i$, we have
$d{x'}^i = \Lambda^i_0 \, c \, dt$. 
Using the relations for $d{x'}^i$ and $dt'$ we first obtain
\begin{equation}
\frac{d{x'}^i}{dt'} = c \frac{\Lambda^i_0}{\Lambda^0_0} = v^i
\quad
{\rm and \ then } \quad
\Lambda^i_0 = -\frac{v_i}{c}\Lambda^0_0 \, .
\label{salt2}
\end{equation}
Substituting (\ref{salt2}) into (\ref{salt0}) we obtain
$\left(\Lambda^0_0\right)^2 = 1 + \displaystyle \sum_{i=1}^3 \left(\frac{v^i}{c}\right)^2 \left(\Lambda^0_0 \right)^2$,
and therefore $\Lambda^0_0 = \gamma$, $\Lambda^i_0 = v^i \gamma/c$ and $\Lambda^0_j = \gamma v_j/c$.
We still need to verify that
$\Lambda^i_j = \delta^i_j + v^i v_j (\gamma - 1)/v^2$. For events $x$ and $y$, the interval measured by observer $O$ is given by
$\Delta{s}^2 = (x^0 - y^0)^2 - \displaystyle\sum_{i=1}^{3} (x^i - y^i)^2$,
whereas in the moving frame $O'$, an observer measures
$\Delta{s'}^2 = (x'^0 - y'^0)^2 - \displaystyle\sum_{i=1}^{3} (x'^i - y'^i)^2$.
The Lorentz transformation that takes us to the moving frame is
given by $x'^{\mu} = \Lambda^\mu_\nu x^\nu$, so we can rewrite
$\Delta{s'}^2$ using the difference $x'^\mu - y'^\mu =
\Lambda^\mu_\nu (x^\nu - y^\nu)$ as $\Delta{s'}^2 = (\Lambda^0_\alpha)^2 (x^\alpha - y^\alpha)^2 - \displaystyle\sum_{l=1}^{3} (\Lambda^l_\mu)^2 (x^\mu - y^\mu)^2$.
Pulling the $\mu = 0$ part out of the first term and the $p=0$ part
out of the second term, we can see that this expression becomes
$\Delta{s'}^2  =   (\Lambda^0_0)^2 (x^0  - y^0)^2 +  (\Lambda^0_i)^2 (x^i - y^i)^2 - \displaystyle\sum_{l=1}^{3} (\Lambda^l_0)^2 (x^0 - y^0)^2 - \displaystyle\sum_{l=1}^{3}  (\Lambda^l_q)^2 (x^q-y^q)^2$.
We can now substitute into this expression the terms that were obtained for $\Lambda^0_0, \Lambda^0_j, \cdots$;
\begin{eqnarray}
\Delta{s'}^2    &= & \gamma^2  (x^0-y^0)^2 + \frac{\gamma^2 v_i^2}{c^2} (x^i - y^i)^2 - (x^0-y^0)^2 \displaystyle\sum_{l =1}^{3} \left( \frac{\gamma v_l}{c}\right)^2 - \displaystyle\sum_{l=1}^{3} (\Lambda^l_q)^2 (x^q - y^q)^2 \nonumber \\
    &= & \gamma^2  (x^0-y^0)^2 (1 - \frac{v^2}{c^2}) + \frac{\gamma^2 v_i^2}{c^2} (x^i - y^i)^2 - \displaystyle\sum_{l=1}^{3} \left( \delta^l_q + (\gamma -1) \frac{v^l v_q}{v^2} \right)^2 (x^q - y^q)^2
\end{eqnarray}
Recognizing that $\gamma^2 = (1- v^2/c^2)^{-1}$, we get
\begin{eqnarray}
\Delta{s'}^2 & = & (x^0-y^0)^2 + \displaystyle\sum_{i=1}^3\frac{\gamma^2
v_i^2}{c^2} (x^i - y^i)^2 - \displaystyle\sum_{l,q=1}^{3} \left( \delta^l_q
+ (\gamma -1) \frac{v^l v_q}{v^2} \right)^2 (x^q - y^q)^2
\label{deltasp}
\end{eqnarray}
where sums over spatial indices $i$ and $q$ are now explicit. Thus
we need to show that the terms that involve the spatial
coordinates of $x$ and $y$ amount to $-(x^1-y^1)^2- (x^2-y^2)^2 -
(x^3-y^3)^2$. By working out the expression in the last summation
we obtain
\begin{eqnarray}
    \displaystyle\sum_{l,q=1}^{3} \left( \delta^l_q + (\gamma -1)
    \frac{v^l v_q}{v^2} \right)^2 (x^q - y^q)^2 &=&
    \displaystyle\sum_{l,q=1}^{3} \left[ (\delta^l_q)^2 +
     2\delta^l_q \frac{v^l v_q}{v^2} (\gamma -1)
    + (\gamma-1)^2 \left(\frac{v^l v_q}{v^2}\right)^2\right] (x^q - y^q)^2 \nonumber \\
    &=& \displaystyle\sum_{l=1}^{3} \left[ 1 + (\gamma -1)^2 \frac{1}{v^2}v_l^2  +
    2 \frac{v_{l}^2}{v^2} (\gamma-1) \right](x^{l}-y^{l})^2\nonumber \\
    &=& \displaystyle\sum_{l=1}^{3} \left[ 1 + \frac{v_{l}^2}{v^2} (\gamma^2-1) \right]
    (x^{l}-y^{l})^2\, .
\end{eqnarray}
Thus the last two terms in the right hand side of Eq.~(\ref{deltasp}) can be
written as
\begin{eqnarray}
    \displaystyle\sum_{q=1}^3\left[
    \frac{\gamma^2 v_q^2}{c^2} -
    \displaystyle\sum_{l=1}^{3} \left( \delta^l_q + (\gamma -1)
    \frac{v^l v_q}{v^2} \right)^2\right] (x^q - y^q)^2
    &=& - \sum_{l=1}^3 \left[1 + \frac{v_l^2}{v^2}(\gamma^2-1) -
    \frac{\gamma^2}{c^2}v_l^2\right] (x^l-y^l)^2\, .
\end{eqnarray}
All that is left is to show that
\begin{equation}
  1 + \frac{v_l^2}{v^2}(\gamma^2-1) - \frac{\gamma^2}{c^2}v_l^2 = 1 \,,
\end{equation}
which is indeed true since
\begin{equation}
\frac{v_l^2}{v^2}(\gamma^2-1) - \frac{\gamma^2}{c^2}v_l^2 =
\frac{v_l^2}{v^2}(\gamma^2-1  - \gamma^2 \frac{v^2}{c^2}) =
\frac{v_l^2}{v^2}\left(\gamma^2 (1-\frac{v^2}{c^2}) - 1\right) = 0 \, .
\end{equation}
Therefore, $\Delta{s'}^2 = \Delta{s}^2,$
which proves that the transformation $\Lambda$ is an invariant of the interval, and
is therefore a Lorentz transformation.\\

6.3~Let us first write down the transformation rules for the components of the displacement:
$d\bar x^0 = \gamma (dx^0 - \beta dx^1)$, $d\bar x^1 = \gamma (dx^1 - \beta dx^0)$, $d\bar x^2 = dx^2$,   $d\bar x^3 = dx^3$. Since the proper time is invariant (your own clock goes the same, it does not matter whether you are in $S$ or $\overline S$), we have: $d\bar x^0/d\tau = \gamma (dx^0/d\tau - \beta dx^1/d\tau)$, $d\bar x^1/d\tau = \gamma (dx^1/d\tau - \beta dx^0/d\tau)$, $d\bar x^2/d\tau = dx^2/d\tau$,  $d\bar x^3/d\tau = dx^3/d\tau$.\\

6.4~{\it (i)}~If $x^\mu \, x_\mu > 0$ in the frame $S$, then  in any other frame $\overline S$ we have $\bar x^\mu \bar x_\mu >0$, because $x^\mu x_\mu$ is a Lorentz invariant. The condition $x^\mu x_\mu = x_0 x^0 - x_i \, x^i >0$ implies 
$|\vec x| < |x^0|$. Now,  assume  that $x^0 <0$ in the frame $S$ and that the associated value on $\overline S$ is obtained from $S$ by a boost in the $x$-direction.  Thus, $\bar x^0 = \gamma (x^0 - \beta x^1) <0$, since $|\beta| <1$ and $|x^1| \leq |\vec x| < |x^0|$. That is, $\bar x^0 <0$ in $\bar S$. Since $x^0$ is unchanged by any rotation, the same conclusion holds in all frames $\overline S$. {\it (ii)}~A point $x$ in spacetime lies on the forward light cone if and only if $x^\mu x_\mu =0$ and $x^0>0$. We have to show that if these two conditions hold in a frame $S$, they automatically hold in any other frame $\overline S$. Since $x^\mu \, x_\mu$ is Lorentz invariant, it follows that $\bar x^\mu \, \bar x_\mu =0$. To check the second condition, note that because $x^\mu x_\mu = x_0 x^0 - x_i x^i =0$ it follows that $|x^0| = |\vec x|$. Now supose $x^0 >0$ (in frame $S$) and let us consider a frame $\overline S$ related to $S$ by the standard boost in the $x$-direction.  In $\overline S$ we have $\bar x^0 = \gamma (x^0 - \beta x^1) >0$, because $|\beta| <1$ and $|x^1| \leq |\vec x| = x^0$. That is $\bar x^0>0$ in $\overline S$. Since $x^0$ is unchanged by any rotation, the same conclusion holds in all frames $\overline S$.\\

6.5~{\it (i)}~If $\bm{q}$ is time-like, then $\bm{q} \cdot \bm{q} = q_0 \, q^0 - q_i \, q^i >0$ which implies that $|\vec q| < |q^0|$. First rotate the coordinates so that $\bm{q}$ points along the $x$ axis and $\bm{q} = (q^0, q^1, 0, 0)$, with $|q^1| < |q^0|$. Now apply a boost in the $x$-direction to give $\bar q^1 = \gamma (q^1 - \beta q^0)$. We can choose $\beta = q^1/q^0$ (since $|q^1|<|q^0|$, this makes $|\beta| <1$, as it has to be) and then $\bar q^1 =0$ yielding $\bm{\bar q} = (\bar q^0, 0,0,0)$. {\it (ii)}~A vector $\bm{q}$ is forward time-like if and only if $\bm{q} \cdot \bm{q} >0$ and $q^0 >0$. The first condition is Lorentz invariant and implies that $|\vec q| < q^0$. Now assume that the second condition holds in a frame $S$ and imagine applying a boost in the $x$-direction so that, in the new frame $\overline S$, we have $\bar q^0 = \gamma (q^0 - \beta q^1)$. Now, $|\beta|<1$ and  $|q^1| \leq |\vec q| < q^0$. Therefore $\bar q^0 >0)$  and the second condition is valid in $\overline S$ too. This conclusion would certainly not be changed if we made any rotation of our coordinates, and since any Lorentz transformation can be built up from boosts and rotations, $\overline q^0>0$ in any inertial frame. Therefore, a vector that is forward time-like in one frame is forward time-like in all frames.\\

7.1~The force in the frame $S$ is $\vec F = d\vec p/dt$ and that in $S'$ is $\vec F' = d{\vec p}^{\, \prime}/dt'$.  To relate these we have to use the Lorentz transformation: $dp'_x = \gamma (dp_x - \beta dE/c)$, $dp'_y = dp_y$, $dp'_z= dp_z$, and $dt' = \gamma (dt - \beta dx/c)$. Therefore,
\begin{equation}
F'_x = \frac{dp'_x}{dt'} = \frac{\gamma (dp_x - \beta dE/c)}{\gamma (dt - \beta dx/c)} = \frac{F_x - \beta (dE/dt)/c}{1 -\beta u_x/c} = \frac{F_x - \beta \vec F \cdot \vec u/c}{1 - \beta u_x/c} \,, 
\end{equation}
where in the last equality we have used (\ref{vecfdotu}). Similarly
\begin{equation}
F'_y = \frac{dp'_y}{dt'} = \frac{dp_y}{\gamma (dt - \beta dx/c)} = \frac{F_y}{\gamma ( 1 -\beta u_x/c)} \,, 
\end{equation}
with a similar result for $F'_z$.\\

7.2~Consider that the light emitted in the galaxy's rest frame $S$ has a wave vector $k^\nu = (\omega/c, \vec k)$, with $\omega = 2 \pi \nu = 2\pi c/\lambda$, $\vec k = (k,0,0)$, and $|\vec k| = 2 \pi/\lambda$. On Earth we have ${k'}^\mu = {\Lambda^\mu}_\nu k^\nu = {\Lambda^\mu}_0 k^0 + {\Lambda^\mu}_1 k^1$. Then $\lambda' = 2 \pi c/\omega' = 2 \pi/{k'}^0 = 2 \pi/({\Lambda^0}_0 k^0 + {\Lambda^0}_1 k^1) = 2 \pi c/[({\Lambda^0}_0 + {\Lambda^0}_1) \omega] = \lambda/({\Lambda^0}_0 + {\Lambda^0}_1)$. If the galaxy is moving in straight line, chosing the axes $x$ and $x'$ in the direction of the movement we have
\begin{equation}
{\Lambda^\mu}_\nu = \left(\begin{array}{cccc} 
\gamma & - \gamma \beta & 0 & 0 \\
-\gamma \beta & \gamma & 0 & 0 \\
0 & 0 & 1 & 0 \\
0 & 0 & 0 & 1 \end{array} \right), \quad  \gamma = (1 - v^2/c^2)^{-1/2}, \quad \beta = v/c \, .
\end{equation}
Substituting the elements of the Lorentz transformation it follows that $\lambda' = \lambda/[\gamma (1 -\beta)]$, or equivalently $(1 - \beta)^2/(1 -\beta^2) = 1-\beta/ (1 + \beta) = (\lambda/\lambda')^2$. This leads to  $1-\beta = (\lambda/\lambda')^2 + (\lambda/\lambda')^2\beta$, or else $\beta = [1 - (\lambda/\lambda')^2]/[1 +(\lambda /\lambda')^2]$. Taking $\lambda = 4,870~\dot {\rm A}$ and $\lambda' = 7,300~\dot {\rm A}$, we obtain $\beta = 0.384$. If the velocity is constant then $v t = 5 \times 10^9 c~{\rm yr}$ and so $t = 5 \times 10^9\, c~{\rm yr}/v = 5 \times 10^9~{\rm yr}/\beta = 1.3 \times 10^{10}~{\rm yr}$.\\

8.1~{\it (i)}~A four vector $\bm{q}$ is forward time-like if and only if $|\vec q|< q^0$. The 4-momentum of a massive particle is defined by (\ref{4momento}), and since $m>0$ and $|u| < c$ we have $|\vec p|<p^0$, showing that $\bm{p}$ is forward time-like. {\it (ii)}~If $\bm{p}$ and $\bm{q}$ are forward time-like, $|\vec p| < p^0$ and $\vec q < q^0$. It follows that $|\vec p + \vec q| \leq |\vec p| + |\vec q| < p^0 + q^0 = (p+q)^0$. Therefore $\bm{p} + \bm{q}$ is forward time-like. {\it (iii)}~From {\it (i)} and {\it (ii)} it follows that the total 4-momentum $\bm{p}_{\rm tot}$ is forward time-like and so $|\vec p_{\rm \, tot}| < p^0_{\rm tot}$. By rotating the axes if necessary we can put $\bm{p}_{\rm tot}$ along the positive $x$ axis, so that $\bm{p}_{\rm tot} = (p^0_{\rm tot}, p^1_{\rm tot}, 0, 0)$ with $p^1_{\rm tot} , p^0_{\rm tot}$. Now consider a boost in the $x$ direction to a frame $\overline S$ in which
$\bar p_{\rm tot}^1 = \gamma (p_{\rm tot}^1 - \beta p_{\rm tot}^0)$. If we choose $\beta = p_{\rm tot}^1/p_{\rm tot}^0$ (which is less than 1 because $p^1_{\rm tot} < p^0_{\rm tot}$), then in the frame $\overline S$ the total 3-momentum is zero. {\it (iv)}~It is already clear from {\it (i)}, {\it (ii)}, and {\it (iii)} that in the original frame $S$ the velocity of the CM frame has to be given by $\beta_{\rm CM} = \vec p_{\rm tot}/p_{\rm tot}^0 = \sum_k \vec p_k c/\sum_k E_k$ and therefore $\gamma = \sum_k E_k/E_{\rm CM}$. \\

8.2~The conservation of the total  4-momentum, $\bm{p}_a + \bm{p}_b$, implies that the total energy $E_a + E_b$ and the total 3-momentum $\vec p_a + \vec p_b$ are conserved.  In the CM frame $\vec p_{a,i} = - \vec p_{b,i}$, so the two initial momenta are equal in magnitude, i.e. $|\vec p_{a,i}| =  |\vec p_{b,i}| = p_i$. By conservation of momentum, the final total momentum is also zero, so the same argument applies to the final momenta and $|\vec p_{a,f}| = |\vec {p_b,f}| = p_f$. Now the initial total energy is $E_ i  = \sqrt{(p_i c)^2 + (m_a c^2)^2} + \sqrt{(p_i c)^2 + (m_b c^2)^2}$, with a similar expression for $E_f$. Note that $E_i$ is a monotonically increasing function of $p_i$ (and likewise $E_f$).  Thus, conservation of energy ($E_f = E_i$) requires that $p_f = p_i$ and so that $\vec{p}_{a,f}  = - \vec{p}_{a,i}$. We conclude that in the CM frame the 3-momemntum of particle $a$ (and likewise $b$) simply reverses itself.\\

8.3~We have to determine the opening angle for the process $\pi^0 \to 2 \gamma$. 
In the CM frame, $E_a = E_b = qc$. We use (\ref{problema84}) twice to get
\begin{equation}
\tan \phi_a = \frac{\sin \theta}{\gamma (\cos \theta + \beta)} \quad {\rm and} \quad \tan \phi_b = \frac{\sin \theta}{\gamma (-\cos \theta + \beta)} \,,
\end{equation}
so
\begin{equation}
\tan (\phi_a + \phi_b) = \frac{\tan \phi_a + \tan \phi_b}{1 - \tan \phi_a \ \tan \phi_b }= \frac{2 \beta \gamma \sin \theta}{(\gamma^2 -1) \sin^2 \theta -1} ,
\end{equation}
where $\beta$ and $\gamma$ refer to the parent $\pi^0$ and $\theta$ is the angle shown in Fig.~\ref{fig:Subir}.\\

8.4~Using the useful relation (\ref{mcsq}) we can find the energies of the two final particles $E_a = \sqrt{p_a^2c^2 + m_a^2 c^4} = 2.06~{\rm GeV}$ and $E_b = \sqrt{p_b^2 c^2 + m_b^2 c^4} = 1.80~{\rm GeV}$. By conservation of energy and momentum, the original particle had $E = E_a + E_b = 3.86~{\rm GeV}$ and $\vec p = \vec p_a + \vec p_b = 2.5~{\rm GeV}/c$, at an angle of $53^\circ$ with the $x_1$-axis. Finally, $M = \sqrt{E^2 - p^2 c^2}/c^2 = 2.95~{\rm GeV}/c^2$ and $\beta = pc/E = 0.65$.\\

8.5~{\it (i)}~In the CM frame (the rest frame of the original particle $X$), the two final particles move with equal and opposite 3-momenta and equal energies, $E_{a1} = E_{a2} = m_X c^2/2 = 5 m_a c^2/4$. Thus either $a$ particle has 3-momentum of magnitude given by
$|\vec p_a| c = \sqrt{E_a^2 - m_a^2 c^4} = 3 m_a c^2 /4$, and speed $v_a = |\vec p| c^2 /E_a = 0.6 c$. {\it (ii)}~The two $a$ particles travel in opposite directions with velocities  $u' = \pm 0.6c$ along the $x$ axis of the CM frame, and the CM frame travels at speed $v=0.5c$ relative to the frame $S$. Therefore, the velocities relative to $S$ are given by the inverse velocity transformation as
\begin{equation}
u_{a1} = \frac{|u'| + v}{1 + |u'| v} = 0.85 c \quad {\rm and } \quad u_{a2} = \frac{-|u'| + v}{1 - |u'| v} = -0.14c \, .
\end{equation}  

\vspace{0.25cm}

8.6~In the lab frame we have $\bm{p}_a = ( E_a^{\rm lab}/c, \vec p)$ and $\bm{p}_b = (m_b c, \vec 0)$, which leads to 
$s = c^2 (\bm{p}_a + \bm{p}_b)^2 =  c^2 [(E_a^{\rm lab}/c + m_b c) ,  \vec p_a) \cdot  (E_a^{\rm lab}/c + m_b c, -  \vec p_a)] =  (E_a^{\rm lab} + m_b c^2)^2 - c^2 |\vec p_a|^{2}$. Now, using (\ref{mcsq})  we write  $(E_a^{\rm lab})^2 - |\vec p_a|^2 c^2 = m_a^2 c^4$, which leads to $s = (m_a^2 + m_b^2) c^4 + 2 m_b E_a^{\rm lab} c^2$.  By definition, in the CM frame we have $s = E_{\rm CM}^2$, and so it follows that $E_{\rm CM}^2 = (m_a^2 +m_b^2) c^4 + 2 m_b c^2 E_a^{\rm lab} $. If $E_a^{\rm lab} \gg m_{a,b}$, we obtain the desired result $E_{\rm CM} \sim \sqrt{2 m_b c^2 E_a^{\rm lab}}$.\\

8.7~In the lab frame $S$, the total 4-momentum is $\bm{p} = (E_a/c + m_b c, \vec p_a)$. The velocity in the CM frame $S'$ relative to $S$ is the velocity of a boost that makes $\vec p =0$ and this is easily seen to be $\vec v = \vec p_a c^2/(E_a + m_b c^2)$. {\it (ii)}~Let us denote the two final 4-momenta by $\bm{q}_a$ and $\bm{q}_b$ in the lab frame (and, of course, $\bm{q}_a^*$ and $\bm{q}_b^*$ in the CM frame). Then by the inverse Lorentz transformation $q_{ax} = \gamma (q^*_{ax} + v E_{q^*_a}/c^2)$ and $q_{ay} = q_{ay}^*$. Dividing the second of these equations by the first, we have
\begin{equation}
\tan \theta = \frac{q_{ay}}{q_{ax}} = \frac{q_{ay}^*}{\gamma (q_{ax}^* + v E_{q^*_a}/c^2)} =  \frac{|\vec q_a^{\,*}| \sin \theta^*}{\gamma (|\vec q^{\,*}_a | \cos \theta^* + v |\vec q^{\,*}_a| / u_a^*) }= \frac{\sin \theta^*}{\gamma (\cos \theta^* + v/u_a^*)} \,,
\label{Eesta}
\end{equation}   
where for the third equality we used the fact that $ \vec u^*_a = \vec q^{\,*}_a c^2 /E_{q^*_a}$ so that $E_{q^*_a} /c^2 = |\vec q^{\,*}_a|/u^*_a$. {\it (iii)}~In the nonrelativistic limit, $\gamma \to 1$. In addition, since $v = u^*_b$, we have $v/u^*_a = u^*_b/u^*_a$, which in the nonrelativistic limit is just the mass ratio $\lambda = m_a/m_b$. Hence, (\ref{Eesta}) becomes $\tan \theta = \sin \theta^*/(\lambda + \cos \theta^*)$. {\it (iv)}~As we have already seen, $v = u_b^*$, which in the case of equal masses also equals $u_a^*$, yielding $v/u^*_a =1$.  An argument parallel to that of part (ii) gives $\tan \psi = \sin \theta^*/\gamma (1- \cos \theta^*)$. {\it (v)}~For $m_a = m_b$, (\ref{Eesta}) reduces to  $\tan \theta = \sin \theta^*/[\gamma (1 + \cos \theta^*)]$. Combining these two results in the formula for $\tan (\theta + \psi)$ and performing a little algebra, we find that
\begin{equation}
\tan (\theta + \psi) = \frac{\tan \theta + \tan \psi}{1 - \tan \theta \tan \psi} = \frac{2}{\gamma \beta^2 \sin \theta^*} \, .
\end{equation}
In the nonrelativistic limit, $\gamma \to 1$ and $\beta \to 0$; therefore, $\tan (\theta + \psi) \to \infty$ and $(\theta + \psi) \to 90^\circ$.\\

8.8~The basic idea behind this problem is to show that in order for a
photon to decay into an electron and a positron, another particle must
be present to absorb some of the photon's momentum. Suppose that the process $\gamma \rightarrow e^+e^-$ occurs, with the
two particles moving at an angle $\theta$ away from the initial
trajectory of the photon. Conservation of four-momentum demands that $p_i^\mu = p_f^\mu$,  so that
\begin{eqnarray}
	p_i^\mu =
	\left(
		\begin{array}{c}
			E/c
			\\ E/c
			\\ 0
			\\ 0
		\end{array}
	\right) &=& 
	\left(
		\begin{array}{c}
			E_+/c
			\\ \cos \theta \ |\vec p_{+}|
			\\ \sin\theta \ |\vec p_{+}|
			\\ 0
		\end{array}
	\right)
	+ 
		\left(
		\begin{array}{c}
			E_-/c
			\\\phantom{-} \cos \theta \  |\vec p_{-}|
			\\ - \sin \theta \ |\vec p_{-}|
			\\ 0
		\end{array}
	\right) 
	= \bm{p}_+ + \bm {p}_- 
         =  p_f^\mu \, .
\end{eqnarray}
Equating components of these four-vectors, we see that  
$E  =  E_+ + E_- $,
$E/c  =  \cos \theta \ |\vec p_+| +  \cos \theta \ |\vec p_-|,$  
$0  =  \sin \theta \ (|\vec p_+| +|\vec p_-|)$.
For the latter relation to hold for any value of $\theta$, we require
$|\vec p_+|=|\vec p_-|$, so we may also write the first relation as 
$E/c = 2\, \cos\theta \ |\vec p_+|$.
Now, with (\ref{mcsq}) in mind  this means that
$
	E = E_{+} + E_{-} 
           =  \sqrt{|\vec p_{+}|^{2} c^2+m_e^{2} c^{4}} + \sqrt{|\vec p_{-}|^{2}c^{2}+m_e^{2}c^{4}} 
	= 2\sqrt{|\vec p_{+}|^{2}c^{2}+m_e^{2}c^{4}} \,,
$
 but using the expression of $E$ from above, we have that 
$
 	E  =   2\sqrt{|\vec p_{+}|^{2}c^{2}+m_e^{2}c^{4}} 
           =  2 c \cos \theta |\vec p_{+}|$,
yielding
$
\cos^2 \theta \ c^2 |\vec p_+|^2 =  |\vec p_{+}|^{2}c^{2}+m_e^{2}c^{4}$, or equivalently
$ \cos^2 \theta  =  1 + m_e^2c^2/|\vec p_+|^2$, that leads to
$\cos\theta  =  \sqrt{1+ m_e^2c^2/|\vec p_+|^2}$. Further, since $
	 \cos \theta = \sqrt{1+ m_e^2c^2/|\vec p_+|^2} \leq 1$
we have
$ 1+ m_e^2c^2/|\vec p_+|^2 \leq 1$.
 Indeed, $1+ m^2c^2/|\vec p_+|^2$ must \emph{always} be greater
 than 1 for any particles of non-zero mass. It follows that the
 process $\gamma \rightarrow e^+e^-$ cannot occur. That any process involving the decay of a single massless particle into any number of massive particles must be forbidden is easy to see even without elaborate algebra. If all particles in the final state are massive, then one can always transform to the CM frame of the final state particles where, by definition, the total 3-momentum of the final state is zero. However, no such ``rest frame'' exists for the massless particle in the initial state.\\

8.9~{\it (i)}~For definiteness consider $p\gamma_{\rm CMB} \to n\pi^+$. The total 4-momentum is conserved $\bm{p}_\gamma + \bm{p}_p = \bm{p}_n + \bm{p}_\pi$. The threshold energy is found most easily in the CM frame, where the 3-momenta of the colliding particles are equal and opposite. The threshold energy occurs when the initial energies are just enough to lead to a pion and a proton at rest. At threshold in the CM frame the total energy is $E_n^{CM} + E_\pi^{CM} = (m_n +m_\pi) c^2$. The total 3-momentum is zero by definition, $\vec p_n^{\rm \, CM} + \vec p_\pi^{\rm \, CM} = 0$. To what energy $E_p^{\rm CMB}$ does this correspond in the CMB frame where photons have typical energy $E_\gamma^{\rm CMB} \approx 6 \times 10^{-4}~{\rm eV}$? This question can be efficiently answered by utilizing the fact that the length of a 4-vector is a Lorentz invariant. Evaluating $(\bm{p}_n + \bm{p}_\pi)^2$ at threshold in the CM frame gives $(m_n +m_\pi)^2 c^2$. The conservation of 4-momentum means that this is the same as $(\bm{p}_\gamma + \bm{p}_p)^2$. Computing the square using  $\bm{p}_p^2 = m_p^2 c^2$ and $\bm{p}_\gamma^2 =0$ gives
\begin{equation}
\left(\bm{p}_\gamma + \bm{p}_p \right)^2 =
2 \bm{p}_\gamma \cdot \bm{p}_p + m_p^2 c^2 = (m_n + m_\pi)^2 c^2 \, .
\label{GZK1}
\end{equation}
 This relation does not depend on the frame but can be evaluated in terms of the components of the 4-mementa in the CMB frame. Assume that the proton with energy $E_p^{\rm CMB}$ is traveling along the $x$-axis to collide with a photon of energy $E_\gamma^{\rm CMB}$ traveling in the opposite direction. The CMB frame components are 
\begin{equation}
\left(p_p^{\rm CMB} \right)^\mu = \left(E_p^{\rm CMB}/c , \sqrt{(E_p^{\rm CMB})^2 - m_p^2 c^4}/c , 0 , 0 \right) \approx  \left(E_p^{\rm CMB}/c, E_p^{\rm CMB}/c \right) \, {\rm and} \, \left(p_\gamma^{\rm CMB}\right)^\mu = \left(E_\gamma^{\rm CMB}/c, - E_\gamma^{\rm CMB}/c, 0, 0\right), \nonumber
\end{equation}
 where the 3-momenta have been expressed in terms of energies using (\ref{mcsq}) and the approximation $E_p^{\rm CMB} \gg m_p c^2$. The scalar product (\ref{GZK1}) can be computed in terms of this components and the resulting relation solved for $E_p^{\rm CMB}$. The result simplifies using the approximation $m_n \approx m_p$ (more than adequate for present purposes) to give
\begin{equation}
E_p^{\rm CMB} \approx \frac{m_p m_\pi c^4}{2 E_\gamma^{\rm CMB}} \left( 1 + \frac{m_\pi}{2 m_p}\right) \approx 2.3 \times 10^{11}~{\rm GeV} \ .
\end{equation}
{\it (ii)}~The proton mean-free-path on the CMB is $\lambda = 1/(n_\gamma \sigma) \approx 10^{25}~{\rm cm} \approx 10^7~{\rm lyr}$, where $\sigma$ is the cross section and $n_\gamma$ the number density of CMB photons.\\

8.10~Taking the 1/2 into the logarithm and making it a power, you can write $y$ as
\begin{equation}
y = \ln \sqrt{\frac{E+ p_z c}{E- p_z c}} = \ln \left( \frac{E + p_z c}{\sqrt{E - p_z c} \sqrt{E + p_z c}} \right)  = \ln \left( \frac{E+ p_z c}{\sqrt{E^2 - p_z^2 c^2}}\right) = \ln \left(\frac{E + p_z c}{M_T c^2}\right) \, .
\end{equation}
The next expression for the rapidity is found using $\tanh \theta = (e^\theta - e^{-\theta})/(e^\theta + e^{-\theta})$; namely
\begin{eqnarray}
y & = & \tanh^{-1} \left\{\tanh \left[ \ln \left(\frac{E+ p_z c}{M_T c^2} \right)\right]\right\} = \tanh^{-1} \left[ \frac{\exp \left(\ln \frac{E+p_z c}{M_T c^2} \right) - \exp \left(- \ln\frac{E+ p_z c}{M_Tc^2}\right)}{\exp \left(\ln \frac{E+p_z c}{M_T c^2} \right) + \exp \left(- \ln\frac{E+ p_z c}{M_Tc^2}\right)} \right] = \tanh^{-1} \left( \frac{\frac{E+p_z c}{M_T c^2}  - \frac{M_Tc^2}{E+ p_z c}}{\frac{E+p_z c}{M_T c^2}  + \frac {M_Tc^2}{E+ p_z c}} \right) \nonumber \\
& = & \tanh^{-1} \left[\frac{\frac{(E+p_zc)^2 - M_T^2 c^4}{M_T c^2 (E + p_z c)}}{\frac{(E+ p_z c)^2 + M_T^2 c^4}{M_T c^2 ( E+ p_z c)}} \right] = \tanh^{-1} \left[\frac{(E + p_zc)^2 - M_T^2 c^4}{(E+p_zc)^2 + M_T^2 c^4}\right] = \tanh^{-1} \left(\frac{E^2 + 2 E p_z c + p_z c^2 - M_T^2 c^4}{E^2 + 2 Ep_z c + p_z c^2 + M_T^2 c^4} \right) \nonumber \\
& = & \tanh^{-1} \left(\frac{2E p_z c + 2 p_z^2 c^2}{2E^2 + 2 Ep_z c}\right) = \tanh^{-1} \left[\frac{p_z c (E+p_z c)}{E ( E + p_z c)} \right] = \tanh^{-1} \left(\frac{p_z c}{E} \right) \, .
\end{eqnarray}
{\it (ii)}~No we show how rapidity transforms under Lorentz boosts parallel to the $z$-axis. Start with (\ref{rapidity}) and perform a Lorentz boost on $E/c$ and $p_z$
\begin{eqnarray}
y' & = & \frac{1}{2} \ln \left(\frac{\gamma E/c - \beta \gamma p_z + \gamma p_z - \beta \gamma E/c}{\gamma E/c - \beta \gamma p_z - \gamma p_z + \beta \gamma E/c} \right) = \frac{1}{2} \ln \left[ \frac{\gamma (E/c+p_z) - \beta \gamma (E/c + p_z)}{\gamma(E/c - p_z) + \beta \gamma (E/c-p_z)}\right] = \frac{1}{2} \ln \left(\frac{E/c +p_z}{E/c -p_z} \ \frac{\gamma - \beta \gamma}{\gamma + \beta \gamma} \right)\nonumber \\
& = & \frac{1}{2} \ln \left(\frac{E+ p_z c}{E-p_z c} \right) + \ln \sqrt{\frac{1-\beta}{1+\beta}} = y + \ln \sqrt{\frac{1-\beta}{1+\beta}} \, .
\label{rapidity-lorentz}
\end{eqnarray}
This can be further simplified by noting that
\begin{eqnarray}
\ln \sqrt{\frac{1-\beta}{1 + \beta}} & = & \tanh^{-1} \left[\tanh \left(\ln \sqrt{\frac{1-\beta}{1 +\beta} } \right) \right] = \tanh^{-1} \left( \frac{\sqrt{\frac{1-\beta}{1+\beta} } -\sqrt{ \frac{1+\beta}{1-\beta}}} {\sqrt{\frac{1-\beta}{1+\beta} } +\sqrt{ \frac{1+\beta}{1-\beta}}}\right) = \tanh^{-1}  \left[\frac{(1-\beta) - (1 +\beta)}{(1-\beta) + (1+\beta)} \right]  \nonumber \\
& = & - \tanh^{-1} \beta \, .
\label{betarel}
\end{eqnarray}
Substituting (\ref{betarel}) into (\ref{rapidity-lorentz}) we obtain the desired result.\\

8.11~Using (\ref{mcsq}) we rewrite (\ref{rapidity}) as
\begin{equation}
y = \frac{1}{2} \ln \left( \frac{\sqrt{p^2 c^2 + m^2 c^4} + p_z c}{\sqrt{p^2 c^2 + m^2 c^4} - p_z c} \right) \, .
\end{equation}
Knowing that for a highly relativistic particle, $pc$ is far bigger than $mc^2$, we factor $pc$ out of each square root and use a binomial expansion to approximate what is left inside
\begin{equation}
y = \frac{1}{2} \ln \left[ \frac{pc ( 1 + \frac{m^2 c^4}{p^2 c^2})^{1/2} + p_z c}{pc ( 1 + \frac{m^2 c^4}{p^2 c^2})^{1/2} - p_z c} \right] \simeq \frac{1}{2} \ln \left[ \frac{pc + p_z c + \frac{ m^2 c^4}{2pc} + \cdots }{pc - p_z c + \frac{ m^2 c^4}{2pc} + \cdots } \right] = \frac{1}{2} \ln \left( \frac{1 + \frac{p_z}{p} + \frac{m^2c^4}{2 p^2 c^2} + \cdots}{1 - \frac{p_z}{p} + \frac{m^2c^4}{2 p^2 c^2} + \cdots} \right) \, .
 \label{nontrigo-p}
\end{equation}
Now, $p_z/p = \cos \theta$, where $\theta$ is the angle made by the particle trajectory with the beam pipe, and hence we have
\begin{equation}
1 + \frac{p_z}{p} = 1 + \cos \theta = 1 + [\cos^2 (\theta/2) - \sin^2 (\theta/2)] = 2 \cos^2 (\theta/2) \, .
\label{trigo-p1}
\end{equation}
Similarly 
\begin{equation}
1 - \frac{p_z}{p} = 1 - \cos \theta = 1 - [\cos^2 (\theta/2) - \sin^2 (\theta/2)] = 2 \sin^2 (\theta/2) \, .
\label{trigo-p2}
\end{equation}
Substituting (\ref{trigo-p1}) and (\ref{trigo-p2}) into (\ref{nontrigo-p}) we obtain the desired result
\begin{equation}
y \simeq \frac{1}{2} \ln \left[\frac{ \cos^2 (\theta/2)}{\sin^2 (\theta/2)} \right] \simeq - \ln \left(\tan \frac{\theta}{2} \right) \ . 
\end{equation}

\appendix

\section{Vectors, one-forms, and the metric tensor}
\label{appendixA}

A {\it vector} is a quantity with magnitude and direction. More specifically,
it is an element of a vector space $\vec{\mathbb {V}}$, that is to say: {\it(i)}~if $\vec V$ is a vector $\in \vec {\mathbb{V}}$ and $a$ a real number (scalar) then $a \vec V$ is a vector $\in \vec{\mathbb{V}}$ with the same direction (or the opposite direction, if $a<0$) whose magnitude is multiplied by $|a|$; {\it (ii)}~if $\vec V$ and $\vec W$ are vectors  $\in \vec{\mathbb {V}}$ then so is $\vec V + \vec W$; {\it (iii)}~the commutative $\vec V + \vec W = \vec W + \vec V$ and associative $\vec V + (\vec W + \vec Z) = (\vec V + \vec W) + \vec Z$ laws of vector addition are satisfied; {\it (iv)}~there exists an element $\vec 0 \in \vec{\mathbb {V}}$ such that $\vec V + \vec 0 = \vec V$; {\it (v)} to every element $\vec V \in \vec{\mathbb{V}}$ there corresponds an inverse element $-\vec V$, such that $-\vec V + \vec V = \vec 0$; {\it (vi)}~the associative law
of scalar multiplication $(a b) \vec V = a (b \vec V)$ and distributive laws of scalar multiplication $(a+b) \vec V = a \vec V + b \vec V$ and $a (\vec V +\vec W) = a \vec V + a \vec W$ are satisfied; {\it (vii)}~there exists the identity element of scalar multiplication $1$, such that $1 \vec V = \vec V$.

A {\it one-form} is defined as a linear scalar function of a vector. That is, a one-form takes a vector as input and outputs a scalar. For the one-form $\tilde P$, $\tilde P(\vec V)$ is also called the scalar product and may be denoted using angle brackets: $\tilde P(\vec V) = \langle \tilde P, \vec V\rangle$. The one-form is a linear function, meaning that for all scalars $a$ and $b$,  vectors $\vec V$ and $\vec W$, the one form satisfies the following relations
\begin{equation}
\tilde P (a \vec V + b \vec W) = \langle \tilde P, a \vec V + b \vec W \rangle = a \langle \tilde P, \vec V\rangle + b \langle \tilde P, \vec W \rangle = a \tilde P (\vec V) + b \tilde P (\vec W) \, .
\label{tensor2}
\end{equation}
Just as we may consider any function $f( )$ as a mathematical entity independently of any particular argument, we may consider the one-form $\tilde P$ independently of any particular vector $\vec V$. We may also associate a one-form with each spacetime point, resulting in a one-form field $\tilde P = \tilde P_{\bf x}$. Now, the distinction between a point and a vector is crucial: $\tilde P_{\bf x}$ is a one-form at point ${\bf x}$ while $\tilde P (\vec V)$ is a scalar, defined implicitly at point ${\bf x}$. The scalar product notation with subscripts makes this more clear: $\langle \tilde P_{\bf x}, \vec V_{\bf x} \rangle$. One-forms obey their own linear algebra distinct from that of vectors. Given any two scalars $a$ and $b$ and one-forms $\tilde P$ and $\tilde Q$, we may define the one-form $a\tilde P + b \tilde Q$ by
\begin{equation}
(a \tilde P + b \tilde Q) (\vec V) = \langle a \tilde P + b \tilde Q, \vec V \rangle = a \langle \tilde P , \vec V \rangle + b \langle \tilde Q, \vec V \rangle = a \tilde P (\vec V) + b \tilde Q (\vec V) \, .
\label{tensor3}
\end{equation}
Comparing equations (\ref{tensor2}) and (\ref{tensor3}), we see that vectors and one-forms are linear operators on each other, producing scalars. It is often helpful to consider a vector as being a linear scalar function of a one-form. Thus, we may write 
\begin{equation}
\langle \tilde P, \vec V\rangle = \tilde P (\vec V) = \vec V(\tilde P) \, .
\label{tensor4}
\end{equation}
The set of all one-forms is a vector space distinct from, but complementary to, the linear vector space of vectors. The vector space of one-forms is called the dual vector (or cotangent) space to distinguish it from the linear space of vectors (tangent space).

Having defined vectors and one-forms we can now define tensors. A {\it tensor of rank $(m, n)$}, also called an  $(m, n)$ tensor, is defined to be a scalar function of $m$ one-forms and $n$ vectors that is linear in all of its arguments. It follows at once that scalars are tensors of rank $(0, 0)$, vectors are tensors of rank $(1, 0)$ and one-forms are tensors of rank $(0, 1)$. The scalar product is a tensor of rank $(1,1)$ defined by (\ref{tensor4}).
We have seen that the scalar product (\ref{tensor4}) requires a vector and a one-form. Is it possible to obtain a
scalar from two vectors or two one-forms? From the definition of tensors, the answer is
clearly yes. Any tensor of rank $(0, 2)$ will give a scalar from two vectors and any tensor
of rank $(2, 0)$ combines two one-forms to give a scalar. However, there is a particular
$(0, 2)$ tensor field ${\bf  g_x}$ called the {\it metric tensor} and a related $(2, 0)$ tensor field ${\bf g_x}^{-1}$ called 
the inverse metric tensor for which special distinction is reserved. The metric tensor is a symmetric bilinear scalar function of two vectors. That is, given vectors $\vec V$ and $\vec W$, ${\bf g}$ returns a scalar called the {\it dot product}:
\begin{equation}
{\bf g}(\vec V, \vec W) = \vec V \cdot \vec W = \vec W \cdot \vec V = {\bf g}(\vec W, \vec V) \, .
\label{tensor7}
\end{equation}
Similarly, ${\bf g}^{-1}$ returns a scalar from two one-forms $\tilde P$ and $\tilde Q$, which we also call the dot product:
\begin{equation}
{\bf g}^{-1} (\tilde P, \tilde Q) = \tilde P \cdot \tilde Q = \tilde Q \cdot \tilde P = {\bf g}^{-1} (\tilde Q , \tilde P) \, .
\end{equation}
Although a dot is used in both cases, it should be clear from the context whether ${\bf g}$ or ${\bf g}^{-1}$ is implied. One of the most important properties of the metric is that it allows us to convert vectors to one-forms. If we forget to include $\vec W$ in equation (\ref{tensor7}) we get a quantity, denoted $\tilde V$, that behaves like a one-form:
\begin{equation}
\tilde V ( \cdot) \equiv {\bf g} (\vec V, \cdot) = {\bf g} (\cdot, \vec V) \,,
\end{equation}
where we have inserted a dot to remind ourselves that a vector must be inserted to give a scalar. In summary, the metric tensor ${\bf g}$ is a mapping from the space of vectors to the space of one-forms, ${\bf g}: \vec {\mathbb {V}} \to \tilde{ \mathbb {V}}$. By definition, the inverse metric ${\bf g}^{-1}$ is the inverse mapping, ${\bf g}^{-1}: \tilde{\mathbb {V}} \to \vec{\mathbb{V}}$. (The inverse always exists for nonsingular spaces.) 

We will now give  concret examples that show how this abstract machinery really works. For a 3-dimensional Euclidean space, ${\bf g} \equiv g_{ij} = {\rm diag} (1,1,1)$, whereas for Minkowski spacetime ${\bf g} \equiv {\rm diag} (1,-1,-1,-1)$. Then, the dot product of 3-dimensional vectors $\vec v$ and $\vec w$ is 
\begin{equation}
g_{ij}  \, v^i \, w^j = v_j \, w^j = v_1 \, w^1 + v_2 \, w^2 + v_3 \, w^3 \, .
\end{equation}
and the dot product of 4-dimensional vectors $\bm{v}$ and $\bm{w}$ is
\begin{equation}
g_{\mu \nu} \, v^\mu \, w^\nu = v_\nu \, w^\nu = v_0 \, w^0 + v_1 \, w^1 + v_2 \, w^2 + v_3 \, w^3 \, .
\end{equation}
Note that in the Euclidean space $v_i = v^i$, and in Minkowski spacetime $v_0 =v^0$ but $v_i = - v^i$.

For a comprehensive discussion on tensor calculus see e.g.~\cite{Misner:1974qy,Carroll:1997ar}.

\end{document}